\documentclass[1p]{elsarticle}

\usepackage{supertabular}
\usepackage{longtable}
\usepackage{booktabs}
\usepackage{changes}
\usepackage{hyperref}
\usepackage[T1]{fontenc}

\newcommand{\code}[1]{\texttt{#1}}

\newcommand{\EE}[2]{\ensuremath{{#1}\times 10^{#2}}}
\newcommand{\EEerr}[3]{\ensuremath{{#1}\pm{#2}\times 10^{#3}}}

\newcommand*{\starlib}{\code{STARLIB} \cite{iliadisChargedparticleThermonuclearReaction2010b}}

\graphicspath{{./}{}}

\begin{document}


\title{Nuclear Physics of X-ray Bursts}

\author[1]{Y. Xu}
\ead{yi.xu@eli-np.ro}

\author[2,3]{H. Schatz}
\ead{}

\author[4] {R. Lau}
\ead{kityulau.rita@gmail.com}

\author[5]{Z. Meisel}
\ead{zachary.meisel@us.af.mil}

\author[6]{P. Mohr}
\ead{mohr@atomki.hu}

\affiliation[1]{organization={Extreme Light Infrastructure - Nuclear Physics (ELI-NP), "Horia Hulubei" National Institute for R\&D in Physics and Nuclear Engineering (IFIN-HH)},
addressline={30 Reactorului Street},
city={Magurele},
postcode={077125},
state={Ilfov},
country={Romania}}

\affiliation[2]{organization={Facility for Rare Isotope Beams,
Michigan State University},
addressline={South Shaw Lane},
city={East Lansing},
postcode={48864},
state={MI},
country={USA}}

\affiliation[3]{organization={Department of Physics and Astronomy,
Michigan State University},
addressline={South Shaw Lane},
city={East Lansing},
postcode={48864},
state={MI},
country={USA}}

\affiliation[4]{organization={HKU SPACE PO LEUNG KUK STANLEY HO COMMUNITY COLLEGE},
addressline={66 Leighton Road, Causeway Bay},
city={Hong Kong},
country={Hong Kong}}

\affiliation[5]{organization={Department of Engineering Physics, Air Force Institute of Technology},
city={Wright-Patterson Air Force Base},
postcode={45433},
state={OH},
country={USA}}

\affiliation[6]{organization={HUN-REN Institute for Nuclear Research (ATOMKI)},
addressline={P.O.\ Box 51},
city={Debrecen},
postcode={H-4001},
country={Hungary}}

\begin{abstract}
Thermonuclear X-ray bursts from the surface of accreting neutron stars are the most common astrophysical explosions in our galaxy. They provide a unique window into the physics of neutron stars, the physics of matter under extreme conditions, and the physics of astrophysical thermonuclear explosions. X-ray bursts are powered by a broad range of nuclear reactions that need to be understood to interpret observations. The relevant nuclei are mostly neutron deficient and unstable, and thus experimental information and theoretical understanding is limited and an active area of research in nuclear science. We review the current status of the nuclear physics of X-ray bursts, with special emphasis on new experimental and theoretical information on a large number of reaction rates. As such we provide an overview of the broad experimental and theoretical methods currently used to advance the nuclear physics of X-ray bursts. The new information is used to update the public JINA REACLIB database with 32 new reaction rates based on experimental information, and a new dataset of theoretical statistical model reaction rates where no experimental information is available. Using several models for X-ray bursts that are powered by mixed hydrogen and helium burning, we take advantage of the updated nuclear data to review the current understanding of the nuclear reaction sequences in such X-ray bursts, the modeling of light curves, and predictions of the composition of nuclear ashes.

\end{abstract}

\maketitle
\section{Introduction} 
\label{sec:intro}

Thermonuclear X-ray bursts are the most frequent astrophysical explosions in our Galaxy (see \cite{strohmayerNewViewsThermonuclear2006, schatzXrayBinaries2006a, parikhNucleosynthesisTypeXray2013b, joseStellarExplosionsHydrodynamics2016, meiselNuclearPhysicsOuter2018a, Galloway2021} for reviews). There are more than 110 sources that burst with typical recurrence times of hours to days. With energies of typically 10$^{39} - 10^{40}$ erg, the bursts are easily observed by all current space-based X-ray observatories and large databases containing 1000s of bursts are available \cite{gallowayMultiINstrumentBurstARchive2020}. X-ray bursts originate from the surface of neutron stars that accrete matter from a companion star in a stellar binary system. The accumulated accreted layer undergoes explosive nuclear burning triggered by the thin-shell instability \cite{hansenSteadystateNuclearFusion1975a} and aided by mild electron degeneracy. There is a broad variety of bursts owing to differences in ignition conditions and nuclear fuel composition. Ignition conditions depend on system parameters such as accretion rate,  local gravity, thermal state of the underlying neutron star. The fuel composition depends on the composition of the outer layers of the companion star that are accreted, as well as on any modifications of the accreted composition by nuclear reactions prior to burst ignition. Such nuclear reactions include spallation of heavy nuclei upon impact on the neutron star atmosphere \cite{bildstenFateAccretedCNO1992a, randhawaSpallationalteredAccretedCompositions2019}, stable hydrogen burning via the hot CNO cycle facilitated by CNO nuclei in the accreted composition, and nuclear burning through previous X-ray bursts. Observationally, one can distinguish faint pure hydrogen flashes \cite{castenHydrogentriggeredXRayBursts2023a}, short pure He flashes (duration 1-10~s) \cite{gallowayThermonuclearBurstObservations2017}, mixed H/He bursts (duration 10-100~s) \cite{intzandNeutronStarCooling2017, gallowayThermonuclearBurstObservations2017}, intermediate long bursts (deep ignition of a helium layer, duration several hours) \cite{alizaiCatalogueUnusuallyLong2023}, superbursts (deep ignition of a carbon layer, duration several hours to a day) \cite{zandUnderstandingSuperbursts2017, alizaiCatalogueUnusuallyLong2023}, and, indirectly, possible hyperbursts (very deep ignition of oxygen or neon, duration 100s of days) \cite{pageHyperburstMAXIJ05562022}. 

Here we focus on the nuclear physics of mixed H/He bursts that have been modeled successfully with 1D hydrodynamic models \cite{woosleyModelsTypeXRay2004a, fiskerExplosiveHydrogenBurning2008, joseHydrodynamicModelsType2010a, meiselConsistentModelingGS2018,Zhen2023}. These bursts are relatively long and are characterized by a 10-100~s tail in the X-ray light curve. Unlike most types of bursts, where the burst duration is determined by the radiation transport timescale from the burning layer to the surface, the extended tail of mixed H/He bursts is produced by ongoing hydrogen burning via the rapid proton capture process (rp-process), which is slowed by the many $\beta^+$-decays in the reaction chain \cite{wallaceExplosiveHydrogenBurning1981a, schatzRpprocessNucleosynthesisExtreme1998b, schatzEndPointRp2001a}. These bursts are common; \cite{intzandNeutronStarCooling2017} analyzed 1254 bursts from the MINBAR burst archive \cite{gallowayMultiINstrumentBurstARchive2020} and found an rp-process tail in about half of them. Mixed H/He bursts are of particular interest as they offer unique opportunities to extract physics from observations, however, as the burst tail is determined by nuclear physics and not the thermal timescale, accurate nuclear physics is a pre-requisite.

The long tails of mixed H/He bursts are particularly sensitive to the surface redshift, which can be extracted from observations by comparison with model light curves \cite{zamfirConstraintsNeutronStar2012, meiselInfluenceNuclearReaction2019a}. This provides constraints on neutron star compactness and the mass radius relationship, and thus the nuclear matter equation of state \cite{xieImpactNewlyRevised2024}. The possibility to use observed burst tails to extract the precise amount of hydrogen present at ignition offers opportunities to probe ignition conditions and the thermal state of the underlying neutron star. Compared to other burst types, mixed H/He bursts also exhibit a much more extended nucleosynthesis, producing elements potentially up to Cd \cite{schatzEndPointRp2001a} or even beyond \cite{koikeFinalProductsRpProcess2004}. As it appears possible that small amounts of material get ejected into space \cite{schatzRpprocessNucleosynthesisExtreme1998b, weinbergExposingNuclearBurning2006b, herreraMasslossCompositionWind2023b}, the nucleosynthesis of mixed H/He bursts needs to be understood to predict possible observational signatures in X-ray spectra or any contributions to Galactic nucleosynthesis such as the light p-nuclei. Most of the burst nuclear ashes is predicted to remain on the neutron star and should have replaced the original neutron star crust in most systems. As such, the broad range of elements produced in mixed H/He bursts has a particularly strong impact on the thermal conductivity of the neutron star crust, which is dominated by impurity scattering and thus by the spread in atomic number of the composition. The range of isotopes in the ashes of mixed H/He bursts also enables a uniquely broad range of additional deep nuclear reactions in the crust that heat and cool \cite{guptaHeatingAccretedNeutron2007a, schatzStrongNeutrinoCooling2014a, lauNuclearReactionsCrusts2018}. Reliable predictions of the composition of the burst ashes of mixed H/He bursts is therefore essential to extract information about the neutron star interior from observations of cooling neutron stars in transiently accreting systems once the accretion shuts off \cite{meiselNuclearPhysicsOuter2018a}. The sensitivity of the interpretation of such data to the X-ray burst ashes composition has recently been demonstrated in models \cite{Jain2025}. 

The importance of nuclear physics in modeling mixed H/He burst observables has been shown in a number of nuclear sensitivity studies. \cite{thielemannElementSynthesisStars2001a, fisker15O19NeReaction2007, parikhEffectsVariationsNuclear2013, cyburtDependenceXRayBurst2016a, schatzDependenceXRayBurst2017a, lamImpactNew65As2022b} demonstrate the impact of nuclear physics on burst light curves and final composition of the burst ashes. Some reactions may leave particularly strong signatures in the light curves that have been proposed to possibly explain sometimes observed multi-peaked burst profiles \cite{fiskerNuclearReactionWaiting2004a, songDoublepeakedTypeXray2024}. \cite{meiselInfluenceNuclearReaction2019a} identified key reactions that most strongly affect the extraction of the surface redshift from burst observations. 


Here we review the current status of the nuclear physics of X-ray bursts from an experimental and theoretical point of view. The main foci of the paper are new theoretical and updated experimental reaction rates available as part of the JINA REACLIB nuclear data base \cite{cyburtJINAREACLIBDatabase2010a}. We also explore the impact of these updates in X-ray burst models and assess the current understanding of burst conditions, reaction sequences, and predictions of the burst ashes.

\section{Nuclear Physics of X-ray Bursts}
\label{sec:nuc}

\begin{figure*}
\includegraphics[width=1.0\textwidth]{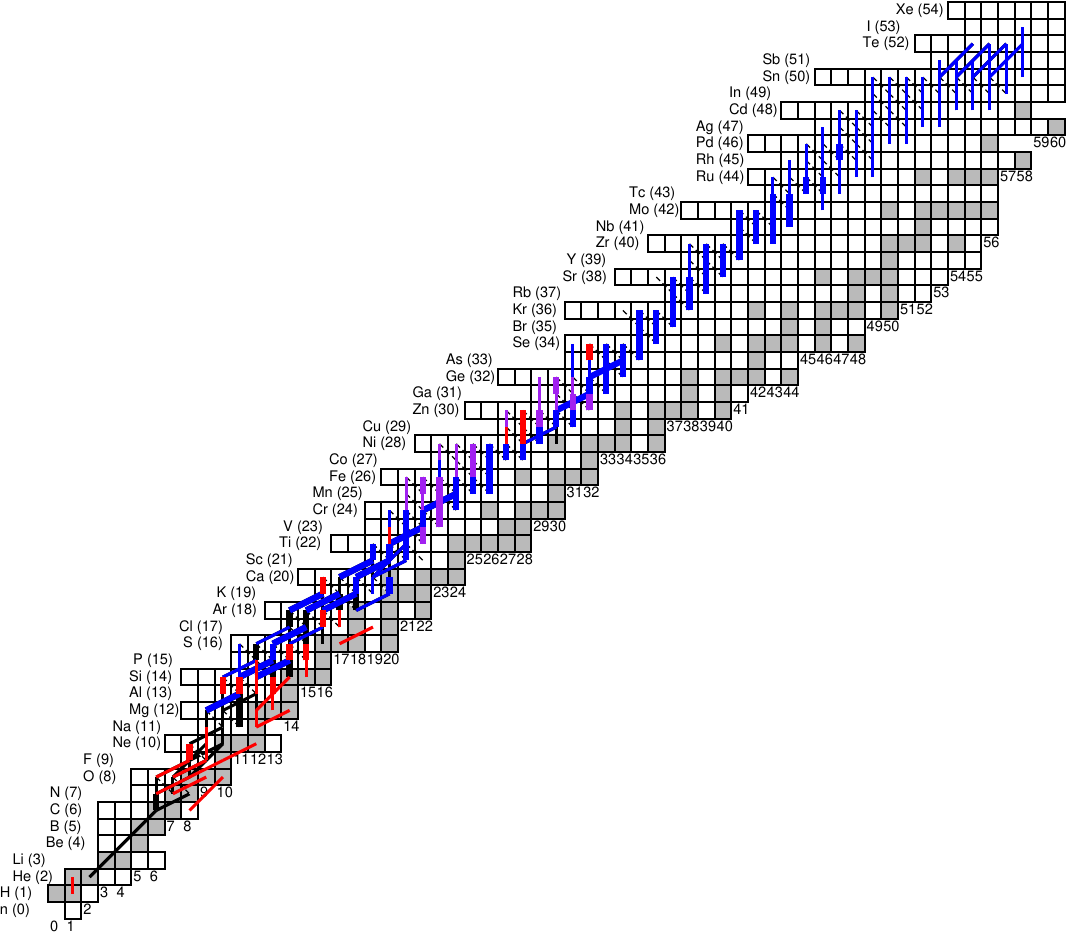}
\caption{Sources of charged particle reaction rates in the updated JINA REACLIB and used in our X-ray burst model simulations. Red lines mark reactions that were updated in this work using some experimental information as discussed in Sec.~\ref{subsec:individualrate}. Also shown are the sources of additional rates that are relevant for the main path of an extended rp-process \cite{schatzEndPointRp2001a}. Black lines mark other reactions in JINA REACLIB where some experimental information has been used in the past to derive a rate and that have not been updated in this work (Sec.~\ref{subsec:nonupdaterate}). Blue lines mark reactions determined from statistical model calculations, which have also been updated in this work as discussed in  Sec.~\ref{subsubsec:HFmodel}. Violet lines mark reactions previously determined from shell model calculations and included in prior versions of JINA REACLIB as discussed in Sec.~\ref{subsec:shellmodel}.
All reactions that have been identified as important in the sensitivity study of \cite{cyburtDependenceXRayBurst2016a} are marked as thick lines.\label{fig:reaclib_sources}}
\end{figure*}

The typical nuclear reaction sequence powering a mixed H/He X-ray burst is shown in Fig.~\ref{fig:reaclib_sources}. Prior to burst ignition, in the inter-burst phase, hydrogen is burned via the hot CNO cycle. The hot CNO cycle burn rate is governed by the $\beta^+$ decays of $^{14}$O and $^{15}$O and is thus temperature independent. As such it does not contribute to the thermonuclear instability, but reduces the hydrogen abundance and increases the $^{4}$He abundance over time. As temperature and density at the bottom of the freshly accreted layer steadily increase, a thermonuclear runaway is eventually triggered by the temperature sensitive 3-$\alpha$ rate producing $^{12}$C. In the run-up to ignition, $^{12}$C production via the 3-$\alpha$ reaction may already become significant. There is a potential feedback between the resulting increase in CNO nuclei, leading to a more efficient CNO cycle, which increases the $^4$He abundance, and consequently the 3$\alpha$-rate further \cite{cyburtDependenceXRayBurst2016a}. It is in this regime where the $^{15}$O($\alpha$,$\gamma$)$^{19}$Ne CNO breakout reaction may play a role, as it counteracts this feedback loop by removing CNO nuclei. The details of this mechanism have not been delineated in detail, and may depend sensitively on the details of the astrophysical model. This may explain the differences in sensitivity to the $^{15}$O($\alpha$,$\gamma$)$^{19}$Ne reaction found in different burst models \cite{fisker15O19NeReaction2007,davidsInfluenceUncertainties15O2011,cyburtDependenceXRayBurst2016a}. Once the burst is ignited, the temperature rises within seconds to peak temperatures of 1--2 GK, and the main reaction flow proceeds via 3-$\alpha$, $^{12}$C(2p,$\gamma$)$^{14}$O($\alpha$,p)$^{17}$F(p,$\gamma$)$^{18}$Ne, before breaking out of the CNO cycle via the $^{18}$Ne($\alpha$,p)$^{21}$Na reaction. $^{15}$O is entirely bypassed as its population would require a $^{14}$O $\beta^+$-decay with a half-life of 70.6~s that is comparable to or longer than the burst timescale. Therefore, at the onset of the burst the $^{15}$O($\alpha$,$\gamma$)$^{19}$Ne breakout reaction rapidly feeds the initial $^{15}$O abundance into the rp-process, and is not expected to play a role during the remainder of the X-ray burst. 

The $^{18}$Ne($\alpha$,p)$^{21}$Na breakout reaction marks the beginning of the $\alpha$p-process \cite{wallaceExplosiveHydrogenBurning1981a}, a sequence of alternating ($\alpha$,p) and (p,$\gamma$) reactions. The $\alpha$p process does not consume any hydrogen and thus represents a rapid He burning process, though small amounts of hydrogen need to be present as a catalyst. The main $\alpha$p process path continues from $^{18}$Ne with alternating (p,$\gamma$) and ($\alpha$,p) reactions, resulting in ($\alpha$,p) reactions on $^{22}$Mg, $^{26}$Si, $^{30}$S, $^{34}$Ar and so on \cite{schatzNucleiRpprocessAstrophysical1999}. While (p,$\gamma$) proton capture rates always exceed ($\alpha$,p) reaction rates, proton capture is hampered on these nuclei, making ($\alpha$,p) reactions competitive. This is due to the relatively low proton binding energies of the nuclei formed by proton capture that at the high temperatures during an X-ray burst lead to strong inverse ($\gamma$,p) photodisintegration that quickly removes any captured proton. The increasing Coulomb barrier for $\alpha$-induced reactions strongly limits the $\alpha$p-process. This leads to the formation of waiting points during the burst rise, where the $\alpha$p reaction flow waits for the temperature to rise sufficiently to proceed with the next ($\alpha$,p) reaction. Waiting points at $^{22}$Mg, $^{32}$S, and $^{34}$Ar have been proposed to explain sometimes observed multi-peaked burst profiles \cite{fiskerNuclearReactionWaiting2004a, songDoublepeakedTypeXray2024}. 

While the process waits for the temperature to rise to enable the next ($\alpha$,p) reaction, some residual proton capture flow via sequential 2p-capture can occur \cite{schatzRpprocessNucleosynthesisExtreme1998b}. In the 2p capture process, strong inverse ($\gamma$,p) photodisintegration creates a (p,$\gamma$)-($\gamma$,p) equilibrium that heavily favors the waiting point nucleus. However, proton capture on the low equilibrium abundance of the photodisintegrating isotone can lead to non-negligible leakage out of the equilibrium towards heavier nuclei. In addition, some $\beta$-decay occurs at the waiting points as the reaction path proceeds through a region of relatively short half-lives. Both, 2p capture and $\beta$-decay lead to parallel $\alpha$p process branches. These parallel branches tend to be less important but can still play a significant role, as indicated by the $^{28}$S($\alpha$,p), $^{29}$S($\alpha$,p), $^{24}$Si($\alpha$,p), and $^{25}$Si($\alpha$,p) reactions appearing in sensitivity studies \cite{cyburtDependenceXRayBurst2016a}. For example, in the one-zone model of \citet{cyburtDependenceXRayBurst2016a}, at the $^{22}$Mg waiting point, about 50\% of the reaction flow proceeds via 2p-capture (based on the here recommended new $^{23}$Al(p,$\gamma$)$^{24}$Si rate). This leads to additional $\alpha$p process pathways at Si via $^{24}$Si($\alpha$,p)$^{27}$P. The endpoint of the $\alpha$p process strongly depends on the peak temperature, and thus on the exact nature of the burst in terms of initial fuel composition and ignition depth. In most models, the last ($\alpha$,p) reaction occurs somewhere between $^{22}$Mg and $^{38}$Ca.

The endpoint of the $\alpha$p process marks the beginning of the rp-process. In the rp-process, rapid proton captures drive the reaction flow very quickly to isotopes near the proton drip line, where strong ($\gamma$,p) reactions impede the reaction flow \cite{wallaceExplosiveHydrogenBurning1981a, schatzRpprocessNucleosynthesisExtreme1998b, fiskerExplosiveHydrogenBurning2008}. Typically a single (p,$\gamma$)-($\gamma$,p) equilibrium pair of isotopes forms, where the reaction flow waits for the slow $\beta^+$ decay to occur. The pattern then repeats in the next isotonic chain. The endpoint of the rp-process depends sensitively on the hydrogen mass fraction at ignition and the endpoint of the $\alpha$p process and thus the peak temperature. Note that the endpoint of the $\alpha$p process plays a very strong role, as it not only defines the starting point of the rp-process, but also the proton to seed ratio, as a more extended $\alpha$p-process will form fewer seed nuclei for a given He abundance. A rough estimate of the mass number for the endpoint of the rp-process $A_{\rm rp}$ can be obtained by assuming all He is burned into seed nuclei quickly, and all the remaining hydrogen is captured on these seed nuclei: $A_{\rm rp} = A_{\alpha {\rm p}}*X/(1-X)$ with $A_{\alpha {\rm p}}$ the endpoint of the 
$\alpha$p-process (mostly determined by the peak temperature) and $X$ the hydrogen mass fraction at ignition \cite{schatzRapidProtonProcess1999a}. Under most favorable conditions (close to solar hydrogen mass fraction at ignition, $X=0.7$, and high peak temperature with extended $\alpha$p-process $A_{\alpha {\rm p}}$=38) one obtains $A_{\rm rp}=126$. However, it has been shown that the rp-process is then terminated at Te ($A=108$), where it encounters $\alpha$-unbound nuclei and thus ends in a Sn-Sb-Te cycle \cite{schatzEndPointRp2001a, joseHydrodynamicModelsType2010a}. For less favorable but not uncommon conditions such as $A_{\alpha {\rm p}}=26$ and $X=0.6$ one finds, for example, $A_{\rm rp}=64$. Therefore, the endpoint of the rp-process will vary significantly from system to system. The hydrogen burn rate of the rp-process is mostly defined by the slow $\beta^+$ decays along the path. The most prominent bottlenecks occur early on when crossing the $Z=20$ shell closure, and at late times at the relatively long-lived isotopes of $^{64}$Ge, $^{68}$Se, and $^{72}$Kr. The long-lived isotopes of $^{56}$Ni and $^{60}$Zn are also on the rp-process path but are typically less important as their decays can be effectively bypassed by proton capture. 

The rp-process reaction flow is further complicated by the potential formation of cycles at nuclides where there is a significant (p,$\alpha$) reaction branch \cite{vanwormerReactionRatesReaction1994a,fiskerExplosiveHydrogenBurning2008}. Such cycles reduce the synthesis of heavier nuclei not only by impeding the reaction flow, but also by producing He, which will increase seed production via the 3-$\alpha$ reaction, reduce the proton to seed ratio, and thus accelerate hydrogen consumption. In current burst simulations, using currently predicted reaction rates, most cycles are relatively weak. For example, in the \code{ONEZONE} model of \cite{cyburtDependenceXRayBurst2016a} there are two significant cycles, the Ni-Cu cycle via $^{59}$Cu(p,$\alpha$)$^{56}$Ni and the Zn-Ga cycle via $^{63}$Ga(p,$\alpha$)$^{60}$Zn, with integrated branchings of 1\% and 0.6\%, respectively. All other cycles are orders of magnitude weaker. However, the cycle branchings depend sensitively on nuclear reaction rates that are not well known. 

In summary, the important nuclear physics in X-ray bursts are the 3$\alpha$-reaction, especially its temperature sensitivity, the $^{15}$O($\alpha$,$\gamma$) reaction, a broad range of (p,$\gamma$), ($\gamma$,p), ($\alpha$,p), (p,$\alpha$) reactions, and $\beta^+$ decays between stability and the proton drip line up to $A \approx 108$. While stable nuclei are mostly by-passed by the reaction flow, reactions on stable nuclei can play a role as they process initially present heavy seed nuclei \cite{thielemannElementSynthesisStars2001a}. Nuclear masses play a critical role in determining negative Q-value reactions from their inverse positive Q-value reactions via the detailed balance principle, and for defining the abundance distribution within (p,$\gamma$)-($\gamma$,p) equilibrium pairs \cite{schatzRpprocessNucleosynthesisExtreme1998b}. Masses are also an essential ingredient for theoretical reaction rate predictions, and for estimating reaction rates from experimentally determined nuclear level schemes. In the following we summarize the status of the available nuclear data.

\subsection{Charged Particle Reactions} 
\label{subsec:reaction}

Proton and $\alpha$ induced charged particle reactions play an important role in mixed H/He X-ray bursts. The number of reactions per second and target nucleus (with target nucleus referring to the heavy nucleus interacting with the proton or $\alpha$-particle) induced by particles $p$ can be expressed as $\lambda=Y_p \rho N_A <\sigma v>$ with mass density $\rho$, particle abundance $Y_p$ typically in mole per gram, and the product of cross section $\sigma$ and relative velocity $v$ averaged over the temperature dependent Maxwell Boltzmann relative velocity distribution of the particles in the stellar plasma $<\sigma v>$. The expression $N_A <\sigma v>$ is typically referred to as reaction rate and is what is tabulated, for example in the JINA REACLIB database \cite{cyburtJINAREACLIBDatabase2010a}. The rate per particle pair $<\sigma v>$ is sometimes referred to as reactivity (e.g. \cite{rauscherRelevantEnergyRanges2010}).

An important consideration for determining reaction rates is the relevant temperature range. A lower bound arises from the requirement that the reaction timescale $\tau = 1/\lambda$ be smaller than the burst timescale $\tau_{\rm burst}$ (this requirement would be alleviated for reactions operating prior to burst ignition such as the hot CNO cycle or $^{15}$O($\alpha$,$\gamma$)). For $\tau_{\rm burst}=10~s$, $Y_p=0.5$, and $\rho=10^6$~g/cm$^3$ one finds a minimum relevant reaction rate of $\EE{2}{-6}$~cm$^3$/s/mole. For X-ray burst models, accurate reaction rates are only needed for temperatures where the rate exceeds this value. The corresponding minimum temperature depends on the Coulomb barrier and thus on the element number of projectile and target nucleus. For currently tabulated reaction rates one finds for example for $^{19}$Ne(p,$\gamma$) $T >  0.2$~GK, for $^{56}$Ni(p,$\gamma$) $T >  0.5$~GK, and for $^{99}$In(p,$\gamma$) $T >  1.3$~GK. The upper bound arises from the peak temperature of the burst, which is typically in the 1-2~GK range, and is expected to vary depending on system parameters such as accreted composition, accretion rate, neutron star surface gravity, and thermal state of the neutron star. In many cases the rate will tend to be most important in the lower range of the relevant temperatures, where it will significantly impede the reaction flow. The exception to this are branchings, in particular between (p,$\gamma$) and (p,$\alpha$) reactions. 

\subsubsection{Reaction Rates from Isolated Resonances}
\label{subsubsec:isores}

For the relevant temperatures, X-ray burst charged particle reaction rates 
are typically dominated by resonances, though direct capture may play a role in some cases. Owing to the significant Coulomb barrier, resonance widths are typically narrow and interference effects negligible, though there are important exceptions to this, especially for lighter target nuclei with $A<20$. In the case of isolated narrow resonances, where changes in partial widths and changes in the velocity distribution of the incident particles over the width of the resonance are negligible, the velocity integration of $<\sigma(v) v>$ can be carried out analytically, and the contribution of the resonance to the reaction rate can be expressed with the narrow resonance equation \cite{IliadisBook}: 
\begin{eqnarray}
    N_A<\sigma v> = 1.54 \times 10^{11} (AT_9)^{(-3/2)} \ \omega \gamma [{\rm MeV}]\exp(-11.6045 \ E_{\rm r}[{\rm MeV}]/T_9)
  \label{eq:narrowresonance}
\end{eqnarray}

with the reduced mass in atomic mass units $A$ and temperature $T_9$ in GK. The most important ingredients that determine the rate are then the resonance energy $E_r$ and the resonance strength $\omega \gamma$, which depends on the projectile and ejectile partial widths $\Gamma_p$ and $\Gamma_e$, respectively,
\begin{equation}
\omega \gamma = \frac{(2J_r+1)}{(2J_p+1)(2J_T+1)} \frac{\Gamma_p \Gamma_e}{\Gamma_{\rm total}}
\label{eq:omegagamma}
\end{equation}
with $J_p$, $J_T$, and $J_r$ being the spins of projectile, target, and resonance state, respectively. The most straight forward experimental determination of a resonant reaction rate contribution is a direct measurement of resonance energy and resonance strength by impinging a radioactive beam on a target that is thick enough to induce sufficient energy loss for all beam particles to pass through the narrow resonance energy range of interest, but not through any other resonances (e.g., \cite{dauria21NaPg2004}). If that is not feasible, for example owing to limited available beam intensities, the partial widths in Eq.~\ref{eq:omegagamma} can be determined either directly, through observations of the various decays of the resonant state, or indirectly, by constraining the structure of the resonant state, for example using transfer or pickup reactions that are sensitive to spectroscopic factors or Asymptotic Normalization Coefficients (ANCs) \cite{iliadisSpectroscopicFactorsDirect2004,bertulaniNuclearAstrophysicsRadioactive2010,tribbleIndirectTechniquesNuclear2014a,bardayanTransferReactionsNuclear2016}, which can then be related to particle widths. Direct measurements of partial widths in principle determine the reaction rate completely, provided all relevant resonances are identified and characterized. The challenge is that often one of the partial widths is much smaller than the other. In that case $\Gamma_{\rm large} \Gamma_{\rm small} / \Gamma_{\rm total} \approx \Gamma_{\rm small} $ and thus the rate is entirely determined by the smaller width, which is often difficult to observe.  Indirect measurements do not suffer from this limitation, but result in additional uncertainties from the underlying reaction theory models, as well as from deficiencies in knowledge of all relevant properties of the state (e.g. \cite{tribbleIndirectTechniquesNuclear2014a, bardayanTransferReactionsNuclear2016, hebbornOpticalPotentialsRareisotope2023}). 

For isolated resonances with no or incomplete experimental information, the shell model has been quite successful in predicting resonance properties of nuclei that can be described within the employed model spaces of the sd- and lower fp-shell up to around Ge \cite{herndlProtonCaptureReaction1995, fiskerShellModelBased2001}. The main challenge for shell model predicted reaction rates are the predictions of precise resonance energies. Typical uncertainties of 100s of keV translate into many orders of magnitude of uncertainty in reaction rates (e.g. \cite{clementNewApproachMeasuring2004a}). A common strategy for rp-process reaction rates is to reduce this uncertainty by using experimentally known states in the isospin mirror nucleus where neutron and proton numbers are swapped.  In that case, the models only need to predict the energy shift due to Coulomb effects \cite{herndlProtonCaptureReaction1995}. However, excitation energy uncertainties tend to still be of the order of 100~keV~\citep{HendersonMirrorLevels2020}. For resonances where the excitation energy is experimentally known, the use of shell model predictions for the partial widths can lead to relatively reliable reaction rate estimates (e.g. \cite{brownShellmodelStudiesAstrophysical2014,Cl34pgRichter2020}). This approach requires to match the observed state with the shell model state, which is straightforward at low level densities, but can be challenging for higher level densities, for example in heavier nuclei.

\subsubsection{Reaction Rates from the Statistical Model}
\label{subsubsec:HFmodel}

For cases where a large number of resonances directly contribute to the reaction rate, for example for heavier nuclei with high level densities, or at high temperatures, the reaction rate does not depend anymore on the individual resonance properties and can be calculated using the statistical or Hauser-Feshbach model \cite{rauscherAstrophysicalReactionRates2000,gorielyImprovedPredictionsNuclear2008}. While such models strictly speaking require large numbers of overlapping resonances, of which most are experimentally indistinguishable, one can take advantage of the fact that astrophysical reaction rates at a given temperature sample a relatively broad energy range and, as such, enough resonances may contribute to justify a statistical approach.  \cite{rauscherNuclearLevelDensity1997} surveyed the applicability of the statistical model for astrophysical calculations by using a level density model to determine the minimum temperature at which there are of the order of 10 resonance states participating for reactions across the chart of nuclides. Uncertainties of typically factors of 2-3 have been obtained for proton capture reactions with heavy nuclei, where a large number of isolated states is contributing to the rate (see for example \cite{simonSystematicStudyReactions2013}). For $\alpha$-induced reactions uncertainties tend to be somewhat larger, though progress related to $\alpha$ optical potentials has recently been achieved \cite{mohrSuccessfulPredictionTotal2020}. The applicability of the statistical model is a particular challenge for the rp-process. Near the proton drip line, excitation energies of resonances are low and level densities can be very small. In addition, not all levels participate equally in reactions typically dominated by s- or p-wave resonances. $\alpha$-induced reactions populate mainly natural parity states, and cluster structure can enhance the contribution from certain individual states, which further limits the applicability of a statistical approach~\citep{LongIndirectAlphaP2017}. 

Nevertheless, because the statistical model is the only theoretical approach that produces a complete set of astrophysical reaction rates spanning the entire chart of nuclides and all relevant reactions, it is used as a default for all cases where there is no experimental or shell model information available. Currently this is the vast majority of reaction rates in X-ray burst reaction networks. This includes many cases where the statistical model is strictly speaking not applicable (e.g. $^{64}{\rm As}(p,\gamma)$~\citep{MeiselGe64Bypass2025}). In such cases, uncertainties are expected to be significantly larger and difficult to estimate. For astrophysical model calculations that rely simply on a recommended best central value based on the statistical distribution of all uncertainties, statistical model rates may even be preferable over shell model rates, as for the latter an off prediction of the energy of a single resonance can produce extreme outliers in the reaction rate. 

The Hauser-Feshbach model is built on the fundamental Bohr independence hypothesis that the exit channel loses the memory of the entrance channel by means of the intermediary formation of a compound nucleus reaching a state of thermodynamic equilibrium \citep{Bohr1937,Wolfenstein1951,Hauser1952}. Within the Hauser-Feshbach model, the binary reaction cross section of $A$($a$,$b$)$B$ by compound nucleus capture (CNC) can be written as
\begin{equation}
\sigma^{\mathrm{CNC}}(E)=\sum^{B}_{x=0}\sigma^{\mathrm{CNC}}_{A+a\rightarrow B^x+b}.
\end{equation}
The summation $\sum^{B}_{x=0}$, where the energy-level scheme is represented by the $x$$^{\rm th}$ excited state ($x$ = 0 is the ground state), covers the ground and all the possible excited states of the residual nucleus $B$. Each state is characterized by a spin $I^x_B$, a parity $\pi^x_B$, and an excitation energy $E^x_B$ for the residual nucleus $B$. Note that in the condition of laboratory measurement, the target $A$ is mostly in the ground state, though while in astrophysical environment, the thermal target characterized by the Boltzmann population weight is considered~\citep{AstrophysicalRatesSargood1982}.

The cross section $\sigma^{\mathrm{CNC}}_{A+a\rightarrow B^x+b}(E)$ can be expressed as \cite{gorielyImprovedPredictionsNuclear2008,PhysRevC.90.024604}
\begin{eqnarray}
&&\sigma^{\mathrm{CNC}}_{A+a\rightarrow B^x+b}(E) =\frac{\pi}{k^2}\sum^{l_{max}+I_A+I_a}_{J=mod(I_A+I_a,1)}\sum^1_{\Pi=-1}\frac{2J+1}{(2I_A+1)(2I_a+1)}\times \nonumber \\
&&\sum^{J+I_A}_{J_a=\left|J-I_A\right|}\sum^{J_a+I_a}_{l_i=\left|J_a-I_a\right|}
\sum^{J+I^x_B}_{J_{b}=\left|J-I^x_B\right|}\sum^{J_{b}+I_{b}}_{l_f=\left|J_{b}-I_{b}\right|} \delta^{\pi}_{C_a}\delta^{\pi}_{C_{b}}\frac{\left\langle T^J_{C_a,l_i,J_a}(E)\right\rangle \left\langle T^J_{C_{b},l_f,J_{b}}(E_{b})\right\rangle}{\sum_{Clj}\delta^{\pi}_C\left\langle T^J_{C,l,j}(E_{C})\right\rangle}W^J_{C_{a}l_{i}J_{a}C_{b}l_fJ_{b}}.
\label{HF}
\end{eqnarray}
\noindent In Eq.~\ref{HF},
\begin{itemize}
\item $k$ is the wave number of relative motion for the entrance channel,
\item $E$ is the incident energy of the projectile,
\item $E_{b}$ is the energy of the emitted particle,
\item $l_{max}$ is the maximum value of the relative orbital momentum $l_i$ of $A$ + $a$,
\item $l_{f}$ is the relative orbital momentum of the residual nucleus $B$ and $b$,
\item $I_a$, $J_a$ and $\pi_{a}$ are the spin, total angular momentum, and parity of $a$, respectively,
\item $I_b$, $J_b$ and $\pi_{b}$ are the spin, total angular momentum, and parity of $b$, respectively,
\item $J$ and $\Pi$ are the total angular momentum and parity of the compound nucleus, respectively,
\item $C_a$ is the channel label of the initial system $(a+A)$ designated by $C_a=(a,I_a,E,E_{A},I_A,\pi_A)$ with the excitation energy $E_{A}$ ($E_{A} = 0$ for ground state target), spin $I_A$ and parity $\pi_A$ of the target $A$,
\item $C_b$ is the channel label of the final system $(b+B^x)$ designated by $C_b=(b,I_b,E_b,E^x_B,I^x_B,\pi^x_B)$ with the excitation energy $E^x_B$, spin $I^x_B$ and parity $\pi^x_B$ of the residual $B^x$,
\item $\delta^{\pi}_{C_a}=1$ if $\pi_A\pi_a(-1)^{l_i}=\Pi$ and 0 otherwise,
\item $\delta^{\pi}_{C_b}=1$ if $\pi^x_B\pi_b(-1)^{l_f}=\Pi$ and 0 otherwise,
\item $T$ is the transmission coefficient,
\item $\sum_{Clj}\delta^{\pi}_C\left\langle T^J_{C,l,j}(E_{C})\right\rangle$ is the sum of the transmission coefficient for all possible decay channels $C$, and
\item $W$ is the width fluctuation correction factor.
\end{itemize}

The transmission coefficient $T$ is calculated for each level with known energy, spin, and parity. If the excitation energy $E^x$, which is implicit in the definition of the channel $C$, corresponds to a state in the continuum, or in an energy region where not all individual levels are known, an effective transmission coefficient for an excitation-energy bin of width $\Delta E$ is defined by the integral
\begin{eqnarray}
\left\langle T^J_{Cjl}(E_C) \right\rangle=\int^{E^x+\Delta E/2}_{E^x-\Delta E/2}\rho(E,J,\Pi)T^J_{Clj}(E_C)dE
\label{eq:tran}
\end{eqnarray}
\noindent over the nuclear level density (NLD) $\rho(E,J,\Pi)$. The calculation of the transmission coefficients $T$ for the open reaction channels depends on different nuclear ingredients. In particular, $T$ for particle emission relies on the optical model potentials of the two reacting particles, while for photon emission, $T$ is determined by the $\gamma$-ray strength function. The detailed nuclear structure ingredients that are used to obtain the transmission coefficients are described in Section III.

\subsubsection{Direct Capture Reaction Rates}
\label{subsubsec:Dicap}
In principle the direct capture mechanism can contribute to the proton capture rates in the rp-process in X-ray bursts. A significant contribution is only expected for cases where strong resonances are absent. Unlike in the case of neutron capture, Coulomb repulsion for charged particle reactions pushes the relevant compound nucleus excitation to higher energies where some reasonably strong resonances are typically populated. This is especially true for the relatively high relevant temperatures in X-ray bursts. For these reasons, direct capture tends to be relevant only for a small number of rp-process reactions, predominantly in the lower mass region where level densities are lower. 

Indeed, for the individual reaction updates discussed in ~\ref{subsec:individualrate}, direct capture plays only a significant role for X-ray bursts (contribution to the rate above $\EE{2}{-6}$~cm$^3$/s/mole as discussed in Sec.~\ref{subsec:reaction}) in one case, the $^{17}$F(p,$\gamma$)$^{18}$Ne reaction below T = 0.5 GK. \cite{kuvinMeasurement17F18Ne2017}. A previously performed systematic study of the contribution of direct capture relative to the statistical model rate for about 2700 proton capture reactions with 8$\leq$ Z $\leq$ 100 lying between the proton drip line and the valley of $\beta$ stability \cite{PhysRevC.109.014611} indicates no significant contribution in the rp-process region. We note however that this study determines the direct capture contribution under the assumption that the statistical model is applicable. Reactions where direct capture dominates are expected to be characterized by low Compound nucleus level densities where that assumption may break down. 

Nevertheless, we employ in this study a potential model to calculate the direct proton capture contribution to all updated proton capture reaction rates that were calculated with the statistical model. The model is similar to previous potential models built on the basis of pertubation theory \cite{ROLFS197329,Oberhummer1991,1997A&A325414G,PhysRevC.86.045801,PhysRevC.90.024604,PhysRevC.109.014611}. For a reaction $A$($p$,$\gamma$)$B$, 
the model describes the transition from the initial scattering state $A$+$p$ to the final nucleus $B$ with the emission of a $\gamma$-ray. In such model, the incoming proton is scattered directly into a final bound state in nucleus $B$. The allowed electric-dipole ($E1$), electric-quadrupole ($E2$), and magnetic-dipole ($M1$) transitions to the ground state, as well as to all possible excited states in the final nucleus, are taken into account. Final states include known discrete levels at lower energies as well as a continuum described by a level density at higher energies. In this way, the cross section of the direct capture $A$($p$,$\gamma$)$B$ can be written as
\begin{eqnarray}
\sigma^{\mathrm{DIC}}(E) =  \sum^{E^{th}_B}_{x=0}S^x_F \sigma^{\mathrm{DIC}}_{A+p\rightarrow B^x+\gamma}(E) 
+\int^{S_p}_{E^{th}_B} \langle S_F\rangle \sum_{I^x_B,\pi^x_B}  \rho(E,I^x_B,\pi^x_B)\sigma^{\mathrm{DIC}}_{A+p\rightarrow B^x+\gamma}(E)dE,
\label{eq:dicxs}
\end{eqnarray}
\noindent with $E$ the energy of the incident proton; $x$ indexes the experimentally known energy levels (with $x$ = 0 the ground state); $E^{th}_B$ the excitation energy up to which all levels are known experimentally in the residual nucleus $B$. Below $E^{th}_B$, the sum runs over all available discrete final states. $S^x_F$ is the spectroscopic factor describing the overlap between the antisymmetrized wave function of the initial system $A$ + $p$ and the final state of $B^x$. Above $E^{th}_B$, the summation is replaced by a continuous integration over a spin($I^x_B$)- and parity($\pi^x_B$)-dependent nuclear level density $\rho(E,I^x_B,\pi^x_B)$ with the cross section and the averaged spectroscopic factor $\langle S_F \rangle$ for all levels.

The potential model calculates the transition matrix element between the initial and the final states by sandwiching the electromagnetic operators in the long-wavelength limit. Taking into account the $E1$, $E2$, and $M1$ transitions, the partial cross section $\sigma^{\mathrm{DIC}}_{A+p\rightarrow B^x+\gamma}(E)$ for a transition from the initial $A$ + $p$ system to the final state $B^x$ +$\gamma$ can be written as
\begin{eqnarray}
\sigma^{\mathrm{DIC}}_{A+p\rightarrow B^x+\gamma}(E)=\frac{2I^x_B+1}{Ek(2I_A+1)(2I_p+1)} \times \\
\sum_{S_f,J_i,l_i,S_i}\left\{ \frac{2}{9}k^3_{\gamma}(\left|M_{E1}\right|^2+\left|M_{M1}\right|^2)+
		\frac{1}{150}k^5_{\gamma}\left|M_{E2}\right|^2 \right\}.
\label{eq:dic}
\end{eqnarray}
\noindent In Eq. \ref{eq:dic}, $I^x_B$, $I_A$, and $I_p$ are the spin of the nucleus $B^x$, $A$, and proton, respectively, and $k_\gamma$ is the wave number of the emitted photon. The summations run over the channel spins $S_i$, orbital angular momenta $l_i$, and total angular momenta $J_i$ of the initial system $A$ + $p$, and over the channel spins $S_f$ of the final state. The matrix elements consist of two components, the radial moments ($\mathcal{M}_{E1}$, $\mathcal{M}_{M1}$, $\mathcal{M}_{M2}$) and, if any, the internal moments of the nucleus $A$ [$\mathcal{M}^{\mathrm{int}}_{M1}(A)$, $\mathcal{M}^{\mathrm{int}}_{E2}(A)$] and proton [$\mathcal{M}^{\mathrm{int}}_{M1}(p)$, $\mathcal{M}^{\mathrm{int}}_{E2}(p)$], which are given by
\begin{eqnarray}
&& M_{E1}=\mathcal{M}_{E1}, \nonumber\\
&& M_{M1}=\mathcal{M}_{M1}+\mathcal{M}^{\mathrm{int}}_{M1}(A)+\mathcal{M}^{\mathrm{int}}_{M1}(p), \nonumber\\
&& M_{E2}=\mathcal{M}_{E2}+\mathcal{M}^{\mathrm{int}}_{E2}(A)+\mathcal{M}^{\mathrm{int}}_{E2}(p).
\end{eqnarray}
\noindent The complete formula of the matrix elements can be found in Refs. \cite{descouvemont2003theoretical,PhysRevC.86.045801}.

The radial wave function $\psi$ of these matrix elements are obtained by solving the two-body Schr\"odinger equation
\begin{eqnarray}
\left[ \frac{d^2}{dr^2}-\frac{l(l+1)}{r^2}+\frac{2\mu}{\hbar}\left\{ E-V(E,r) \right\} \right]\psi=0
\label{eq:sch}
\end{eqnarray}
\noindent expressed in the relative coordinate $r$. Here, $V(E,r)$ is an energy-dependent central potential that consists of the nuclear and Coulomb parts, $l$ is the relative orbital angular momentum, and $\mu=m_Am_p/(m_A+m_p)$ is the reduced mass. For a clearer notation, $\psi$ is replaced by $\chi_{l}(E,r)$ for the initial scattering system $A$ + $p$  with $E$ $>$ 0, and by $\phi_{nl}(r)$ for the bound state $B^{x}$ $(E<0)$ with $E$ $<$ 0.

For the scattering system $A$ + $p$, the radial wave functions $\chi_l(k,r)$ (Here $E=\hbar^2k^2/2\mu$ with the wave number $k$) behave asymptotically at large distances as
\begin{eqnarray}
\chi_l(k,r) \rightarrow e^{i(\delta^c_l+\delta_l)}[\cos(\delta_l)F_l(kr)+\sin(\delta_l)G_l(kr)],
\end{eqnarray}
\noindent where $\delta_l$ is the phase shift by the nuclear potential, $\delta^c_l$ is the Coulomb phase shift, and $F_l(kr)$ and $G_l(kr)$ are the regular and irregular Coulomb wave functions, respectively.

For the bound states $B^x$, the radial wave functions $\phi_{nl}(r)$ must vanish at infinity and be normalized as
\begin{eqnarray}
\int^{\infty}_0 |\phi_{nl}(r)|^2dr=1,
\end{eqnarray}
\noindent where $n$ stands for the radial quantum number.

\subsubsection{Role of Excited Target States}
\label{subsubsec:Exited}

Nuclei in a stellar plasma are not always in their ground state. Rather a thermal equilibrium population of ground and excited states $Y(E_x)$ is established: 
\begin{equation}
   Y(E_{x})=\frac{(2J+1) \exp( E_{x}/kT)}     {\sum_i (2J_i+1) \exp( E_{xi}/kT)}
\label{eq:thermalpop}
\end{equation}
with excitation energy $E_x$ and level spin $J$. For a typical maximum X-ray burst temperature of 1~GK, $kT$$\approx$86~keV and thus states up to a few 100~keV can be significantly populated. While these are relatively low excitation energies, there are many nuclei with such states along the path of the rp-process. Nuclear reactions in X-ray bursts therefore do not proceed exclusively through the ground state, but through all populated excited states. 

All charged particle reaction rates commonly employed in astrophysical models and included in the JINA REACLIB database take into account reactions on ground and thermally populated excited states. In most cases available experimental information pertains to the target in its ground state. Therefore, the typical procedure is to derive a ground-state rate, which is then corrected with a stellar enhancement factor (SEF) to account for the excited states. The SEF is the ratio of actual rate to the ground state rate and despite its name can be larger or smaller than one. For the vast majority of reactions the SEF is calculated with the statistical model \cite{rauscherAstrophysicalReactionRates2000}, even if the ground state rate is not. 

In some cases, where reaction rates are determined from individual narrow resonances, partial widths for the decays of resonances to excited target states may be available, either from shell model calculations or from experiments. In that case, the total reaction rate on ground and all thermally excited target states can be determined from the data for individual resonances. For example, for particle capture rates
\cite{Vancraeynest98,schatzRevisedResult32Cl2005a}

\begin{equation}
    N_{A}\langle\sigma v\rangle=\sum_{j} N_{A}\langle\sigma v\rangle_{0 j} \frac{1}{G(T)}\left(1+\sum_{i>0}^{E_{i j}>0} \frac{\Gamma_{\mathrm{p} i j}}{\Gamma_{\mathrm{p} 0 j}}\right),
\end{equation}
where $\sum_j N_A \langle\sigma v\rangle_{0 j}=\langle\sigma v\rangle_{\rm gs}$ is the sum of the individual resonant reaction rate contributions from the target ground state to all excited resonance states $j$, $\Gamma_{\mathrm{p} i j}$ the incident particle decay widths from resonance state $j$ to target state $i$ (with $i=0$ being the ground state), 
and $E_{i j}$ the energy difference between resonance state $j$ and target state $i$. $G(T)$ is the normalized temperature $T$ dependent partition function 
\begin{equation}
    G(T)=\left(2 J_{0}+1\right)^{-1} \sum_{i}\left(2 J_{i}+1\right) \exp \left(-E_{i} / k T\right)
    \label{eq:partition}
\end{equation}
with $J_i$ the target spin of state $i$ with excitation energy $E_i$. The corresponding SEF is then given by  $\langle\sigma v\rangle/\langle\sigma v\rangle_{\rm gs}$.

As the statistical model assumes the rate is an average of many different resonances for both, the ground and the excited states, statistical model SEFs tend to be relatively small. Statistical model SEFs likely underestimate the SEF significantly for the many reactions that are governed by a small number of dominant resonances, as ground and excited states can have very different spins and spectroscopic overlaps. An example is the $^{32}$Cl(p,$\gamma$)$^{33}$Ar  reaction, where the statistical model predicts an SEF of 0.8-1, however, a more accurate determination using the shell model resulted in an SEF of up to a factor of 5 already at moderate temperatures below 1 GK \cite{schatzRevisedResult32Cl2005a}. 

In some cases the population of excited target states may deviate from thermal equilibrium further enhancing the role of excited states. This can occur in particular for excited states that are terrestrial isomers. Isomers are characterized by a strongly reduced decay probability and therefore an anomalously long lifetime that may hamper rapid thermalization and lead to out of equilibrium population. For example, a preceding nuclear reaction may populate an isomeric state, which owing to its long lifetime does not thermalize, and would then become the target for the next reaction. In this way, even target states at higher excitation energies can become relevant. However, it is important to note that not every state that is an isomer in the laboratory is also an isomer in the stellar environment. At sufficiently high temperatures, additional pathways for deexcitation via $\gamma$-induced transitions to higher lying states may open up \cite{guptaInternalEquilibrationNucleus2001}. If that is the case, the state may still contribute, however, its population will then follow the thermal population in Eq.~\ref{eq:thermalpop}. Isomeric states that preserve their isomeric character in the astrophysical environment and as such can be populated out of equilibrium have been coined astromers \cite{mischAstromersNuclearIsomers2021}. In order to include non-thermal populations of excited states in model simulations, ground and isomeric state ensembles (all states thermally connected to them \cite{guptaInternalEquilibrationNucleus2001}) must be tracked separately in the reaction network, and transition rates from one to the other must be determined and implemented as well. Until recently such information was only available for $^{26}$Al and its 0$^{+}$ isomeric state, but has now been compiled for a broad range of cases \cite{mischAstromersNuclearIsomers2021}.
These rates have not yet been implemented in reaction networks in general. The relevant cases for X-ray bursts are $^{26}$Al and $^{34}$Cl, and perhaps $^{38}{\rm K}$~\citep{chipps38IsomerProduction2018a}. However, at the relatively high temperatures in X-ray bursts thermal equilibrium will be established in each case. Therefore, for the purpose of this study, we employ only rates for nuclei in thermal equilibrium. 

\subsubsection{Inverse Reaction Rates}
\label{subsubsec:Inverse}

For many reactions in X-ray bursts, both the reaction in the forward and the reaction in the reverse direction are important. The rates of these reactions are directly related to each other via the detailed balance principle. To ensure correct equilibrium behavior when both directions are fast, reactions are only determined in one direction. The inverse reaction rate is then derived from the forward rate using the detailed balance principle. Typically the positive Q-value reaction is the forward reaction. A typical example are positive Q-value particle capture rates $N_A<\sigma v>_{p,\gamma }$ that are determined using the various methods described above. The inverse ($\gamma$,p) photodisintegration reaction rate $\lambda_{\gamma,p}$
can then be derived via detailed balance using 
\begin{eqnarray}
\lambda_{\gamma,p} = \frac{(2J_p+1)(2J_T+1)}{(2J_R+1)} \frac{G_T(T)}{G_R(T)}  \left( \frac{\mu_{\rm pT} k T}{2 \pi \hbar^2} \right) ^{(3/2)} 
     \exp \left( -\frac{Q_{p,\gamma}}{kT} \right) N_A<\sigma v>_{p,\gamma}
     \label{eq:reverse}
\end{eqnarray}
with $J_p$, $J_T$, and $J_R$ the ground state spins of proton, proton capture target, and proton capture residual, respectively; $\mu_{\rm pT}$ reduced mass of proton and target; and $Q_{p,\gamma}$ the capture Q-value. The normalized partition functions $G$ correct the ground state spin factors for the spins of thermally excited states and are calculated as a function of temperature from Eq.~\ref{eq:partition}. 

Eq.~\ref{eq:reverse} applies to nuclei and reactions in full thermal equilibrium, i.e. reactions on thermally excited states must be included in the forward rate. A similar equation applies to particle induced inverse reaction rates, for example in the case of a (p,$\alpha$) reaction:
\begin{eqnarray}
N_A<\sigma v>_{\alpha,p} = \frac{(2J_p+1)(2J_T+1)}{(2J_R+1)(2J_\alpha+1)} \frac{G_T(T)}{G_R(T)} 
\left( \frac{\mu_{\rm pT}}{\mu_{\rm \alpha R}} \right) ^{(3/2)}
     \exp \left( -\frac{Q_{p,\alpha}}{kT} \right) N_A<\sigma v>_{p,\alpha}
     \label{eq:reverse2}
\end{eqnarray}
with $\mu_{\rm \alpha R}$ the reduced mass of residual nucleus and emitted $\alpha$ particle.

It should be noted that inverse reactions in the JINA REACLIB database are marked as such, and do not contain the normalized partition functions $G$ in Eqs.~\ref{eq:reverse},\ref{eq:reverse2}, which need to be applied when the rates are used. Most applications will only use forward rates from the database and determine inverse rates using consistent mass and partition function tables. 

\subsection{Nuclear Masses}
\label{subsec:masses}

Nuclear masses play an important role in X-ray burst model calculations \cite{schatzRpprocessNucleosynthesisExtreme1998b, schatzImportanceNuclearMasses2006a, schatzDependenceXRayBurst2017a}. They enter such calculations in different ways: (1) mass tables serve as direct input in X-ray burst models to determine the nuclear energy generation from the binding energy differences as nuclear reactions change the composition; (2) masses are needed to determine reaction rates from data on individual resonances where no direct reaction rate measurements are available or via the statistical model; (3) masses are key to calculate reaction Q-values, which are needed to determine reverse reaction rates via detailed balance. Accuracy needs and requirements differ considerably for these three cases: 

(1) The accuracy requirement for reliable calculation of nuclear energy generation is modest as all that matters is the total energy generation from all reactions. This aspect is therefore not driving accuracy requirements. 

(2) When reaction rates are determined from the contributions of individual resonances, mass sensitivities dramatically increase. Narrow resonance rates depend exponentially on the resonance energy $E_r$ (Eq.~\ref{eq:narrowresonance}), which in most cases is determined from excitation energies $E_x$ and Q-values, $E_r=E_x-Q$, requiring reliable ground state nuclear masses to calculate the Q-value. The exponential resonance-energy dependence stems from the overlap of resonance energy and the Maxwell Boltzmann energy distribution of the protons, and, for cases of reactions dominated by charged-particle widths, from the exponential energy dependence of the Coulomb penetrability. In extreme cases Q-value changes of a few 10s of keV can lead to order of magnitude changes in reaction rate, and even keV uncertainties lead to non negligible reaction rate errors as shown for example for the $^{23}$Al(p,$\gamma$) reaction rate \cite{puentesHighprecisionMassMeasurement2022}. 
While in some cases resonance energies can be measured directly, for example using the beam energy in a direct resonance strength measurement, or by measuring the energy of charged particles from the decay of a resonance, in the vast majority of cases ground-state masses with accuracies in the 1-10 ~keV range are needed for narrow resonance rate determinations from theory and experiment.

Nuclear masses are also an important ingredient for the calculation of reaction rates that are not experimentally known. With the statistical model assuming that the rate is an average over many resonances, the dependence on masses is somewhat reduced, though it is still significant, 
especially for the charged particle transmission coefficients that depend on the penetrability through the Coulomb barrier and thus exponentially on the channel Q-value (see discussion in Section III). The statistical model determines a reaction rate from the competition between all possible reaction channels. Therefore not just the reaction Q-value, but proton, alpha, and neutron separation energies can all be relevant for a given reaction rate. 

(3) For calculations of reverse reaction rates via detailed balance, Q-values and therefore nuclear masses enter exponentially (Eqs. \ref{eq:reverse},\ref{eq:reverse2}). Reverse rates are particularly important in the rp-process to correctly determine abundance distributions in (p,$\gamma$)-($\gamma,p)$ equilibrium that are critical to determine the reaction flow around waiting points \cite{schatzRpprocessNucleosynthesisExtreme1998b}. \cite{schatzDependenceXRayBurst2017a} investigated the sensitivity of X-ray burst models to this effect, and identified remaining mass uncertainties that affect burst models significantly. The conclusion is that for the purpose of calculating reverse rates, a mass accuracy of 10~keV or better is needed.

\subsection{$\beta$-decay}
\label{subsec:beta}

All $\beta$-decaying isotopes in the rp-process have been studied experimentally at rare isotope facilities, and laboratory decay rates have been measured with sufficient precision~\citep{schatzXrayBinaries2006a,meiselNuclearPhysicsOuter2018a}. However, laboratory decay rates can be significantly modified in the astrophysical environment of X-ray bursts. There are three effects: (1) The capture of bound electrons can be a significant component of laboratory decay rates, but this process is absent in the stellar environment where at the extreme conditions during X-ray bursts atoms are fully ionized. (2) With $kT$ reaching around 80~keV at 1~GK, low lying nuclear states with energies of the order of 100s~keV can be thermally populated significantly. These excited states can have very different decay rates compared to the ground state measured in the laboratory, leading to modifications of the effective decay rate. (3) Nuclei in X-ray bursts, while fully ionized, are embedded into an electron plasma, and thus in principle there is the possibility of continuum electron capture. To account for these effects, X-ray burst models typically do not use the experimental ground state decay rates, for example provided in JINA REACLIB, but instead the temperature and density dependent decay rates from various compilations that combine experimental data and theoretical corrections \cite{fullerStellarWeakInteraction1982b, fullerStellarWeakInteraction1982a, odaRateTablesWeak1994a, pruetEstimatesStellarWeak2003a}. This represents a significant hurdle for implementing updated $\beta^+$-decay rates from new experiments. Detailed decay schemes are important to determine the stellar environment corrections. Theory is needed to estimate decay rates from excited states, and to fill in lack of knowledge about the population of final states needed to remove the laboratory electron capture component. 

\begin{figure}
\includegraphics[width=0.7\textwidth]{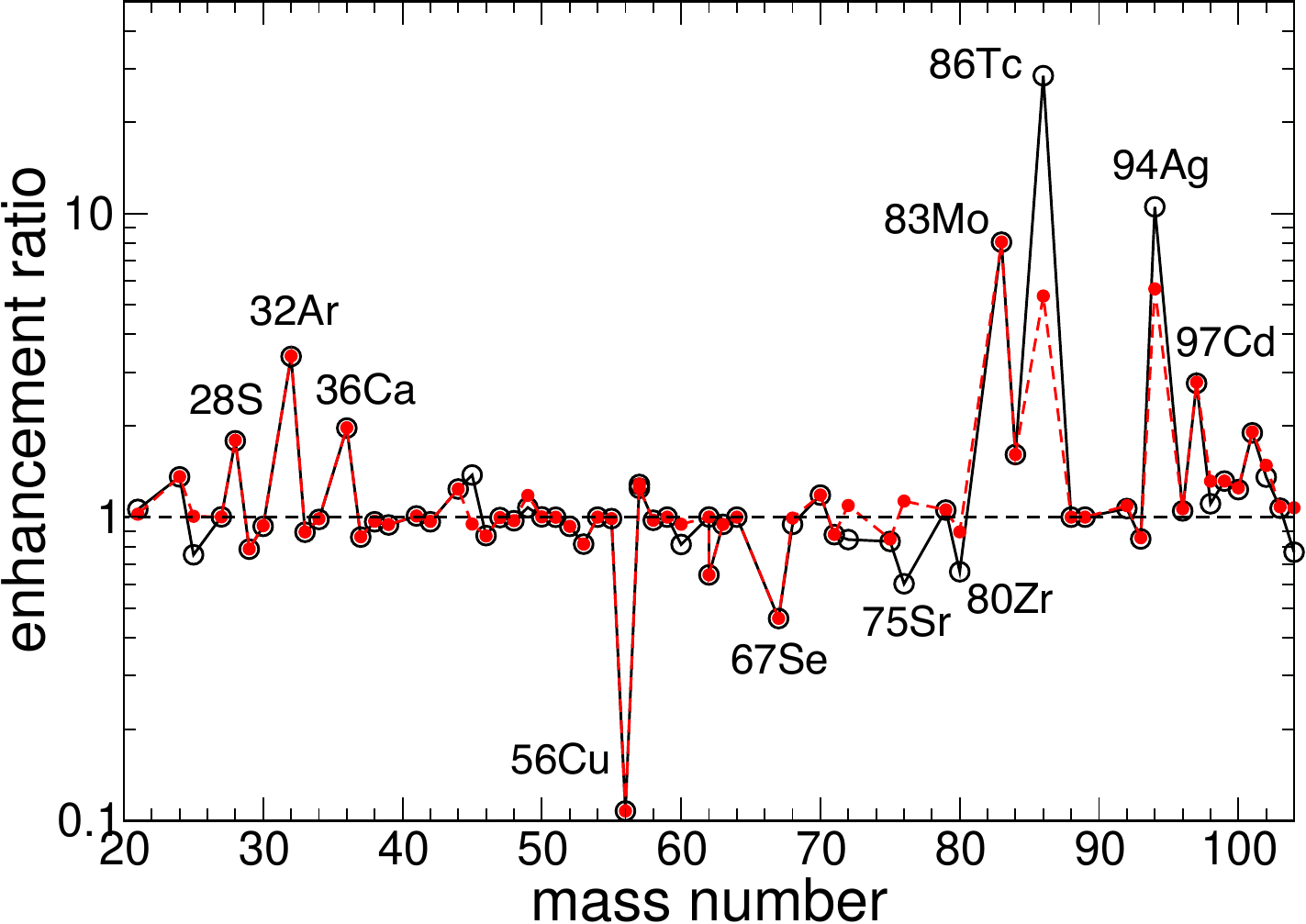}
\caption{Ratio of temperature and density dependent $\beta^+$ and electron capture decay half-lives to experimental ground state half-lives. Shown are enhancement ratios for $T=2$~GK, $Y_{\rm e} \rho$=10$^5$~mol/cm$^{-3}$ (black) and 
 $T=0.01$~GK, $Y_{\rm e} \rho$=10$^1$~mol/cm$^{-3}$ (red, dashed) as functions of parent mass number for the main $\beta$-decaying isotopes along the rp-process path.
\label{fig:beta_ratio}}
\end{figure}

Fig.~\ref{fig:beta_ratio} shows the stellar enhancement ratio of the half-lives of the $\beta$-decaying isotopes along the main rp-process path in a model calculation with an extreme rp-process reaching the $^{100}$Sn region \cite{schatzEndPointRp2001a} compared to experimental ground state rates in JINA REACLIB. To span the range of conditions applicable to X-ray bursts, we show data (in black) for the most extreme rp-process conditions predicted by models, temperatures of 2~GK and electron densities of $\rho Y_{\rm e}$=10$^5$ mole/cm$^3$ (while mixed H/He X-ray bursts can ignite at densities as high as $\rho=10^6$, densities at peak temperature are lower, in the 10$^5$ range, and $Y_{\rm e}$ is reduced rapidly through the $\beta$-decays in the rp-process). For comparison, we also show enhancement ratios for the lowest electron density and temperature conditions in the tables (in red). Cases where the data change with condition clearly indicate changes in rates due to the stellar environment. Cases where the changes are the same for the two conditions shown may indicate environmental effects that kick in at very low densities and temperatures, or, more likely, differences due to updates of experimental rates since the time the rate tables were created. 

In most cases changes are modest, in particular changes due to temperature and density dependence. There are several reasons for this. rp-process nuclei tend to be near the proton drip line where $\beta$-decay Q-values tend to be large. A large Q-value window makes it likely that ground and thermally excited low lying states both have allowed transitions available, reducing differences between decay rates of ground and excited states. Also, the major rp-process waiting points tend to be even even nuclei with relatively high energies for the first excited state. The effect of thermal population on half-lives of major rp-process waiting points has been investigated in the past. \cite{schatzRpprocessNucleosynthesisExtreme1998b} used QRPA calculations for decay rates of the first excited 2$^+$ states to obtain temperature dependent half-lives for the major rp-process waiting points $^{64}$Ge, $^{68}$Se, and $^{72}$Kr and demonstrated that significant changes require temperatures above 2~GK, just beyond the X-ray burst temperature range. Using a different QRPA model \cite{sarrigurenStellarWeakDecay2011a} confirmed this picture, and extended calculations of rp-process waiting points along the $N=Z$ line from $^{64}$Ge to $^{100}$Sn showing that both, temperature and density induced continuum electron capture effects are very small under X-ray burst conditions. A recent calculation using the projected shell model comes to similar conclusions \cite{chenStellarWeakinteractionRates2024}. In principle, magnetic fields can enhance some of the effects impacting decay rates \cite{2025arXiv250217953H}. However, surface magnetic fields in X-ray bursting systems are inferred to be relatively low ($<10^{10}$Gauss \cite{Bildsten2000}). The low field strengths allow the accreted matter to distribute across the neutron star surface, resulting in low local accretion rates and unstable nuclear burning that explains the observations of X-ray bursts. At such low field strength, the impact on nuclear processes is negligible \cite{2025arXiv250217953H}.

Nevertheless Fig.~\ref{fig:beta_ratio} indicates larger than a factor of 2 differences between measured ground state half-lives and calculated stellar half-lives for some cases. For $^{86}$Tc, $^{94}$Ag, and $^{83}$Mo there are strong (p,$\gamma$) branches and while the half-life may affect the local branchings it is unlikely to impact the broader X-ray burst features. For $^{67}$Se the stellar half-life is reduced by about a factor of 2 to 60~ms, however, with or without this change, the rp-process timescale in this region will be dominated by neighboring $^{68}$Se with its long half-life of 34~s. The factor of 10 decrease for the stellar half-life of $^{56}$Cu is interesting as it does reduce the efficiency of the possible bypass of $^{56}$Ni via the $^{55}$Ni(p,$\gamma$)$^{56}$Cu(p,$\gamma$)$^{57}$Zn reaction chain \cite{ongLowlyingLevelStructure2017}. There is also some significant enhancement of stellar half-lives in the lighter mass region, for example the 340~ms half-life of $^{32}$Ar is a factor of 3.4 larger than the current terrestrial laboratory value of 98~ms. While the other changes are smaller, they may also be significant and have non negligible effects on burst light curve modeling. It is however unclear if the differences stem from genuine stellar environment effects, or new experimental information obtained since the stellar reaction rates were calculated (or both). 
However, an update of the weak interaction database using new experimental information is beyond the scope of this paper, which focuses on charged particle reactions. 

$\beta$-delayed proton emission can in principle also affect rp-process nucleosynthesis. While during the main rp-process, any proton emission is quickly negated by subsequent proton capture, $\beta$-delayed proton emission may affect proton abundance and reaction path during freezeout. Many $\beta$-decaying nuclei along the rp-process have significant $\beta$-delayed proton branches. While we are not aware of comprehensive theoretical predictions, there is a large amount of experimental data available \cite{batchelderRecommendedValuesDelayed2020}.  \cite{lorussoVdelayedProtonEmission2012} investigated the impact of $\beta$-delayed proton emission on X-ray burst simulations and found only very small effects. However, the study was limited to the $^{100}$Sn region and branchings were relatively small. \cite{SaxenaZn57betap2022} demonstrated impact of $^{57}{\rm Zn}(\beta p)$ on the bypass of the $^{56}{\rm Ni}$ waiting point, but did not demonstrate an impact on the X-ray burst light curve or ashes. For full consistency, $\beta$-delayed proton emission branchings should be incorporated into the development of temperature and density dependent rate tables as discussed above. No such data set is currently available.

\section{Updates of Experimental and Theoretical Nuclear Physics}
\label{sec:updates_overview}

We summarize here the various approaches and techniques that have been used to provide the data and models for the update of the X-ray burst nuclear physics in JINA REACLIB in this work. This provides an overview of current directions in experimental and theoretical nuclear physics work motivated by X-ray burst applications. Detailed discussions of individual updated reaction rates are provided in the appendices. 

\subsection{Direct Measurements with Radioactive Beams}
\label{subsec:DirMeasureRIB}

The majority of the reactions of interest involve radioactive nuclei near the proton drip line. In principle a direct measurement of the reaction of interest is the most straight forward way to obtain the information needed to determine the astrophysical reaction rate. In this approach a radioactive beam  of the heavy nucleus of interest is impinged on a proton or $^4$He target at the relevant astrophysical energy - typically in the 100~keV to 1~MeV center of mass energy range, and the reaction of interest is studied directly by detecting the reaction products~\citep[e.g.][]{randhawaFirstDirectMeasurement2020a,jayatissaFirstDirectMeasurement2022}. While reaction cross sections are relatively low, they are considerably higher than in most other stellar astrophysical environments owing to the relatively high relevant temperatures in X-ray bursts. However, this advantage is more than negated by the rp-process reaction path proceeding much further away from stability, where the production of sufficiently intense radioactive beams has been a major challenge. While cross section measurements as a function of energy are mostly out of reach, targeted measurements of the resonance strength of selected important narrow and isolated resonances are possible with modest beam intensities in the $10^6-10^9$~pps range. For this approach, resonances and their approximate energies need to be known beforehand. Therefore, even in cases where direct measurements are possible, they have to be carried out in concert with indirect studies that provide guidance. 

For (p,$\gamma$) reactions, the increased selectivity of recoil separators is usually required owing to the larger backgrounds involved in $\gamma$-ray detection. In this approach, the heavy reaction recoil nucleus is separated from the unreacted beam and detected in coincidence with $\gamma$-rays emitted in the capture reaction. This results in a very clean reaction signature. While radioactive beam production continues to be a challenge, a few measurements have been feasible. For the updates in this work we use results from direct measurements of the $^{17}$F(p,$\gamma$)$^{18}$Ne reaction with the DRS recoil separator at ORNL \cite{chippsFirstDirectMeasurement2009b}, and of the $^{18}$F(p,$\gamma$)$^{19}$Ne and $^{19}$Ne(p,$\gamma$)$^{20}$Na reaction with the DRAGON recoil separator at TRIUMF \cite{akersMeasurementRadiativeProton2016, wilkinsonDirectMeasurementKey2017}. At both facilities, a windowless gas target is used. The SECAR recoil separator at FRIB has been designed to cover the mass range relevant for X-ray bursts \cite{bergDesignSECARRecoil2018} and will take advantage of reaccelerated beams as they become available in the future. 

\subsection{Indirect Measurements with Radioactive Beams}
\label{subsec:IndirMeasureRIB}

Given the challenges of direct measurements, a broad range of indirect techniques have been employed that take advantage of the availability of low intensity radioactive beams \cite{iliadisSpectroscopicFactorsDirect2004,bertulaniNuclearAstrophysicsRadioactive2010,tribbleIndirectTechniquesNuclear2014a,bardayanTransferReactionsNuclear2016}.  The most common approach is to use proton or $\alpha$ transfer reactions in inverse kinematics to obtain spectroscopic factors of resonance states that can be used to constrain the particle width, which is often the dominant contributor to the resonant reaction rate. (In principle these are pickup reactions when carried out in inverse kinematics though we refer to them here in their normal kinematics terminology to avoid confusion). The drawback of this approach is that there are additional systematic uncertainties from the use of approximations and limitations in the reaction theory needed to extract the information from the measured cross sections and angular distributions. For proton capture, (d,n) or ($^3$He,d) proton transfers are the most obvious choices that have been frequently used with stable beams in the past. Both are challenging in inverse kinematics, (d,n) because of the difficulties to measure neutrons with sufficient resolution, and ($^3$He,d) because of the need of a $^{3}$He target with minimal or no windows. 

For this update we use results from several (d,n) reaction measurements using radioactive beams impinging on CD$_2$ plastic targets. \cite{kuvinMeasurement17F18Ne2017} use new scintillation detectors in their RESONEUT neutron array at Florida State University to perform a successful measurement of the $^{17}$F(d,n)$^{18}$Ne reaction with a 5.6~MeV/u $^{17}$F beam. \cite{PhysRevLett.122.232701} use the LENDA plastic scintillator array at Michigan State University for a measurement of the $^{23}$Al(d,n)$^{24}$Si reaction with the S800 spectrometer at higher energies of 48~MeV/u. However, the main information on resonance properties is obtained from $\gamma$-spectroscopy, while the neutron detectors provide additional constraints but do not resolve individual states. A number of experiments have taken advantage of the fact that, for certain energies and reactions of interest, the detection of the recoil, together with the protons or alpha particles from the decay of populated resonances in the final nucleus, is sufficient to reconstruct the energy and angle of the undetected neutron. These types of experiments include $^{18}$F(d,n)$^{19}$Ne measurements at ORNL to constrain the $^{18}$F(p,$\alpha$)$^{15}$O reaction \cite{adekolaFirstProtontransferStudy2011a}, as well as measurements of the 
$^{19}$Ne(d,n)$^{20}$Na and $^{25}$Al(d,n)$^{26}$Si reactions at Florida State University to constrain  the $^{19}$Ne(p,$\gamma$)$^{20}$Na and $^{25}$Al(p,$\gamma$) reactions \cite{belargeExperimentalInvestigation192016,peplowskiLowestProtonResonance2009}, respectively. Another approach is to constrain spectroscopic factors from angle integrated cross sections for (d,n) reaction populated resonances via $\gamma$-ray spectroscopy. This approach takes advantage of the fact that the resonances where proton spectroscopic factors are needed are the ones where the proton width is small and hence decay by $\gamma$-ray emission. The angle integrated cross sections can then be obtained from the measured $\gamma$ intensities. This technique has been used at Michigan State University using the GRETINA $\gamma$-ray detection array to probe $^{30}$P(p,$\gamma$)$^{31}$S \cite{kankainenMeasurementKeyResonance2017}, $^{56}$Ni(p,$\gamma$)$^{57}$Cu \cite{kahlSingleparticleShellStrengths2019}, and $^{57}$Cu(p,$\gamma$)$^{58}$Zn \cite{PhysRevLett.113.032502}. 

Similar to proton transfer reactions, $\alpha$-transfer reactions can be used to probe $\alpha$-induced reactions on unstable nuclei. \cite{PhysRevC.80.045805} used ($^6$Li,d) and ($^7$Li,t) reactions at Florida State University to constrain the $^{14}$C($\alpha$,$\gamma$)$^{18}$O reaction. 

As the rp-process proceeds to a large extent at or beyond the $N=Z$ line, isospin symmetry can be used to perform transfer reactions on the isospin mirror. For example, instead of a (d,n) proton transfer, a (d,p) neutron transfer can be performed on the mirror nucleus. This has several advantages: often the isospin mirror nucleus is closer to stability (or even stable) hence beam intensities are higher. In addition, (d,p) offers significant experimental advantages as protons are much easier to detect with sufficient resolution than neutrons. The technique has been used to constrain the $^{26}$Al(p,$\gamma$)$^{27}$Si reaction via $^{26}$Al(d,p) using a radioactive $^{26}$Al beam at TRIUMF and detecting protons with the TUDA array \cite{Lotay2020}. Stable beam (d,p) measurements on $^{26}$Mg have been performed at the Czech Academy of Science accelerator to constrain proton capture on the unstable $^{26}$Si mirror. Stable beam $\alpha$-transfer reaction measurements on $^{13}$C were performed at Orsay to constrain the $^{13}$N($\alpha$,p) reaction \cite{meyerEvaluation13N16O2020}. 

Instead of proton transfer, proton breakup can also be used to probe the wave function of a resonance state for a (p,$\gamma$) reaction. \cite{PhysRevC.86.015806} used the peripheral breakup of a $^{24}$Si radioactive beam on a carbon target to determine asymptotic normalization coefficients for resonances in $^{23}$Al(p,$\gamma$)$^{24}$Si that can be related to particle widths. Resonant elastic proton scattering in inverse kinematics was used by \cite{huExaminationRoleO142014a} using a $^{17}$F beam from the CRIB facility at the University of Tokyo to obtain $^{14}$O($\alpha$,p)$^{17}$F resonance total widths and angular momenta. 

The Trojan Horse method \cite{TuminoTrojanHorseReview2025} is another indirect approach that can be used to obtain information on a proton induced reaction by transferring the proton from another nucleus. In this approach, kinematics is chosen such that the reaction proceeds in the quasi-free regime. While the nucleus bringing in the proton, the Trojan Horse, has energies above the Coulomb barrier, one can extract from the data the reaction cross section of the proton induced reaction at low astrophysical energies. This method has been applied for the first time with a radioactive beam to probe the $^{18}$F(p,$\alpha$)$^{15}$O reaction via the $^2$H($^{18}$F,$\alpha ^{15}$O)n reaction at the CRIB facility at the University of Tokyo \cite{cherubiniFirstApplicationTrojan2015}.

Another approach is to measure the inverse of a proton capture via Coulomb breakup \cite{baurCoulombDissociationSource1986}. A radioactive beam impinges a high $Z$ target, and photons from the Coulomb field of the target induce the inverse ($\gamma$,p) reaction. Rp-process reactions are in principle well suited for this technique because in many cases protons are only weakly bound, leading to very few -- if any -- excited states in the target which are particle-bound. Note that contributions of thermally excited states in the target are not detected in ($\gamma$,p) experiments. For this update we use data from the Coulomb breakup of $^{27}$P obtained at GSI in Darmstadt \cite{r3bcollaborationCoulombDissociation272016} and in a separate experiment at RIKEN in Tokyo \cite{toganoResonanceStates272011}, which constrain the $^{26}$Si(p,$\gamma$)$^{27}$P reaction. 

For nuclei far from stability, excitation energies of proton or $\alpha$ capture resonances are small owing to the low capture Q-values, while $\beta$-decay Q-values are large. $\beta$-decay can therefore be a powerful method to populate resonance states and observe subsequent particle or $\gamma$-decays to determine the decay widths needed for the resonance strength directly (Eq.~\ref{eq:omegagamma}). The challenge is that the most important decay width will be related to the smaller branching, typically smaller by orders of magnitude, and in addition, $\beta$-decay feeding can be relatively weak. Nevertheless this is a powerful method that can be applied to many important cases. For charged particle branchings, an additional challenge is the detection of low energy protons and $\alpha$-particles in the presence of a large $\beta$-particle background.  A variety of techniques have been developed to address this challenge:  Resonances in the $^{25}$Al(p,$\gamma$)$^{26}$Si reaction have been investigated via the $\beta$-decay of $^{26}$P by \cite{thomasBetadecayProperties25Si2004} implanting ions at the edge of a Si detector at GANIL and \cite{janiakEnsuremathBetaDelayed2017} using an optical time projection chamber (TPC) at Dubna. Proton branchings of resonances in $^{26}$Si have been studied by $\beta$-delayed proton emission of $^{27}$S at Lanzhou using a cryogenically cooled stack of Si detectors. The proton branching of a $^{30}$P(p,$\gamma$)$^{31}$S resonance was measured via $\beta$-delayed proton emission of $^{31}$Cl using the gas filled GADGET detector at Michigan State University. In addition,  \cite{bennettClassicalNovaContributionMilky2013} determined the $\gamma$-branching of a resonance in $^{25}$Al(p,$\gamma$)$^{26}$Si by observing $\beta$-delayed $\gamma$-rays from the decay of $^{26}$P implanted into a Ge detector.  

Finally a variety of radioactive beam techniques have been used to simply populate resonance states and constrain excitation energies and spins. With resonance energies entering reaction rate calculations exponentially (Eq.~\ref{eq:narrowresonance}) such measurements can be very powerful, and reduce uncertainties from many orders of magnitude to factors of 2-3 (e.g. \cite{clementNewApproachMeasuring2004a}). Techniques include the use of the (d,p) transfer reactions, which has the advantage of using a beam closer to stability where higher intensities can be achieved. This reaction was used at ORNL to probe resonances in $^{18}$F(p,$\alpha$)$^{15}$O using the windowless JENSA gas target \cite{bardayanFirstScienceResult2015a} and at GANIL probing resonances in the $^{35}$K(p,$\gamma$)$^{36}$Ca reaction using the CRYPTA liquid hydrogen target \cite{lalanneEvaluation35K36Ca2021}. Inflight $\gamma$-spectroscopy at higher energies using a variety of reactions has been used at Michigan State University using the GRETINA $\gamma$-detection array. Excitation energies of resonances in $^{55}$Ni(p,$\gamma$)$^{56}$Cu, $^{23}$Al(p,$\gamma$)$^{24}$Si, and $^{24}$Al(p,$\gamma$)$^{25}$Si have been deduced from measurements of (d,2n), 2n knockout, and 1n knockout reactions, respectively \cite{ongLowlyingLevelStructure2017, PhysRevC.101.031303,PhysRevC.97.054307}. 

\subsection{Indirect Measurements with Stable Beams to Probe Reactions with Radioactive Nuclei}
\label{subsec:IndirMeasureSBforRN}

Up to around Ca the X-ray burst reaction sequence proceeds relatively close to stability. Therefore, various types of stable beam reactions can be used to populate states in neutron deficient radioactive nuclei in the rp-process to determine properties of resonances. In this update we use data from $^{32}$S(p,t)$^{30}$S at RCNP in Osaka for $^{29}$P(p,$\gamma$)$^{30}$S \cite{almaraz-calderonLevelStructure302012}, $^{36}$Ar(d,t)$^{35}$Ar at Garching for $^{34}$Cl(p,$\gamma$)$^{35}$Ar \cite{fryDiscovery34MCl2015}, $^{28}$Si($^3$He,n)$^{30}$S at Notre Dame \cite{almaraz-calderonLevelStructure302012} and Argonne National Laboratory \cite{lotayLevelStructure302012} for $^{29}$P(p,$\gamma$)$^{30}$S, and $^{32}$S($\alpha$,p)$^{35}$Cl to probe the mirror relevant for $^{34}$Cl(p,$\gamma$)$^{35}$Ar \cite{setoodehniaExperimentalStudyCl2019}. The $^{19}$F($^3$He,t)$^{19}$Ne reaction has been used extensively to study $^{19}$Ne, which is the Compound nucleus for $^{18}$F(p,$\alpha$)$^{15}$O and $^{15}$O($\alpha$,$\gamma$)$^{19}$Ne. Experiments have been performed at Argonne National Laboratory using the ORRUBA charged particle detection system \cite{hallKey19NeStates2019} and at Garching using the magnetic spectrograph \cite{lairdGRayEmissionNovae2013, parikhSpectroscopy19NeThermonuclear2015}. In addition, $\gamma$-spectroscopy using fusion evaporation reactions with stable beams at Argonne National Laboratory can be used to populate states of interest in neutron-deficient nuclei and study their spins and excitation energies. $\gamma$-decaying resonances in $^{25}$Mg(p,$\gamma$)$^{26}$Al have been measured with Gammasphere \cite{kankainenDecayKey92keV2021a}, and in $^{33}$Cl(p,$\gamma$)$^{34}$Ar using GRETINA \cite{PhysRevLett.124.252702}.

\subsection{Measurements of Reactions on Stable Nuclei}
\label{subsec:MeasureforSN}

While the majority of important reactions during an X-ray burst involve unstable nuclei, reactions on stable isotopes can play a role at burst ignition, for example proton induced reactions on heavy stable nuclei present in the accreted material or produced by a preceding burst. We include updates of a number of reactions on stable nuclei using data from a variety of techniques. Accelerators deep underground offer the possibility to push direct measurements to lower cross sections by suppressing the cosmic-ray induced background in detectors detecting the reaction products. We include several new results from the LUNA underground accelerator at the INFN Laboratori Nazionali Del Gran Sasso, including new data on $^{17}$O(p,$\alpha$)$^{14}$N \cite{lunacollaborationImprovedDirectMeasurement2016}, $^{23}$Na(p,$\gamma$)$^{24}$Mg \cite{boeltzigDirectMeasurementsLowenergy2019}, and $^{25}$Mg(p,$\gamma$)$^{26}$Al \cite{limataNewExperimentalStudy2010,strieder25Mg26AlReaction2012}. Data from direct measurements of the $^{19}$F(p,$\alpha$)$^{16}$O reaction were obtained at the JUNA underground laboratory in Jinping Mountain \cite{zhangDirectMeasurementAstrophysical2021}. 
The $^{23}$Na($\alpha$,p)$^{26}$Mg reaction was studied directly in inverse kinematics using the MUSIC detector at Argonne National Laboratory \cite{avilaExperimentalStudyAstrophysically2016}. \cite{andersonRemeasurement33Cl2017} used an activation technique at the University of Notre Dame to achieve the sensitivity needed for a direct measurement of the $^{33}$S($\alpha$,p)$^{36}$Cl reaction. 

We also include a number of results from indirect studies using stable beams, including ($^3$He,d) proton transfer measurements at the TUNL Enge split-pole spectrograph related to $^{23}$Na(p,$\gamma$)$^{24}$Mg \cite{marshallNewEnergy133keV2021}, ($^7$Li,$^6$He) proton transfer measurements at the Q3D magnetic spectrometer in Beijing for $^{25}$Mg(p,$\gamma$)$^{26}$Al \cite{liIndirectMeasurement572020}, and a measurement using the Trojan horse method via the $^{6}$Li($^{19}$F,p$^{22}$Ne)d reaction at Zagreb for $^{19}$F(p,$\alpha$)$^{16}$O \cite{DAgata_2018}.

\subsection{Measurements of Reactions on Excited States}
\label{subsec:MeasureforES}

Experimental data on reactions on excited states are extremely limited as the majority of techniques require exciting the nucleus prior to the reaction measurement. An exception are direct measurements of particle decay branches from resonances. \cite{belargeExperimentalInvestigation192016} measured proton branchings from resonances in the $^{19}$Ne(p,$\gamma$)$^{20}$Na reaction to the first excited state in $^{19}$Ne. These branchings can be used to directly determine the resonance strength for proton capture on the $^{19}$Ne first excited state.

 Isomeric states that live long enough to provide a beam or a target offer rare experimental opportunities for direct and indirect measurements on excited states. A beam of $^{26m}$Al in its 228~keV isomeric state with a life time of 6.3~s was produced at Argonne National Laboratory taking advantage of the selectivity of the beam producing $^{26}$Mg(p,n)$^{26m}$Al reaction. The beam was used to measure mirror spectroscopic factors using the (d,p) neutron transfer reaction \cite{almaraz-calderonStudy26272017}. Isomeric beams of $^{34m}$Cl and $^{38m}$K of interest for rp-process studies are under development at FRIB at Michigan State University \cite{shehuExperimentalStudy34mCl2022,chipps38IsomerProduction2018a}. Meanwhile, a novel approach to measurements of reactions on an excited state was employed by \cite{PhysRevLett.126.042701}, who took advantage of the fact that the $^{26}$Si ground state is part of an isospin triplet with $^{26m}$Al. Thus, cross sections from (d,p) transfer reactions on $^{26}$Si can be used to obtain spectroscopic information for $^{26m}$Al(p,$\gamma$).

\subsection{Measurements and Calculations of Masses}
\label{subsec:newmasses}

\begin{figure}
\includegraphics[width=0.5\textwidth]{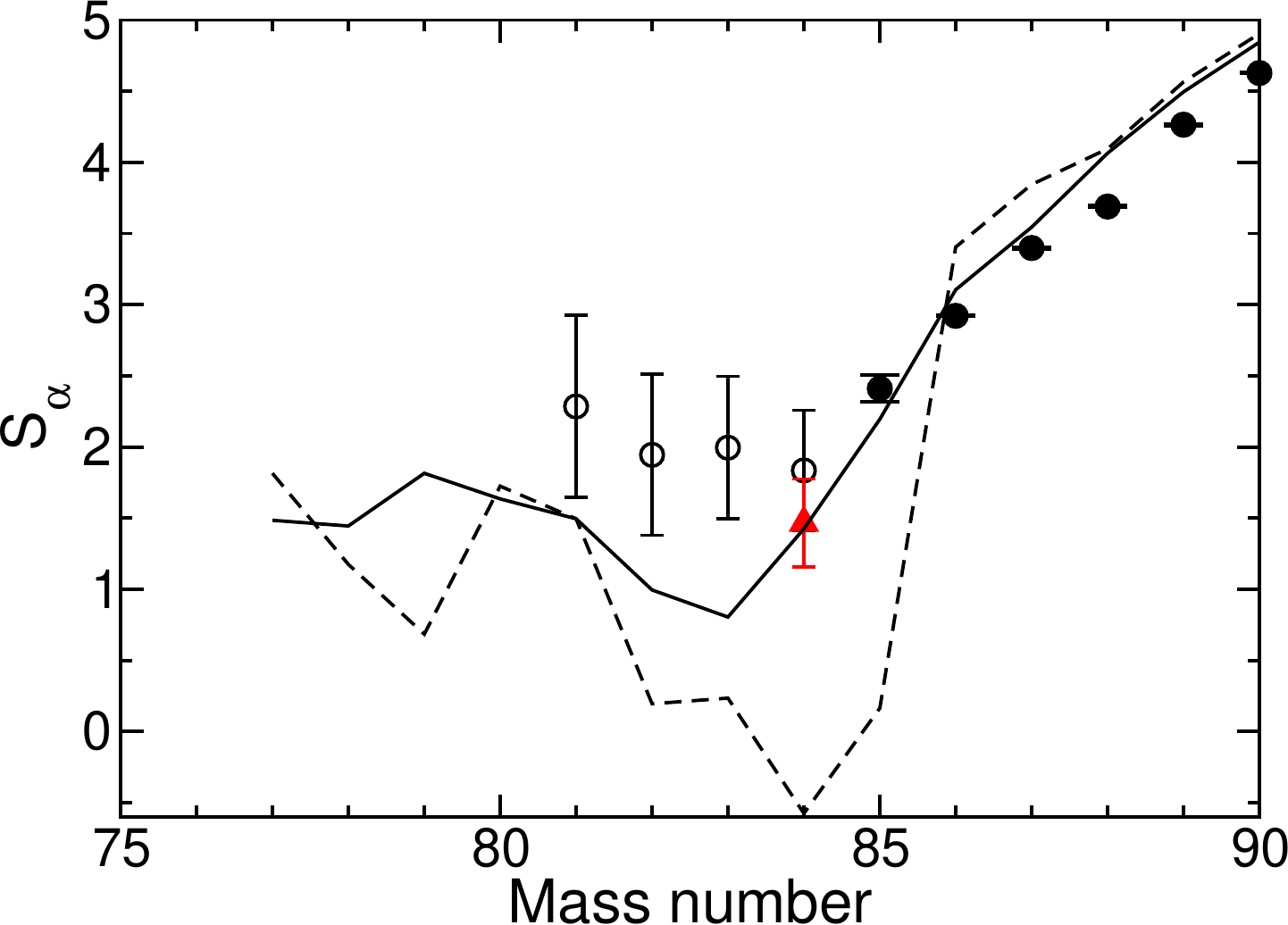}
\caption{$\alpha$-separation energies of Mo isotopes predicted by FRDM2012 (solid line), FRDM1993 (dashed line), AME20 extrapolations (open circles), together with experimental data (solid circles) and a data point for $^{84}$Mo obtained from AME2020 extrapolations for the $^{84}$Mo mass and a new measurement of the $^{80}$Zr mass (red triangle) \cite{hamakerPrecisionMassMeasurement2021}. 
\label{fig:samo}}
\end{figure}

A broad range of techniques has been used to measure ground state masses for rp-process calculations. Penning traps can easily achieve the required accuracy. Recent measurements include a measurement of the $^{24}$Si mass to improve resonance energies in the $^{23}$Al(p,$\gamma$)$^{24}$Si reaction \cite{PhysRevC.106.L012801} and the $^{56}$Cu mass for resonances in $^{55}$Ni(p,$\gamma$)$^{56}$Cu \cite{valverdeHighPrecisionMassMeasurement2018}, both using the LEBIT trap at Michigan State University. Penning trap mass measurements of $^{31}$Cl at JYFLTRAP in Jyv{\"a}skyl{\"a} improved the $^{30}$S(p,$\gamma$)$^{31}$Cl reaction rate \cite{kankainenMassAstrophysicallyRelevant2016a}. Most recently the mass of $^{80}$Zr was measured by LEBIT \cite{hamakerPrecisionMassMeasurement2021}. There is a long standing question, whether an $\alpha$ unbound $^{84}$Mo as predicted by the FRDM 1993 mass model, can lead to the formation of a Zr-Nb-Mo cycle in the rp-process \cite{schatzRpprocessNucleosynthesisExtreme1998b} via a strong $^{84}$Mo($\gamma$,$\alpha$)$^{80}$Zr reaction branch. With the measurement of the $^{80}$Zr mass the $\alpha$-separation energy of $^{84}$Mo $S_\alpha$($^{84}$Mo) can now be more reliably determined. As Fig.~\ref{fig:samo} shows, already the revised FRDM 2012 mass model \cite{mollerNuclearGroundstateMasses2016} predicted a much more modest reduction of $S_\alpha$($^{84}$Mo). The new measurement is somewhat lower but agrees within uncertainties compared to the AME20 mass extrapolations \cite{wangAME2020Atomic2021}, and confirms the predictions of the FRDM2012, which is in excellent agreement. Such a modest reduction in  $S_\alpha$($^{84}$Mo) would exclude the formation of a Zr-Nb-Mo cycle in the rp-process. However, a measurement of $^{84}$Mo would be desirable to confirm this conclusion. 

An alternative technique uses multi-reflection time-of-flight spectrometers (MR-TOF). TRIUMF's TITAN facility used this technique to measure the masses of Ga isotopes in the rp-process \cite{paulMassMeasurementsGa63602021}. With accuracies of 10-30~keV the method is somewhat less precise than the Pennning Trap (though for low statistics common in the rp-process Penning Trap accuracies are similar) but has the advantage of being more sensitive, for example a large isobaric contamination in the incoming radioactive beam can be removed effectively. 

Storage ring mass measurements are another technique that has recently been applied successfully to masses of interest for rp-process calculations. Implemented at GSI and Lanzhou, the method is less precise than Penning trap measurements, but can be performed with lower beam intensities and many isotopes can be measured simultaneously. In particular measurements have been performed on $^{65}$As at the Cooler Storage Ring CSRe at Lanzhou. Together with the well known $^{64}$Ge mass, the $^{65}$As mass is needed to determine its proton separation energy $S_p(^{65}$As), which plays a critical role in X-ray burst models defining the impedance imposed by the important $^{64}$Ge waiting point affecting strongly the composition of the produced isotopes and the burst light curve. The measurement resulted in $S_p(^{65}$As)=-90(85)~keV \cite{tuDirectMassMeasurements2011b}. A recent re-measurement using the same technique obtained $S_p(^{65}$As)=-221(42)~keV \cite{zhouMassMeasurementsShow2023a} thus confirming the earlier result. With this $S_p$ value, $^{64}$Ge $\beta$-decay and proton capture branch are expected to be of the same order of magnitude for the most typical rp-process conditions ($T=1.5$~GK, $Y_p \rho$=10$^5$~g/cm$^3$, see Fig. 2 in \cite{tuDirectMassMeasurements2011b}), indicating the $^{64}$Ge waiting point will impede the rp-process significantly, though the exact effect will depend strongly on the particular burst properties and burst model \cite{tuDirectMassMeasurements2011b,zhouMassMeasurementsShow2023a}. 

$\beta$-delayed proton emission measurements can be used to determine directly proton separation energies by measuring the proton energy, provided the proton is emitted from a known state. The method is particularly useful to measure proton separation energies of proton unbound nuclei beyond the neutron drip line, which are needed to determine 2p capture rates at major waiting points \cite{schatzRpprocessNucleosynthesisExtreme1998b}. These nuclei have lifetimes of much less than 1~ms and can thus not be trapped or measured by time-of-flight. In this case, the $\beta$-decay populates the proton decaying ground state. The method has been used successfully at Michigan State University to determine $S_p(^{69}$Br) for 2p capture on the $^{68}$Se waiting point using $\beta$-delayed proton emission of $^{69}$Kr \cite{delsantoVdelayedProtonEmission2014} and $S_p(^{73}$Rb) for 2p capture on the $^{72}$Kr waiting point using $\beta$-delayed proton emission of $^{73}$Sr \cite{hoffInfluence73RbAshes2020}. $\beta$-delayed proton emission has also been used at Lanzhou to determine the masses of $^{27}$P from $\beta$-delayed proton  emission of $^{27}$S. In this case, the ground state mass is obtained from the measurement of proton and $\gamma$-decay energies of the populated state in $^{27}$P, thus connecting the known $^{26}$Si mass with the unknown $^{27}$P mass. 

\begin{figure*}
\includegraphics[width=1.0\textwidth]{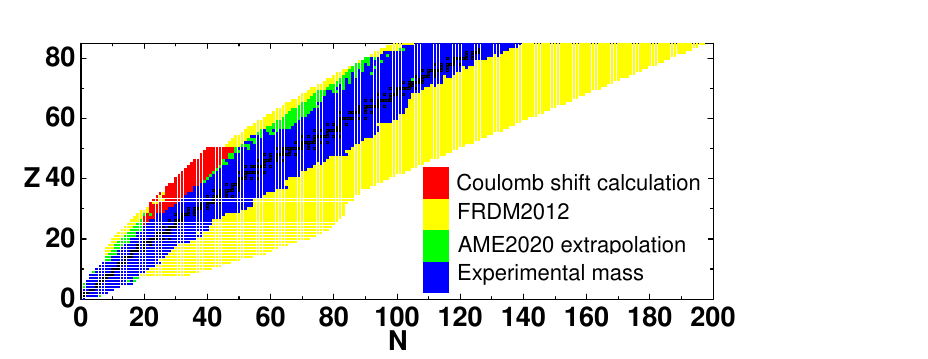}
\caption{Nuclear masses used in the present study. \label{fig:mass_sources}}
\end{figure*}

In this work we use a mass table as direct input in the X-ray burst simulation and to calculate updated statistical model rates. The majority of the needed masses are now experimentally determined and we employ the 2020 Atomic Mass Evaluation (AME) \cite{wangAME2020Atomic2021}. For unknown masses beyond the $N=Z$ line we use the Coulomb shift calculations from \cite{brownProtonDriplineCalculations2002a} that relate them to experimentally known masses of their isospin mirrors and have been shown to be accurate to at least 100~keV. This leaves only a small number of masses above $Z=40$ between the boundary of experiments and the $N=Z$ line. To minimize systematic effects from model switching, we employ the AME2020 mass extrapolations for this region, instead of a global mass model. With these choices, global mass models are not needed anymore for X-ray burst models. Fig.~\ref{fig:mass_sources} summarizes the sources for masses. For the individual reaction rate updates we use the masses employed in the works that calculated the reaction and resonance properties we use. 

\subsection{Shell Model Calculations}
\label{subsec:shellmodel} 

The nuclear shell model can also be used to make predictions for reaction rates dominated by single isolated resonances. It is of particular importance in the case of incomplete experimental data to fill in missing information to obtain the complete data set needed for reaction rate calculations. For the rate evaluations implemented in this update, data from an sd-shell model using the UDSB-cdpn interaction \cite{brownNewUSDHamiltonians2006} have been used for $^{23}$Al(p,$\gamma$)$^{24}$Si \cite{wolfConstrainingNeutronStar2019a}, $^{24}$Al(p,$\gamma$)$^{25}$Si \cite{longfellowMeasurementKeyResonances2018}, and $^{25}$Al(p,$\gamma$)$^{26}$Si \cite{liangSimultaneousMeasurementDelayed2020}. For the heavier sd-shell nuclei, data from more recent combined sd- and fp-shell model spaces have been used, such as the WBP Hamiltonian \cite{brownShellmodelStudiesAstrophysical2014} in the \code{NuShellX} model \cite{brownShellModelCodeNuShellXMSU2014} for $^{30}$P(p,$\gamma$)$^{31}$S and $^{34}$Cl(p,$\gamma$)$^{35}$Ar, as well as another sdfp-shell interaction used for 
$^{35}$K(p,$\gamma$)$^{36}$Ca \cite{lalanneEvaluation35K36Ca2021}. In the fp-shell the most common shell model calculations use the GXPF1A interaction \cite{honmaShellmodelDescriptionNeutronrich2005} with \code{NuShellX}. Data from this approach have been used for the $^{55}$Ni(p,$\gamma$)$^{56}$Cu \cite{valverdeHighPrecisionMassMeasurement2018}, $^{56}$Ni(p,$\gamma$)$^{57}$Cu \cite{kahlSingleparticleShellStrengths2019}, and $^{57}$Cu(p,$\gamma$)$^{58}$Zn \cite{PhysRevLett.113.032502} reactions. \cite{lamReactionRates64Ge2016} used the same shell model to calculate the resonant $^{65}$As(p,$\gamma$)$^{66}$Se reaction rate. 

In principle, the shell model could also be used to predict reaction rates for which no experimental information is available, in parts of the nuclear chart where the respective model space is applicable. This is especially of interest for cases where the applicability of the statistical model may be questionable. However, no new shell model data sets are available at this point. We continue to recommend in REACLIB reaction rate predictions from the fp-shell model \code{ANTOINE} \citep{fiskerShellModelBased2001}  for cases where no new information is available (see Fig.~\ref{fig:reaclib_sources}).

\subsection{Nucleon Potentials}
\label{subsec:npomp} 

The optical model potential (OMP) is an essential input for the calculation of cross sections and reaction rates with both the Hauser-Feshbach and the potential models. Usually, the OMP is either described with a phenomenological Woods-Saxon (WS) expression or constructed by a microscopic folding approach.

We use here a global phenomenological WS OMP for the system of nucleon + target that has been developed and described in detail in \cite{KONING2003231}. This WS OMP has been thoroughly verified with extensive experimental data for nucleon induced reactions with incident energies from 1 keV up to 200 MeV and target mass numbers from $A$ = 24 to $A$ = 209. This WS OMP specifically provides a smooth and unique functional form for the energy dependence of the potential depths and the physically constrained geometry parameters. The explicit expression reads
\begin{eqnarray}
U(E,r)&=&-V_{v}(E,r)-iW_{v}(E,r)-iW_{s}(E,r)+V_{s.o.}(E,r)+V_{c}(r),
\label{ppot1}
\end{eqnarray}
\noindent where $V_{v}$ and $W_{v,s}$ are the real and imaginary components of the volume-central (v) and surface-central (s) potentials, respectively; $V_{s.o.}$ is the spin-orbit potential; and $V_{c}$ is the Coulomb potential. 

The central potentials are separated into energy-dependent well depths and energy-independent form factor, namely
\begin{eqnarray}
V,W_{v}(E,r)=V,W_{v}(E)\times f(r,R_{v},a_{v}),
\label{ppot2}
\end{eqnarray}
\noindent and
\begin{eqnarray}
W_{s}(E,r)=-4a_{s}W_{s}(E)\times d(f(r,R_{s},a_{s}))/dr.
\label{ppot3}
\end{eqnarray}
\noindent The form factor $f$ is given by the WS shape
\begin{eqnarray}
f(r,R_{i},a_{i})=(1+\exp[(r-R_{i})/a_{i}])^{-1},
\label{ppot4}
\end{eqnarray}
\noindent where the geometry parameters are the radius $R_{i}$ = $r_{i}A^{1/3}$ with $A$ being the atomic mass number and the diffuseness $a_{i}$.

A possible alternative would be the Bruy\`eres-le-Ch\^ atel re-normalization \cite{PhysRevC.63.024607} of the Jeukenne-Lejeune-Mahaux potential \cite{PhysRevC.16.80}, referred to as JLMB. JLMB is a global semi-microscopic nucleon-nucleus OMP adjusted on experimental data of $A$ = 30-240 nuclei and for the energies ranging from 10 keV up to 200 MeV \cite{PhysRevC.61.034306,PhysRevC.63.014604}. The JLMB potential has been renormalized in Refs. \cite{PhysRevC.58.1118,PhysRevC.63.014604} to improve the agreement between experimental and predicted observables for a large set of data and is also available for the calculations of nucleon capture reaction rates, especially neutron capture \cite{PhysRevC.86.045801,PhysRevC.90.024604}.

Previous studies \cite{XuEPJwebconf2017,PhysRevC.98.054601} demonstrated that for most (p,$\gamma$) and ($\gamma$,p) reactions involving nuclei lying between the proton drip line and the valley of $\beta$ stability, the ratios of the reaction rates computed by the WS OMP and those computed by the JLMB OMP agree within a factor of 2. This indicates that such calculations are not drastically sensitive to the choice of OMP that determines the transmission coefficient of the proton channel. Moreover, the WS OMP is available and suitable for the calculations of most nuclei, of which the parameters can be flexibly adjusted to reproduce the experimental data in a wider energy range. For these reasons, the phenomenological WS OMP \cite{KONING2003231} with its parameterization as a function of $Z$, $A$ and incident energy $E$ is adopted in the present study.

\subsection{$\alpha$-particle Potential} 
\label{subsec:aomp} 

The $\alpha$-nucleus OMP (AOMP) is the most important input for the Hauser-Feshbach model calculation of $\alpha$-induced reaction cross sections and astrophysical reaction rates. Starting with the pioneering work of Somorjai {\it et al.}\ \cite{Somorjai_AaA1998_sm144ag} on the $^{144}$Sm($\alpha$,$\gamma$)$^{148}$Gd reaction, it was realized that the widely used AOMP by McFadden and Satchler \cite{McFadden_NPA1966_aomp} overestimates the cross sections for heavy target nuclei by orders of magnitude at energies far below the Coulomb barrier (i.e., in the astrophysically relevant energy region around the Gamow window). Conversely, it has been shown that the AOMP by McFadden and Satchler is able to reproduce the average cross sections for lighter targets with masses below approximately $A = 50$ with much smaller deviations of the order of a factor of two \cite{Mohr_EPJA2015_A20-50}.

The simple and energy-independent AOMP by McFadden and Satchler uses a purely phenomenological Woods-Saxon parametrization in the real and imaginary parts of the AOMP, which is obtained from data for elastic scattering of $\alpha$ particles on nuclei from O to U at 24.7 MeV. Such a phenomenological approach has been replaced by a microscopically founded double-folding approach for the real part of the AOMP in many subsequent studies, see e.g.,\ \cite{Demetriou_NPA2002_aomp,Avrigeanu_NPA2003_aomp,Mohr_ADNDT2013_atomki-v1}. Based on these studies, three versions for a global AOMP have been suggested by Demetriou {\it et al.}\ \cite{Demetriou_NPA2002_aomp} which were obtained from a fine-tuning of the parameters in the real part and an adjustment of the imaginary part to the available experimental reaction cross sections. Avrigeanu {\it et al.}\ \cite{Avrigeanu_PRC2014_aomp} followed a similar strategy, but in practice both the real and the imaginary part of the AOMP were finally parameterized by Woods-Saxon potentials with a large number of mass- and energy-dependent parameters. Overall, a good description of the available reaction cross sections was obtained in \cite{Avrigeanu_PRC2014_aomp}. The AOMPs by Demetriou {\it et al.}\ and by Avrigeanu {\it et al.}\ have been implemented as options in the \code{TALYS} code, the latter being the default AOMP since several versions.

Contrary to the above AOMPs by Demetriou {\it et al.}\ \cite{Demetriou_NPA2002_aomp} and Avrigeanu {\it et al.}\ \cite{Avrigeanu_PRC2014_aomp} whose parameters had to be adjusted to experimental reaction cross sections, the new Atomki-V2 AOMP \cite{Mohr_PRL2020} uses a different approach without any adjustment to experimental reaction cross sections. The real part of the Atomki-V2 AOMP is also based on the double-folding approach. The potential strength is adjusted using an analysis of elastic scattering data around the Coulomb barrier \cite{Mohr_ADNDT2013_atomki-v1}. The imaginary potential is a deep and short-ranged Woods-Saxon potential. This combination of a well-founded real part and a short-ranged imaginary part ensures that the Atomki-V2 AOMP behaves similar to simple barrier transmission models. The transmission of the incoming $\alpha$-particle through the Coulomb barrier is governed by the real part of the AOMP and is calculated with small uncertainties. The deep and short-ranged imaginary part is responsible for the absorption of all $\alpha$-particles that are able to tunnel through the barrier. This leads to a very robust prediction of the total cross sections of $\alpha$-induced reactions with typical deviations far below a factor of two at energies below the Coulomb barrier for practically all heavy nuclei under study. In particular, this approach avoids complications with the tail of the imaginary part of the AOMP at large radii (far outside the colliding nuclei) which may become numerically dominant in the calculation of the total reaction cross sections at deep sub-Coulomb energies for the other AOMPs. Astrophysical reaction rates from the Atomki-V2 AOMP were provided in \cite{Mohr_ADNDT2021} for targets above $Z \ge 26$ using an earlier version of \code{TALYS} (V1.80). For the present study, the Atomki-V2 AOMP is adopted in a later version of \code{TALYS} (V1.96).
The Atomki-V2 folding potential is used for heavier nuclei, and the phenomenological WS OMP by McFadden and Satchler \cite{McFadden_NPA1966_aomp} is considered for lighter nuclei where the Atomki-V2 potential is not available. Further information on the properties of various AOMPs can be found in the recent review by Mohr {\it et al.}\ \cite{Mohr_EPJA2025_aomp}.

\subsection{$\gamma$-ray Strength Function and Nuclear Level Scheme}
\label{subsec:gsfnld} 

The $\gamma$ strength function ($\gamma$SF) defines the probability of excitation and decay of the nucleus via $\gamma$-ray transitions. As it determines the transmission coefficient of the photon channel, the $\gamma$SF is a necessary nuclear structure ingredient in the Hauser-Feshbach reaction model calculations. Generally, $\gamma$SF can be derived from microscopic models, e.g. the Hartree-Fock-Bogoliubov plus quasi-particle random phase approximation (HFB+QRPA) \cite{KHAN2000253,GORIELY2004331,PhysRevC.94.014304,PhysRevC.104.044301}, or phenomenological approaches \cite{BRINK1957215,PhysRev.126.671,PhysRevC.41.1941,PhysRevC.99.014303}.

In the present study, the simple modified Lorentzian (SMLO) representation of the electric and magnetic dipole (E1 and M1) $\gamma$SF \cite{PhysRevC.99.014303} is used. The SMLO $\gamma$SF is developed based on existing experimental information related to the dipole resonance. Inspired by microscopic models like the quasi-particle random-phase approximation and the shell-model, the SMLO E1 and M1 $\gamma$SF are systematically derived from a global and phenomenological description of the giant dipole resonance. In particular, for the M1 component, the spin-flip and the scissors modes, as well as the contribution from the zero-energy limit, are taken into account. Furthermore, the average radiative widths and Maxwellian-averaged cross sections computed with the SMLO $\gamma$SF has been compared with the available experimental data \cite{PhysRevC.99.014303}, and the good agreement between theoretical calculations and experimental data demonstrates the reliability of the SMLO approach.

The nuclear level scheme is another crucial ingredient for the reaction model calculations, as shown by Eqs. \ref{HF}, \ref{eq:tran}, and \ref{eq:dicxs}. For most nuclei, the nuclear level scheme is constructed by considering the available discrete experimental levels up to an excitation energy where this information is considered complete, plus for higher excitation energies a level density predicted by theoretical nuclear level density models. Several such models are available. In the Fermi Gas model (FGM) \cite{doi:10.1139/p65-139}, excited states with equal level spacing are constructed based on the single particle level scheme, and the nuclear collective effect is phenomenologically taken into account with an explicit expression. The Constant Temperature Model (CTM) \cite{doi:10.1139/p65-139} divides excitation energies into a lower and a higher energy part at a critical energy $E_{M}$. Below $E_{M}$, the constant temperature expression of FGM is applied, while in the higher energy range above $E_{M}$, the normal FGS remains. The microscopic HFB plus the combinatorial model \cite{PhysRevC.78.064307} is another level density model and has proven its capacity to estimate the non-statistical spin and parity dependence of the nuclear level density and performs well in reproducing the low-lying cumulative number of levels. 
 
A systematic comparison between the experimental level density derived from the Oslo method \cite{GUTTORMSEN1987518,SCHILLER2000498,PhysRevC.83.034315} and six global level density models for 42 nuclei indicates~\cite{PhysRevC.106.044315} that the CTM globally describes the data the best, with the microscopic HFB plus a combinatorial model being the second best with very similar accuracy. Thus in the present study, the NLD derived from the CTM combined with the discrete experimental levels compiled in RIPL-3 library \cite{CAPOTE20093107}, is utilized to generate the nuclear level scheme for the reaction model calculations.

\section{Reaction Rate Evaluation}
\label{sec:evaluation}

JINA REACLIB contains three groups of reaction rates: Individual reaction rates determined from experimental or theoretical information, shell model reaction rates, and statistical model reaction rates. Here we update or add 32 individual reactions with new information as discussed in Sec.~\ref{sec:updates_overview} and in more detail in the Appendices. For the remaining rates we continue to recommend the large fp-shell model data set from the \code{ANTOINE} model \cite{fiskerShellModelBased2001} where available, and for the remaining large number of rates based on statistical model calculations we update JINA REACLIB with new rates calculated with the \code{TALYS} code considering the reaction models and nuclear ingredients discussed in Secs.~\ref{sec:nuc} and~\ref{sec:updates_overview}. We note that statistical model calculations are also used as part of the majority of the individual reaction rate updates to extrapolate the rate to higher temperatures and to determine the stellar enhancement factors. 

Fig.~\ref{fig:reaclib_sources} provides an overview of the different sources of information for the new updated set of reaction rates in X-ray bursts. 
Below Ca, the majority of the relevant reaction rates now come from individual evaluations and generally use at least some experimental information. In that region, the use of statistical model predictions is mostly limited to ($\alpha$,p) reactions. Between Ca and Ge the reaction sequence is carried mostly by proton capture rates, which come mostly from shell model calculations, with a few exceptions. Above Ge the vast majority of reactions are based on the statistical model. In the following we discuss these choices in more detail. In addition we provide a discussion of important reaction rates that were not updated at this time. 

\subsection{Individual Rate Evaluations}
\label{subsec:individualrate} 

We used new experimental information for the update of 32 reaction rates (Tab.~\ref{table:30reactions}), located mostly below Ca, with a few updates in the Ni-Ge range (see Fig.~\ref{fig:reaclib_sources}). As indicated in Fig.~\ref{fig:reaclib_sources} about half of the individually updated reaction rates have been identified as directly affecting X-ray burst model predictions in sensitivity studies \cite{cyburtDependenceXRayBurst2016a}. However, such studies are strongly dependent on the astrophysical model, and the particular burst conditions that vary from system to system. Therefore, we also include in this update a broad range of reactions that are within the range of nuclei populated in X-ray bursts, especially on the proton rich side of the valley of stability.

\begin{table*}
\caption{List of individually evaluated reactions in this work.
\label{table:30reactions}}
\begin{tabular}{ c c }
\hline \hline
Reaction & Reference for evaluation\\
\hline
$^{14}$C($\alpha$,$\gamma$)$^{18}$O & \cite{PhysRevC.80.045805}, \cite{GORRES1992414} \\
$^{13}$N($\alpha$,p)$^{16}$O &  \cite{mccamisStellarRates16O1973}, \cite{caughlanThermonuclearReactionRates1988}, \cite{iliadisChargedparticleThermonuclearReaction2010b}, \cite{meyerEvaluation13N16O2020}, \cite{jayatissaFirstDirectMeasurement2022} \\
$^{14}$O($\alpha$,p)$^{17}$F &  
\cite{hahnStructure18Mathrm1996}, \cite{almaraz-calderonLevelStructure182012}, \cite{fortuneEnsuremathAlphaWidth2012}, \cite{huExaminationRoleO142014a}, \cite{kimMeasurement14Mathrm2015} \\
$^{17}$O(p,$\alpha$)$^{14}$N & \cite{bucknerHighintensitybeamStudy172015}, \cite{lunacollaborationImprovedDirectMeasurement2016},
\cite{rauscherTABLESNUCLEARCROSS2001} \\
$^{17}$F(p,$\gamma$)$^{18}$Ne & \cite{chippsFirstDirectMeasurement2009b}, \cite{kuvinMeasurement17F18Ne2017},
\cite{rauscherTABLESNUCLEARCROSS2001} \\
$^{18}$F(p,$\alpha$)$^{15}$O & \cite{iliadisChargedparticleThermonuclearReaction2010b}, \cite{adekolaFirstProtontransferStudy2011a}, \cite{bardayanFirstScienceResult2015a}, \cite{bardayanSpectroscopicStudy20Ne2017}, \cite{hallKey19NeStates2019}, \cite{hall19NeLevelStructure2020}, \cite{lairdGRayEmissionNovae2013}, \cite{parikhSpectroscopy19NeThermonuclear2015}, \cite{cherubiniFirstApplicationTrojan2015}, \cite{lacognataTrojanHorseApproach2017} \\
$^{18}$F(p,$\gamma$)$^{19}$Ne & \cite{iliadisChargedparticleThermonuclearReaction2010b}, \cite{longlandChargedparticleThermonuclearReaction2010}, \cite{bardayanFirstScienceResult2015a}, \cite{akersMeasurementRadiativeProton2016}, \cite{bardayanSpectroscopicStudy20Ne2017} \\
$^{19}$F(p,$\alpha$)$^{16}$O & \cite{lombardoNewAnalysis192019}, \cite{zhangDirectMeasurementAstrophysical2021}, \cite{zhangDirectMeasurementAstrophysical2022a} \\
$^{19}$F($\alpha$,p)$^{22}$Ne & 
\cite{Pizzone_2017}, \cite{DAgata_2018}, \cite{PhysRevC.77.035801} \\
$^{19}$Ne(p,$\gamma$)$^{20}$Na & \cite{lammLevelStructure20Na1990}, \cite{vancraeynest19MathrmNe1998}, \cite{belargeExperimentalInvestigation192016}, \cite{wilkinsonDirectMeasurementKey2017} \\
$^{23}$Na($\alpha$,p)$^{26}$Mg &  \cite{mohrCrossSectionsAinduced2015a}, \cite{avilaExperimentalStudyAstrophysically2016} \\
$^{23}$Na(p,$\gamma$)$^{24}$Mg & \cite{iliadisChargedparticleThermonuclearReaction2010b}, \cite{cesarattoMeasurement138KeV2013}, \cite{boeltzigDirectMeasurementsLowenergy2019}, \cite{marshallNewEnergy133keV2021}, \cite{boeltzigInvestigationDirectCapture2022} \\
$^{22}$Mg($\alpha$,p)$^{25}$Al &
\cite{randhawaFirstDirectMeasurement2020a}\\
$^{24}$Mg($\alpha$,$\gamma$)$^{28}$Si &  \cite{iliadisChargedparticleThermonuclearReaction2010b}, \cite{adsleyStatus24Mg28Si2020} \\
$^{25}$Mg(p,$\gamma$)$^{26}$Al & \cite{iliadisChargedparticleThermonuclearReaction2010b}, \cite{limataNewExperimentalStudy2010}, \cite{strieder25Mg26AlReaction2012}, \cite{stranieroImpactRevised25Mg2013}, \cite{liIndirectMeasurement572020}, \cite{mischAstromersNuclearIsomers2020a}, \cite{kankainenDecayKey92keV2021a} \\
$^{23}$Al(p,$\gamma$)$^{24}$Si & 
\cite{PhysRevC.74.021303}, \cite{PhysRevC.86.015806}, \cite{PhysRevC.101.031303}, \cite{PhysRevC.106.L012801}, \cite{PhysRevLett.79.3845}, \cite{PhysRevLett.122.232701} \\
$^{24}$Al(p,$\gamma$)$^{25}$Si & 
\cite{KUBONO1997195}, \cite{PhysRevC.97.054307} \\
$^{25}$Al(p,$\gamma$)$^{26}$Si & 
\cite{thomasBetadecayProperties25Si2004}, \cite{peplowskiLowestProtonResonance2009}, \cite{bennettClassicalNovaContributionMilky2013}, \cite{BASUNIA20161}, \cite{PhysRevC.93.035801}, \cite{janiakEnsuremathBetaDelayed2017}, \cite{liangSimultaneousMeasurementDelayed2020} \\
$^{26}$Al(p,$\gamma$)$^{27}$Si & 
\cite{caltechthesis7945}, \cite{PhysRevC.72.065804}, \cite{ruizMeasurement184KeV2006}, \cite{iliadisChargedparticleThermonuclearReaction2010a}, \cite{margerinInverseKinematicStudy2015}, \cite{almaraz-calderonStudy26272017}, \cite{Lotay2020}, \cite{PhysRevLett.126.042701} \\
$^{26}$Si(p,$\gamma$)$^{27}$P &
\cite{iliadisChargedparticleThermonuclearReaction2010b}, \cite{toganoResonanceStates272011}, \cite{r3bcollaborationCoulombDissociation272016}, \cite{ribllcollaborationEnsuremathBetaDecay2019}, \cite{dagata26MathrmSi2021} \\
$^{29}$Si(p,$\gamma$)$^{30}$P &  \cite{iliadisChargedparticleThermonuclearReaction2010b}, \cite{lotaySpectroscopy30Mathrm2020} \\
$^{29}$P(p,$\gamma$)$^{30}$S & 
\cite{lotayLevelStructure302012}, \cite{almaraz-calderonLevelStructure302012}, \cite{almaraz-calderonErratumLevelStructure2013} \\
$^{30}$P(p,$\gamma$)$^{31}$S & 
\cite{rauscherTABLESNUCLEARCROSS2001}, \cite{brownShellmodelStudiesAstrophysical2014}, \cite{kankainenMeasurementKeyResonance2017}, \cite{budnerConstraining30Mathrm2022} \\
$^{33}$Cl(p,$\gamma$)$^{34}$Ar &
\cite{PhysRevLett.124.252702}, \cite{rauscherTABLESNUCLEARCROSS2001} \\
$^{34}$Cl(p,$\gamma$)$^{35}$Ar & 
\cite{rauscherTABLESNUCLEARCROSS2001}, \cite{PhysRevC.72.065804}, \cite{fryDiscovery34MCl2015}, \cite{gillespieFirstMeasurement342017}, \cite{setoodehniaExperimentalStudyCl2019}, \cite{PhysRevC.102.025801}, \cite{mischAstromersNuclearIsomers2020a} \\
$^{36}$Cl(p,$\alpha$)$^{33}$S & 
\cite{andersonRemeasurement33Cl2017} \\
$^{35}$K(p,$\gamma$)$^{36}$Ca & 
\cite{lalanneEvaluation35K36Ca2021} \\
$^{42}$Ti(p,$\gamma$)$^{43}$V & 
\cite{Hou2023} \\
$^{55}$Ni(p,$\gamma$)$^{56}$Cu & 
\cite{ongLowlyingLevelStructure2017}, \cite{valverdeHighPrecisionMassMeasurement2018} \\
$^{56}$Ni(p,$\gamma$)$^{57}$Cu &
\cite{rauscherTABLESNUCLEARCROSS2001}, \cite{kahlSingleparticleShellStrengths2019} \\
$^{57}$Cu(p,$\gamma$)$^{58}$Zn & 
\cite {PhysRevLett.113.032502} \\
$^{65}$As(p,$\gamma$)$^{66}$Se &
\cite{rauscherTABLESNUCLEARCROSS2001}, \cite{tuDirectMassMeasurements2011a}, \cite{lamReactionRates64Ge2016}, \cite{zhouMassMeasurementsShow2023} \\
\hline
\end{tabular}
\end{table*}

We emphasize that these 32 reactions cannot necessarily be considered experimentally known. In many cases incomplete experimental data on cross sections and level properties are combined with shell model predictions or other assumptions. In most cases, these data are only available up to a certain incident energy where the level information is available. Therefore X-ray burst reaction rates often need to be extrapolated to higher temperatures. As level densities increase with excitation energy, the statistical model should become more accurate at higher temperature and is used for this extrapolation. We normalize the statistical model rate to the experimentally determined rate at the highest experimental temperature data point. We also apply the NON-SMOKER stellar enhancement factors \cite{rauscherTABLESNUCLEARCROSS2001} to experimental ground-state rates. 

\begin{figure*}
\includegraphics[width=1.0\textwidth]{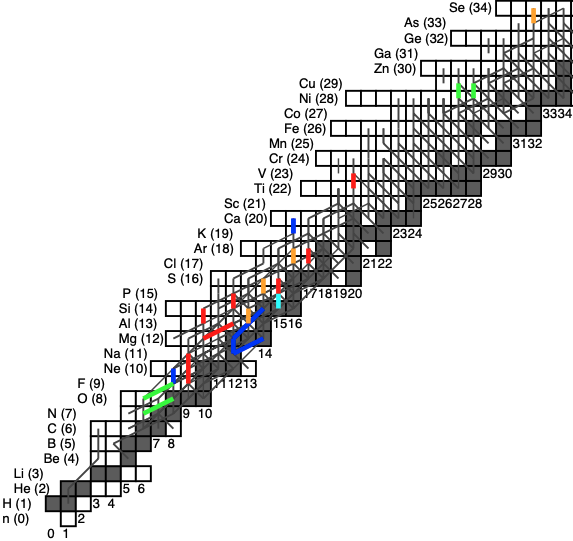}
\caption{Updated reaction rates with experimental information that have changed by more then 10\% in the temperature range relevant for X-ray bursts (thick lines). Colors represent the maximum ratio between updated rate and previous REACLIB result (factors may indicate increases or decreases): factor 1.1-1.5 (dark blue), factor 1.5-2 (light blue), factor 2-3 (green), factor 3-5 (orange), more than factor of 5 (red). Thin lines illustrate a typical reaction flow during an X-ray burst \cite{cyburtDependenceXRayBurst2016a} for illustration.
\label{fig:setI_changes}}
\end{figure*}

Fig.~\ref{fig:setI_changes} shows the maximum ratio of the updated rate to the previous JINA REACLIB rate over the temperature range relevant for X-ray bursts (see Sec.~\ref{sec:nuc}). Most updates led to significant changes in rates of more than a factor of 2. For several reactions such as the (p,$\gamma$) reactions on $^{18}$F, $^{19}$Ne, $^{23}$Al, $^{26}$Si, $^{30}$P and $^{34}$Cl the change is more than a factor of 5.

\subsection{Statistical Model Updates} 

We use the \code{TALYS} 1.97/2.00 Hauser-Feshbach code to update statistical model (discussed in Sec.~\ref{subsec:reaction}) rates for (p,$\gamma$), (p,$\alpha$), ($\alpha$,$\gamma$), and ($\alpha$,p) reactions. Compared to the previous \code{NON-SMOKER} statistical model results used in JINA REACLIB, the new rates take into account updated nuclear masses (Sec.~\ref{subsec:newmasses}), improved optical model potentials (Sec.~\ref{subsec:npomp} and Sec.~\ref{subsec:aomp}), $\gamma$-ray strength functions and nuclear level schemes (Sec.~\ref{subsec:gsfnld}). As \code{TALYS} is open source, the new reaction rates are fully documented and reproducible.

\begin{figure}
\includegraphics[width=1.0\textwidth]{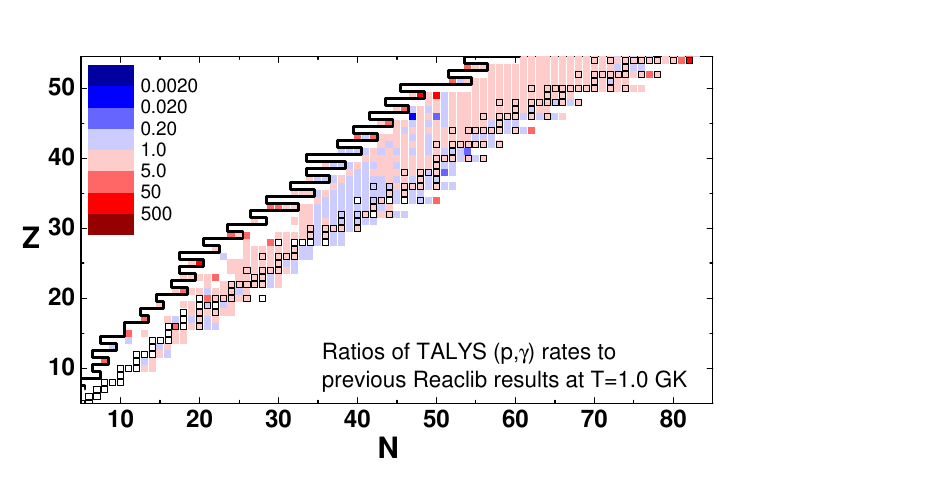}
\caption{The comparison of (p,$\gamma$) reaction rates in the (N,Z) plane between the present \code{TALYS} calculations and the \code{NON-SMOKER} (JINA REACLIB) results at T = 1 GK. The boundary between endothermic and exothermic reactions is marked with the black thick line.}
\label{figm1:TALYS2REACLIB}
\end{figure}

\begin{figure}
\includegraphics[width=1.0\textwidth]{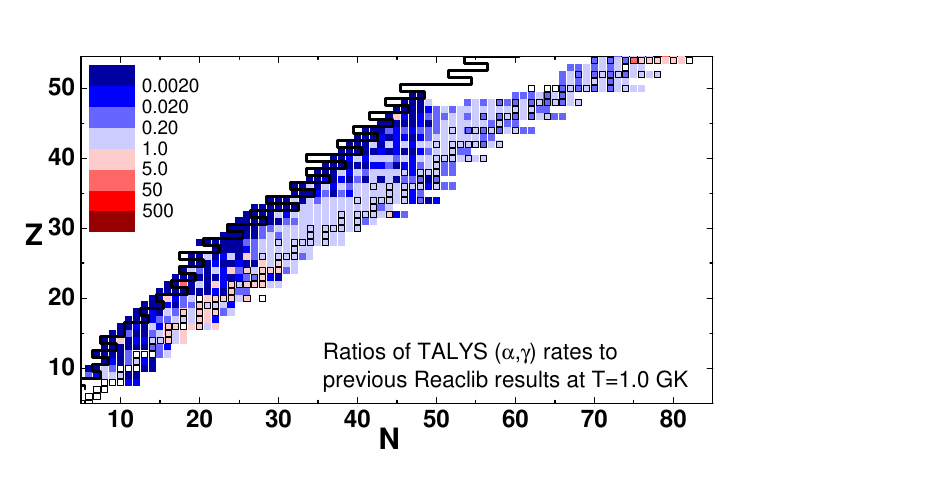}
\caption{The comparison of ($\alpha$,$\gamma$) reaction rates in the (N,Z) plane between the present \code{TALYS} calculations and the \code{NON-SMOKER} results at T = 1 GK.}
\label{figm3:TALYS2REACLIB}
\end{figure}

\begin{figure}
\includegraphics[width=1.0\textwidth]{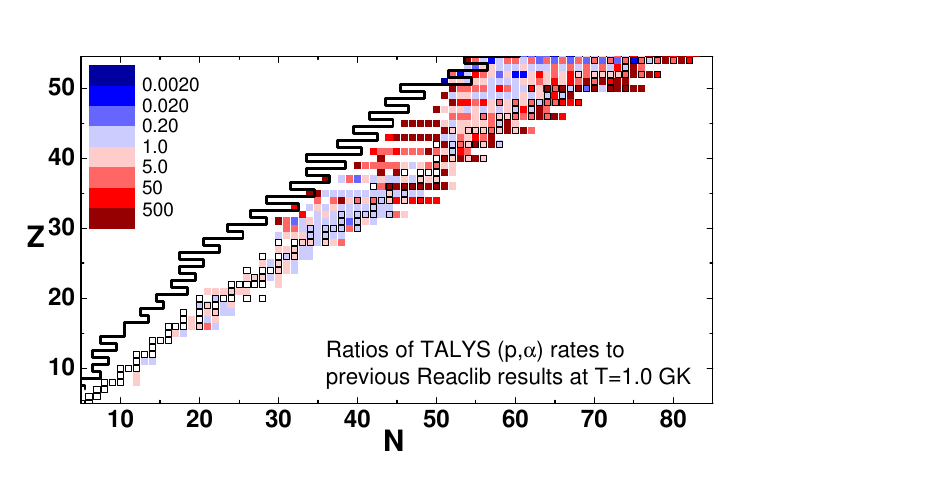}
\caption{The comparison of (p,$\alpha$) reaction rates in the (N,Z) plane between the present \code{TALYS} calculations and the \code{NON-SMOKER} results at T = 1 GK.}
\label{figm4:TALYS2REACLIB}
\end{figure}

\begin{figure*}
\includegraphics[width=1.0\textwidth]{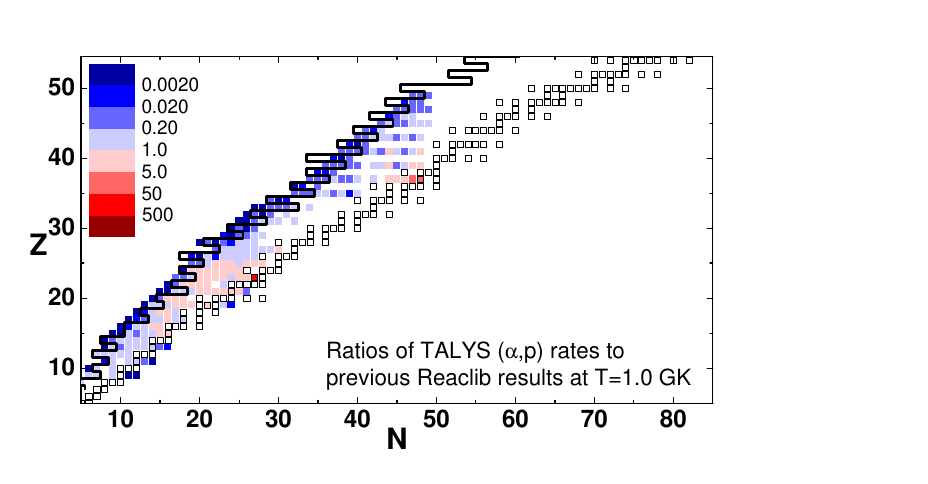}
\caption{The comparison of ($\alpha$,p) reaction rates in the (N,Z) plane between the present \code{TALYS} calculations and the \code{NON-SMOKER} results at T = 1 GK.}
\label{figm5:TALYS2REACLIB}
\end{figure*}

We compare the new \code{TALYS} statistical model rates with the previous \code{NON-SMOKER} rates at a temperature of 1 GK for (p,$\gamma$), ($\alpha$,$\gamma$), 
(p,$\alpha$), and ($\alpha$,p) reactions in Figs.~\ref{figm1:TALYS2REACLIB}, \ref{figm3:TALYS2REACLIB}, \ref{figm4:TALYS2REACLIB},
and \ref{figm5:TALYS2REACLIB}, respectively. 
In the case of (p,$\gamma$), the majority of the rates agree within a factor of 5. Some larger differences near the proton drip line, for example for $^{41}$Ti(p,$\gamma$), are due to changes in nuclear masses, which can have a very large impact around zero proton separation energy. There are a few cases with larger differences that are closer to stability and near the rp-process such as the proton captures on $^{45}$V and $^{68}$Se where differences can reach factors of 20.

For reactions involving $\alpha$ particles differences tend to be larger, likely due to the use of the updated $\alpha$-OMP. 
For ($\alpha$,$\gamma$) reactions, the \code{TALYS} rates tend to be significantly lower than the \code{NON-SMOKER} rates, in many cases approaching a factor of 50 difference near the proton drip line. However, these reactions do not play a significant role in the mixed hydrogen and helium bursts of interest here.

For (p,$\alpha$) reactions Fig.~\ref{figm4:TALYS2REACLIB} shows also large differences between \code{TALYS} and \code{NON-SMOKER} in some cases, especially for odd Z $\ge$ 39 and for $Z=34$ where ratios can reach factors of 5-50 or more. For the large differences, \code{TALYS} tends to predict larger rates. (p,$\alpha$) reactions can play an important role in the rp-process as they create reaction cycles. However, the cases with large discrepancies between the models are not the important ones with their rates being extremely small at X-ray burst temperatures. For the important $^{59}$Cu(p,$\alpha$) and $^{63}$Ga(p,$\alpha$) cases, the new \code{TALYS} rates are at 1~GK factors of 2.3 and 4.6 smaller. While this may not be negligible it can be considered reasonable agreement for statistical models, especially given the differences in AOMP.

For the majority of ($\alpha$,p) rates \code{TALYS} and \code{NON-SMOKER} rates at 1~GK agree within a factor of 5. This includes the important ($\alpha$,p) reaction rates in the $\alpha$p process up to $Z=18-20$. At and beyond the proton drip line differences jump to factors of 50 or more, likely due to differences in masses, however, these rates are not relevant for X-ray bursts.

\begin{figure*}
\includegraphics[width=1.0\textwidth]{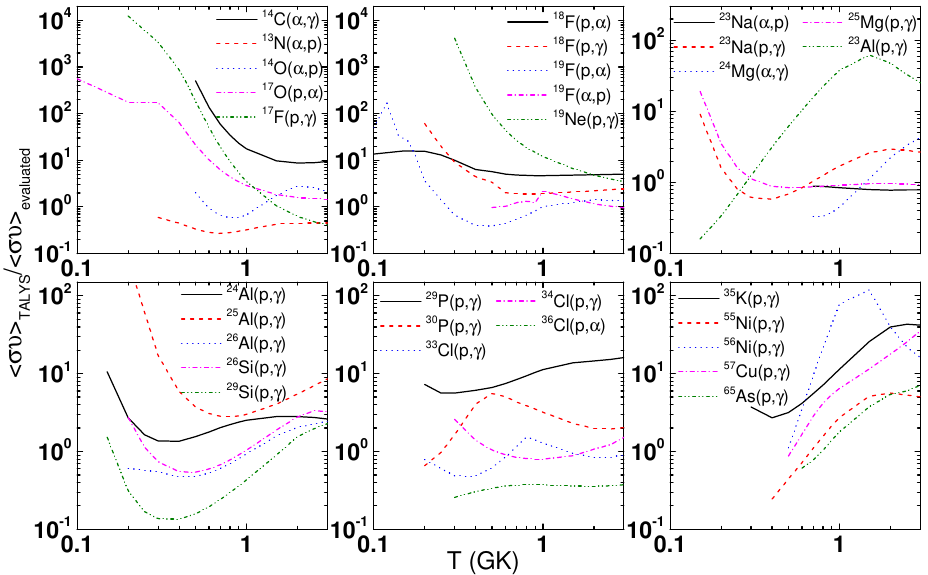}
\caption{The comparison of the individually evaluated reaction rates to the rates systematically calculated with \code{TALYS} as functions of temperature.}
\label{figm:ratecomp}
\end{figure*}

Taking advantage of our 32 reaction rates with updated experimental data, we can compare these rates with the new \code{TALYS} statistical model predictions. While such a comparison is of limited value without detailed estimates of the uncertainties of the experimental reaction rates, it nevertheless provides a measure for how rates used in astrophysical models have changed due to experimental information, and what changes can be expected in the future as more information becomes available for other reactions. Fig.~\ref{figm:ratecomp} shows the rate ratios as a function of temperature. Above 1~GK \code{TALYS} predictions typically agree with the evaluated rates within an order of magnitude. This may be simply due to the fact that statistical model rates are used at high temperatures, or it may indicate that as the relevant level density increases with temperature the statistical model becomes more accurate. Differences clearly increase dramatically as temperatures get lower. Differences mostly stay within a factor of 100, but can be larger in a few cases involving light species ($^{17}$O, $^{17}$F, $^{19}$Ne). This is not surprising as the assumptions of the statistical model, primarily the assumption of a high Compound nucleus level density in the relevant energy range, are expected to break down for light nuclei and at low temperatures (e.g., \cite{rauscherNuclearLevelDensity1997}).

\subsection{Other Key Reactions not Updated}
\label{subsec:nonupdaterate}

The 3$\alpha$ reaction producing $^{12}$C is among the most important reactions in X-ray burst models. Its steep energy dependence at low temperatures triggers the instability responsible for the burst, and its reduced energy dependence at temperatures above 0.4~GK may explain the stable burning modes observed at high accretion rates. We employ the rate by \cite{fynboRevisedRatesStellar2005a}, who constrained the contributions from higher lying states beyond the Hoyle state using $\beta$-delayed $\alpha$-decay studies. The effects are however very small for the temperature range relevant for X-ray bursts. \cite{jinEnhancedTripleaReaction2020} recently demonstrated that the rate can be significantly altered in an astrophysical plasma at high proton densities, as proton scattering can enhance the de-excitation of the Hoyle state. A detailed analysis for X-ray burst conditions and eventual implementation of this effect in burst models is left for future work. The effect may impact ignition conditions and therefore burst behavior as a function of accretion rate.

The $^{15}$O($\alpha$,$\gamma$)$^{19}$Ne reaction and the ($\alpha,p)$ reactions on $^{18}$Ne, $^{22}$Mg, $^{26}$Si, $^{30}$S, and $^{34}$Ar play a very important role in X-ray burst models \cite{cyburtDependenceXRayBurst2016a,keekReactionRateComposition2014a}, in particular $^{15}$O($\alpha$,$\gamma$) as the main CNO breakout pathway during burst ignition when temperatures are too low to trigger $^{18}$Ne($\alpha$,p)$^{21}$Na, and $^{18}$Ne($\alpha$,p)$^{21}$Na as the main CNO breakout pathway after burst ignition. A lot of experimental work has been carried out on these important reactions, however, at this point data have not sufficiently converged. We therefore recommend to continue to use the rate from \cite{davidsInfluenceUncertainties15O2011} for $^{15}$O($\alpha$,$\gamma$) and statistical model rates for the ($\alpha$,p) reactions. Nevertheless we review in the following the progress that has been made in understanding these reaction rates. 

The $^{15}$O($\alpha$,$\gamma$)$^{19}$Ne reaction at CNO breakout temperatures of around 0.33-0.5~GK is determined by the $\alpha$-widths of two single narrow resonances at 502~keV and 612~keV in the center of mass system, corresponding to $^{19}$Ne states at 4.03~MeV and 4.14~MeV with possibly comparable contributions \cite{tanO15Ne19Breakout2007}. The challenge is that these $\alpha$-widths and resulting resonance strengths are very small. Direct measurements would require a very low energy (159 and 196~keV/A) $^{15}$O radioactive beam with an intensity of the order of 10$^{12}$~pps, which is out of reach of current rare isotope beam facilities for the foreseeable future. As the total lifetime of both states has been studied experimentally relatively well, a measurement of the $\alpha$-decay branchings $B_\alpha$ would in principle be sufficient to determine the astrophysical reaction rate. For the 4.03~MeV state \cite{davidsDecayBranchingRatios2003} observed a small excess of $\alpha$-particles using population via $^{21}$Ne(p,t)$^{19}$Ne reaction at KVI in Groningen and reported an upper limit of $B_\alpha <4.3 \times 10^{-4}$. \cite{tanO15Ne19Breakout2007} also observed a small $\alpha$-particle excess using the $^{19}$F($^3$He,t)$^{19}$Ne reaction at the University of Notre Dame and derive $B_\alpha$=2.9$\pm 2.1 \times 10^{-4}$. Both detections are marginal and better data are needed. For their recommended rate \cite{davidsInfluenceUncertainties15O2011} used a probability distribution obtained from the experiment of \cite{davidsDecayBranchingRatios2003} with a peak corresponding to a resonance strength of 8.8~$\mu$eV. For the 4.14~MeV state, data is even more limited, with \cite{tanO15Ne19Breakout2007} only reporting a combined value from the 4.14 and 4.2~MeV states. New experimental approaches are being developed, including the population of the resonances via $\beta$-delayed proton emission of $^{20}$Mg \cite{lundBetadelayedProtonEmission2016,wredeNewPortal15O2017}.

For the $^{18}$Ne($\alpha$,p)$^{21}$Na CNO breakout reaction we use the rate of \cite{mohrThermonuclearReactionRate2014} that is based on combining information about known levels in the $^{22}$Mg compound nucleus, complemented with information from the $^{22}$Ne mirror nucleus and estimates of $\alpha$-strengths. The statistical model appears to provide a good average cross section. A recent direct measurement using a radioactive $^{18}$Ne beam with the ANASEN detection system at Florida State University was limited to energies beyond of what is relevant for X-ray bursts and again excellent agreement with the statistical model was found \cite{anastasiouMeasurement18Ne21Na2022}. A recent $^{24}$Mg(p,t)$^{22}$Mg measurement confirms this picture \cite{brummerProtonDecaysEnsuremath2023}. A direct measurement of the $^{18}$Ne($\alpha$,p)$^{21}$Na reaction at lower energy would be important for a reliable reaction rate. 

The $^{22}$Mg($\alpha$,p)$^{25}$Al reaction has been shown to be particularly important to accurately predict X-ray burst light curves \cite{cyburtDependenceXRayBurst2016a,randhawaFirstDirectMeasurement2020a,huAdvancementPhotosphericRadius2021a, lamImpactNew65As2022b}. 
Indirect studies of the Compound nucleus levels in $^{26}$Si via $^{28}$Si(p,t)$^{26}$Si at RCNP \cite{maticHighprecision28Si26Si2011} identified the relevant states within 1~MeV of the $\alpha$-threshold, covering the Gamow window up to 0.5~GK. With assumptions of typical $\alpha$-widths, the resulting reaction rate estimate agreed well with statistical model predictions for this temperature range. For higher temperatures, additional unidentified higher lying states are expected to dominate and only a lower reaction rate limit could be obtained. However, \cite{maticHighprecision28Si26Si2011} found some evidence for a substantial level density at higher energies, possibly indicating that the statistical model approach may be applicable. A recent direct measurement of the $^{22}$Mg($\alpha$,p)$^{25}$Al  reaction using the Active Target TPC at NSCL at Michigan State University measured the reaction cross section at higher energies at the upper end of the Gamow window for 2~GK and thus mostly beyond the energy range relevant for X-ray bursts \cite{randhawaFirstDirectMeasurement2020a}. The results indicated a factor of 8 overestimation by the statistical model. However, a subsequent direct measurement using the MUSIC detector at Argonne National Laboratory obtained a significantly different cross section that is in excellent agreement with the statistical model \cite{jayatissaStudyMg222023}. More measurements, also towards lower energy bridging the gap between indirect and direct measurements, will be needed to resolve this issue. We included in our updated data set the lower rate from \cite{randhawaFirstDirectMeasurement2020a} so as to be able to explore its impact, however, we recommend to for now to continue to use the statistical model rate. 

For ($\alpha$,p) reactions on heavier targets, estimates based on compound nucleus levels information from stable beam (p,t) experiments have raised questions about the appropriateness of the statistical model. These measurements include $^{32}$S(p,t)$^{30}$S relevant for $^{26}$Si($\alpha$,p)$^{29}$P at RCNP complemented by $^{28}$Si($^{3}$He,n)$^{30}$S measurements at Notre Dame \cite{almaraz-calderonLevelStructure302012},  $^{36}$Ar(p,t)$^{34}$Ar relevant for $^{30}$S($\alpha$,p)$^{33}$Cl at RCNP \cite{longUnboundLevels34Ar2018a}, and $^{40}$Ca(p,t)$^{38}$Ca relevant for $^{34}$Ar($\alpha$,p)$^{37}$K at iThemba LABS \cite{longIndirectStudyStellar2017}. In all cases, reaction rates based on the observed levels and estimated $\alpha$ spectroscopic factors result in reaction rates that are 1-2 orders of magnitude below statistical model calculations. Direct measurements of the inverse $^{33}$Cl(p,$\alpha$)$^{30}$S at Argonne National Laboratory \cite{deibelFirstMeasurement33Cl2011} have also indicated order of magnitude discrepancies with the statistical model. However, a recent direct measurement of the $^{34}$Ar($\alpha$,p)$^{37}$K reaction (together with the $^{34}$Cl($\alpha$,p)$^{37}$Ar reaction due to beam contamination) using the JENSA gas jet target at NSCL at Michigan State University found good agreement within factors of 2-3 with the statistical model, probing directly the high energy end of the $^{40}$Ca(p,t)$^{38}$Ca based results. More direct measurements at the lower energies relevant for X-ray burst temperatures will be needed to determine the reliability of the current statistical model predictions.

\begin{figure*}
\includegraphics[width=1.0\textwidth]{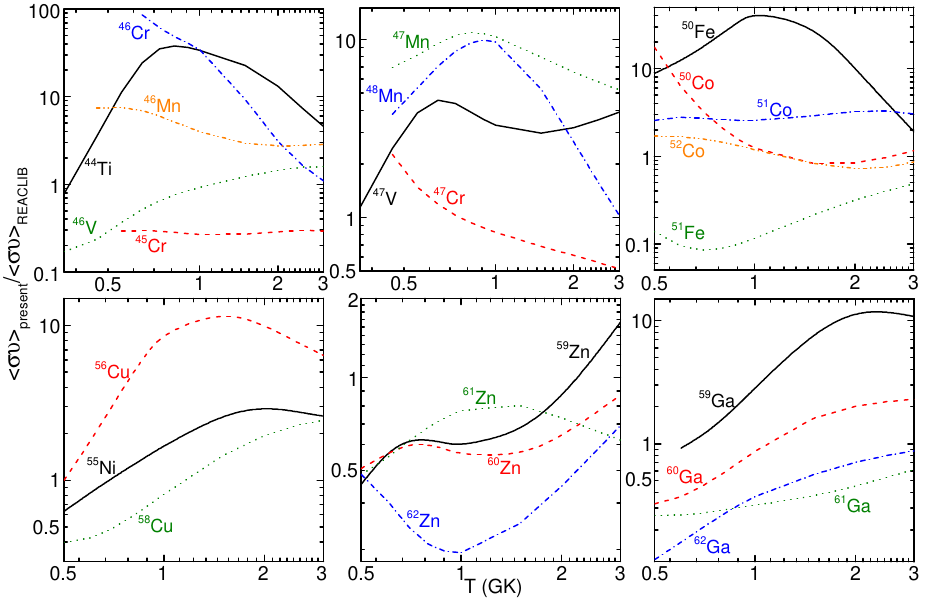}
\caption{Ratio of the fp-shell model proton capture reaction rates \cite{fiskerShellModelBased2001} in JINA REACLIB to the \code{TALYS} reaction rates (Sec.~\ref{subsubsec:HFmodel}) as functions of temperature and labeled with the target isotope.}
\label{fig:TALYSvSM}
\end{figure*}

When available, and when no other experimental information is available, JINA REACLIB continues to recommend the reaction rates determined by the fp-shell model code \code{ANTOINE} \cite{fiskerShellModelBased2001}. The nuclear structure properties including level spectra, proton spectroscopic factors, and electromagnetic transition probabilities are obtained from the shell-model calculation and then used to compute the reaction rates by combining available experimental information. Such shell model reaction rates are expected to be more reliable than those predicted by the statistical model. However, owing to model space limitations, they are restricted to medium-light nuclei with A $<$ 64. 
We compare the fp-shell model reaction rates with the \code{TALYS} results (Sec.~\ref{subsubsec:HFmodel}) in Fig.~\ref{fig:TALYSvSM} to quantify the differences one can expect from these approaches. Though the model approaches are very different, the ratios remain within one order of magnitude for most reactions, indicative of the applicability of the statistical model at that level of accuracy. There are a few exceptions such as proton captures on $^{44}$Ti, $^{48}$Cr, and $^{50}$Fe where discrepancies around temperatures of 1~GK can reach factors of 40. No systematic trends in terms of temperature are observed.

\section{Impact on X-ray Burst Models}
\label{sec:xrb}

We use the updated nuclear reaction library from this work to explore the current understanding of the nuclear reaction sequences during X-ray bursts, and the impact of the revised nuclear reaction rates. To that end, we implemented the new rates into two models, the one-zone model \code{ONEZONE} \cite{cyburtDependenceXRayBurst2016a} and a 1D multi-zone model \cite{meiselConsistentModelingGS2018} based on the \code{MESA} 1D stellar evolution code \cite{paxtonModulesExperimentsStellar2015}.

\subsection{Impact on Composition and Light Curve Predictions}
\label{subsec:complc}

\begin{figure*}
\includegraphics[width=0.5\columnwidth]{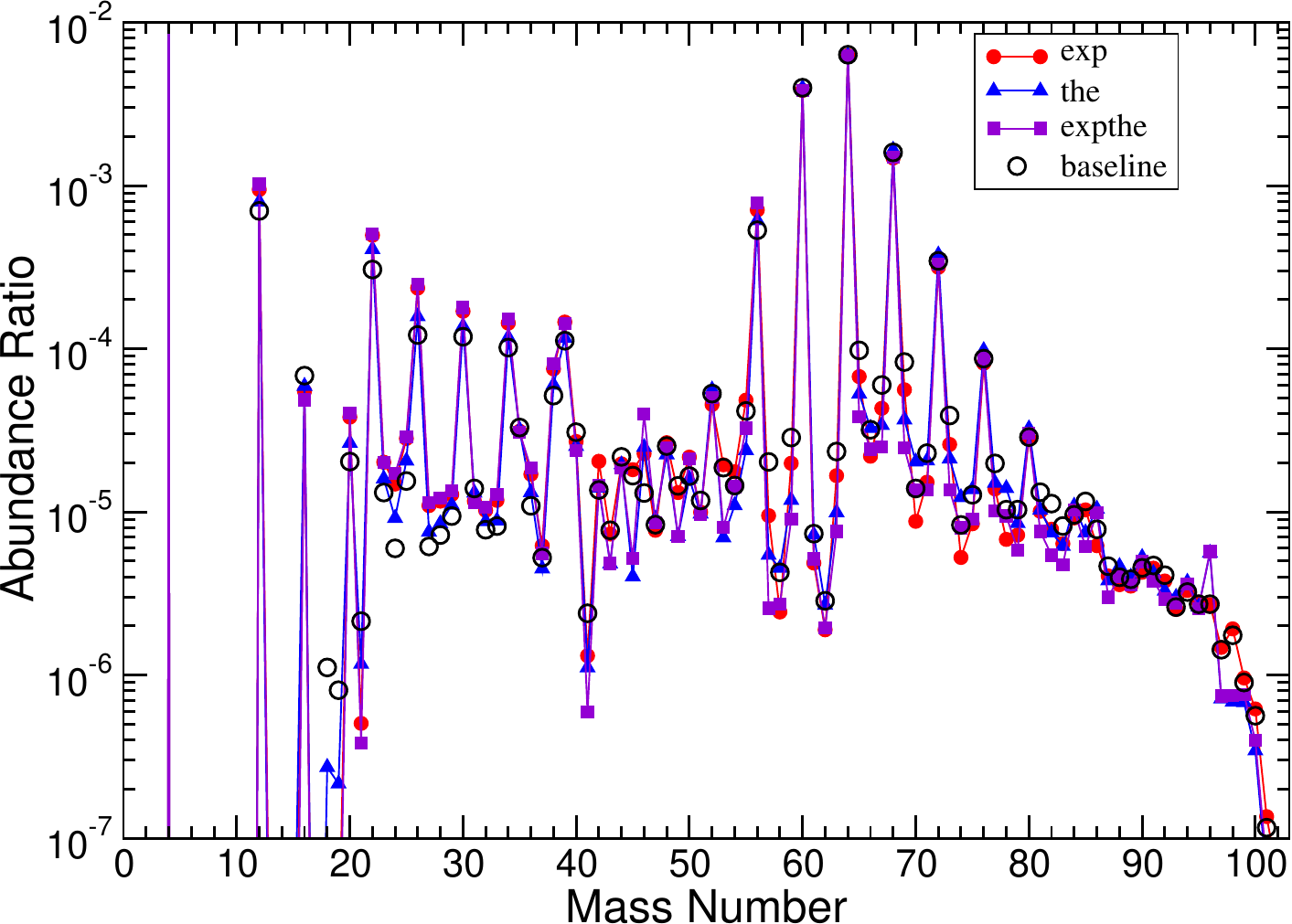}
\includegraphics[width=0.5\columnwidth]{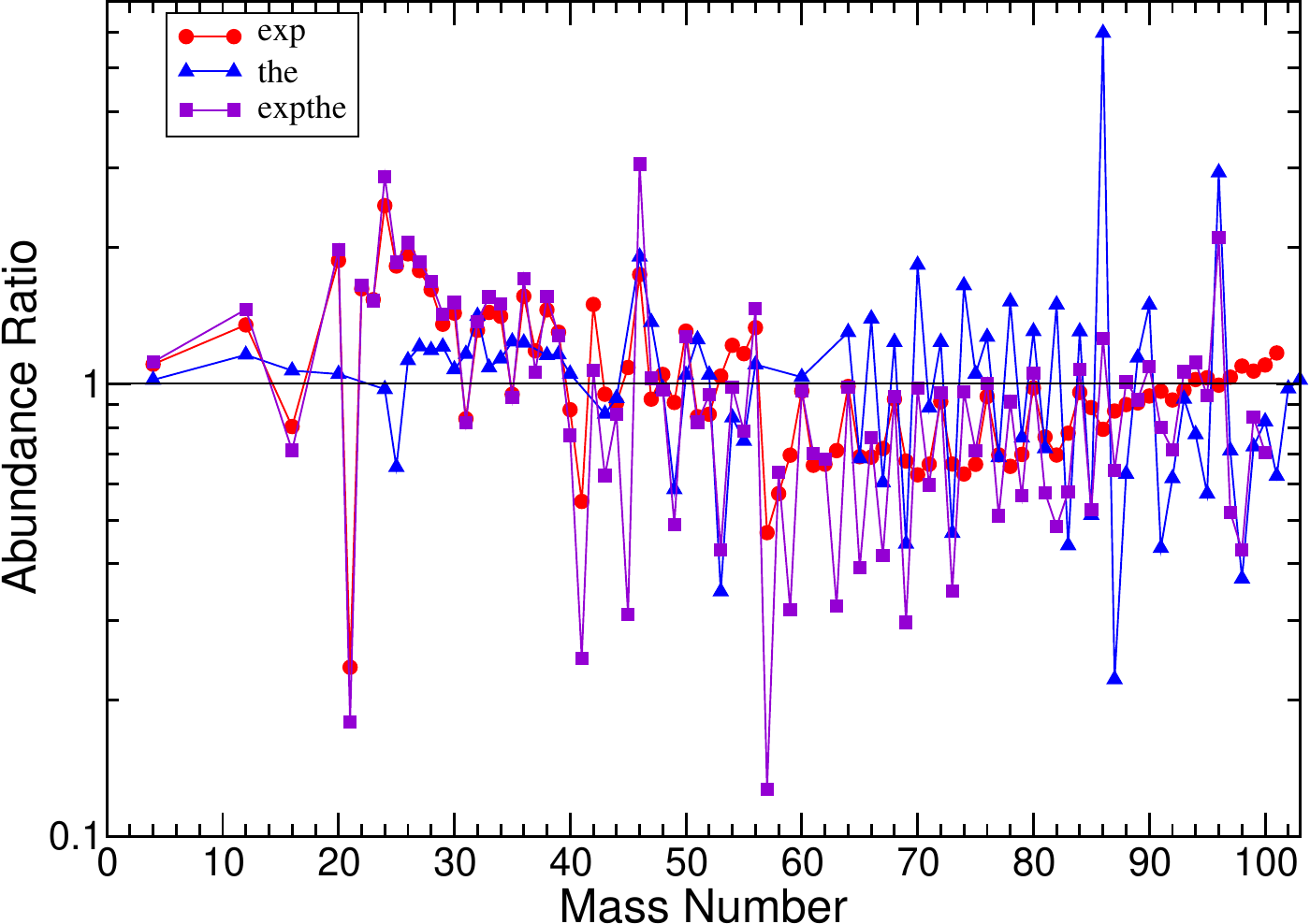}
\caption{Final abundance distribution summed by mass number at the end of the \code{ONEZONE} burst calculation for the baseline prior to our update (open circles), updates of individual rates (red circles), updates of Hauser-Feshbach rates (blue triangle) and all updates (purple squares). Shown are absolute abundances (left) and the ratio to the baseline (right). \label{FigXRB_ONEZONE_ab}}
\end{figure*}

\begin{figure}
\includegraphics[width=0.5\columnwidth]{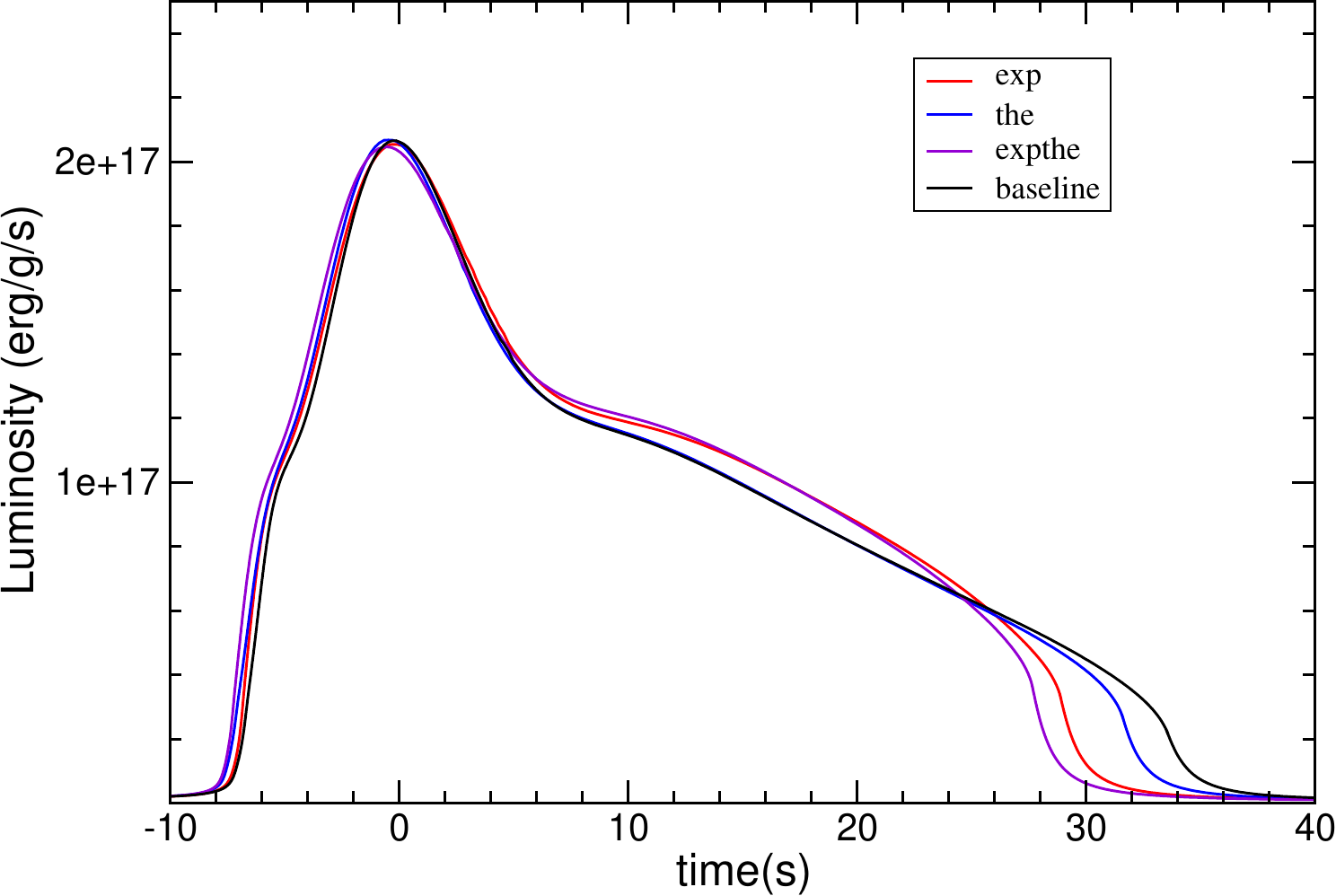}
\caption{Estimated light curves from the \code{ONEZONE} model calculations for the baseline prior to our update (black line), updates of individual rates (red line), updates of Hauser-Feshbach rates (blue line) and all updates (purple line). \label{FigXRB_ONEZONE_lc}}
\end{figure}

\code{ONEZONE} includes a reaction network with 735 nuclei from hydrogen to tellurium and tracks nuclear heating and cooling to determine temperature and density evolution self-consistently. It models a single burst, with ignition conditions determined from the \code{KEPLER} 1D code for a model of the regular mixed H/He bursts observed from the source GS 1826-24 (see \cite{cyburtDependenceXRayBurst2016a} for details). The advantage of \code{ONEZONE} is its speed, which enables the exploration of the impact of each of the individual reaction rates updated here, and the unique and reproducible modeling of a single burst that enables precise evaluation of the influence of nuclear reaction rate changes. The new rates change the composition of the burst ashes significantly (Fig.~\ref{FigXRB_ONEZONE_ab}). While most changes are within a factor of 3, the impact approaches an order of magnitude in some mass chains. New experimental data were mostly available for $A<40$ where indeed most of the larger changes occur. As experimental and shell model based rates dominate the $A<56$ mass range, the impact of the new Hauser-Feshbach rates is larger in the heavier mass region where the model relies on these rates almost exclusively. The impact on the light curve is also significant with an up to 9\% enhancement of the late burst tail, 10-20~s after peak. The more rapid fuel consumption powering this enhancement leads to a slight shortening of the burst and a reduced luminosity at the very end (Fig.~\ref{FigXRB_ONEZONE_lc}). These changes are mainly driven by the individual rate updates. The new Hauser-Feshbach rates have only a small impact on the light curve. 

To isolate the influence of individual rate updates, we carried out calculations for each rate change one by one. The largest impact on composition comes from the new (in order of importance) $^{23}$Al(p,$\gamma$)$^{24}$Si, $^{22}$Mg($\alpha$,p)$^{25}$Al, and $^{26}$Si(p,$\gamma$)$^{27}$P reactions, with smaller contributions from $^{29}$P(p,$\gamma$)$^{30}$S, $^{30}$P(p,$\gamma$)$^{31}$S, and $^{33}$Cl(p,$\gamma$)$^{34}$Ar. The 9\% impact on the burst light curve tail comes mostly from the revised $^{23}$Al(p,$\gamma$)$^{24}$Si rate with 1-2\% contributions from each of the updates of $^{22}$Mg($\alpha$,p)$^{25}$Al, $^{55}$Ni(p,$\gamma$)$^{56}$Cu, and $^{14}$O($\alpha$,p)$^{17}$F. 

\begin{figure*}
\includegraphics[width=0.5\columnwidth]{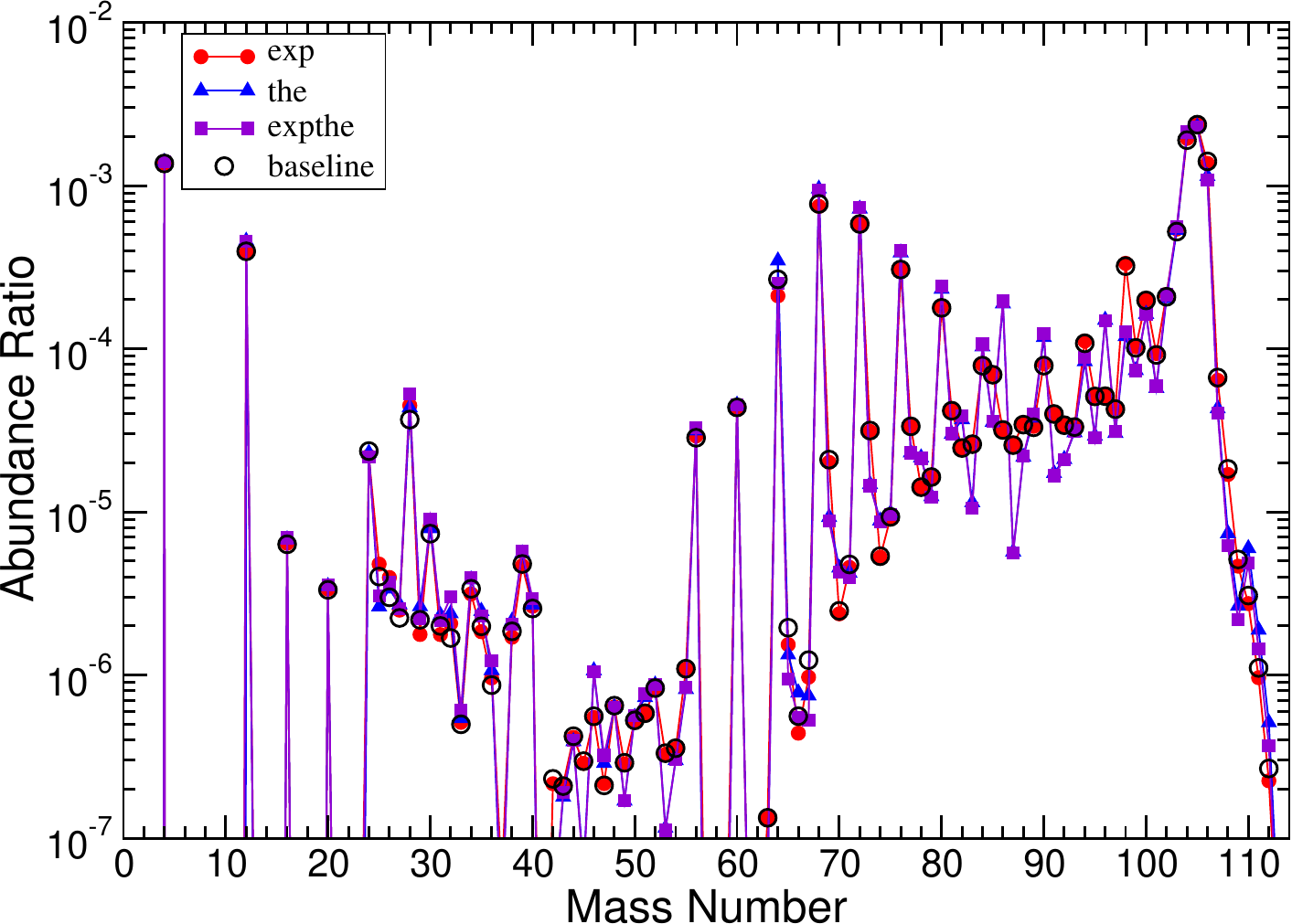}
\includegraphics[width=0.5\columnwidth]{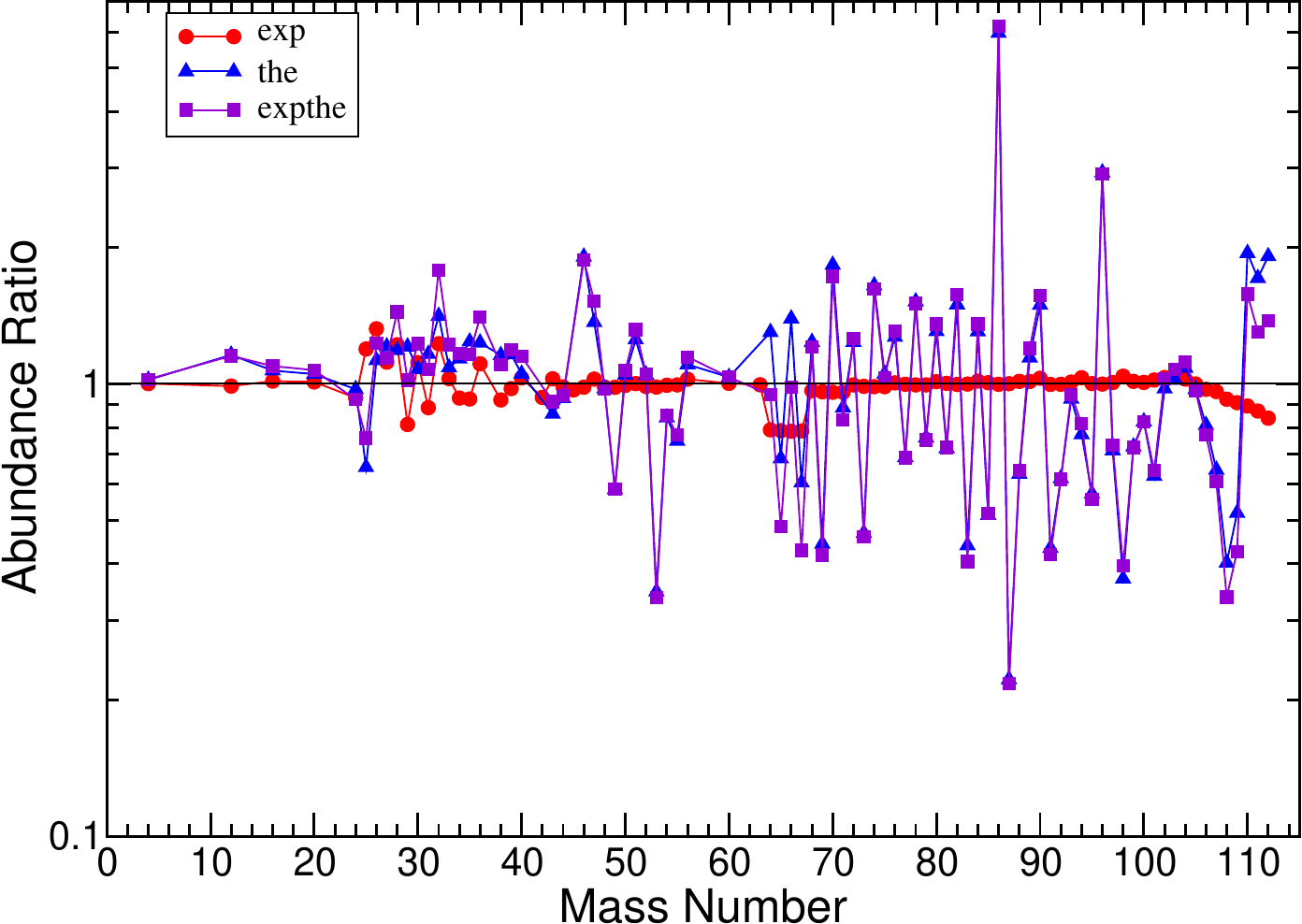}
\caption{Final abundance distribution summed by mass number at the end of the extreme rp-process burst calculation using the \code{ONEZONE} model calculations for the baseline prior to our update (open circles), updates of individual rates (red circles), updates of Hauser-Feshbach rates (blue triangle) and all updates (purple squares). Shown are absolute abundances (left) and the ratio to the baseline (right). \label{FigXRB_AXRB21c_ab}}
\end{figure*}

\begin{figure}
\includegraphics[width=0.5\columnwidth]{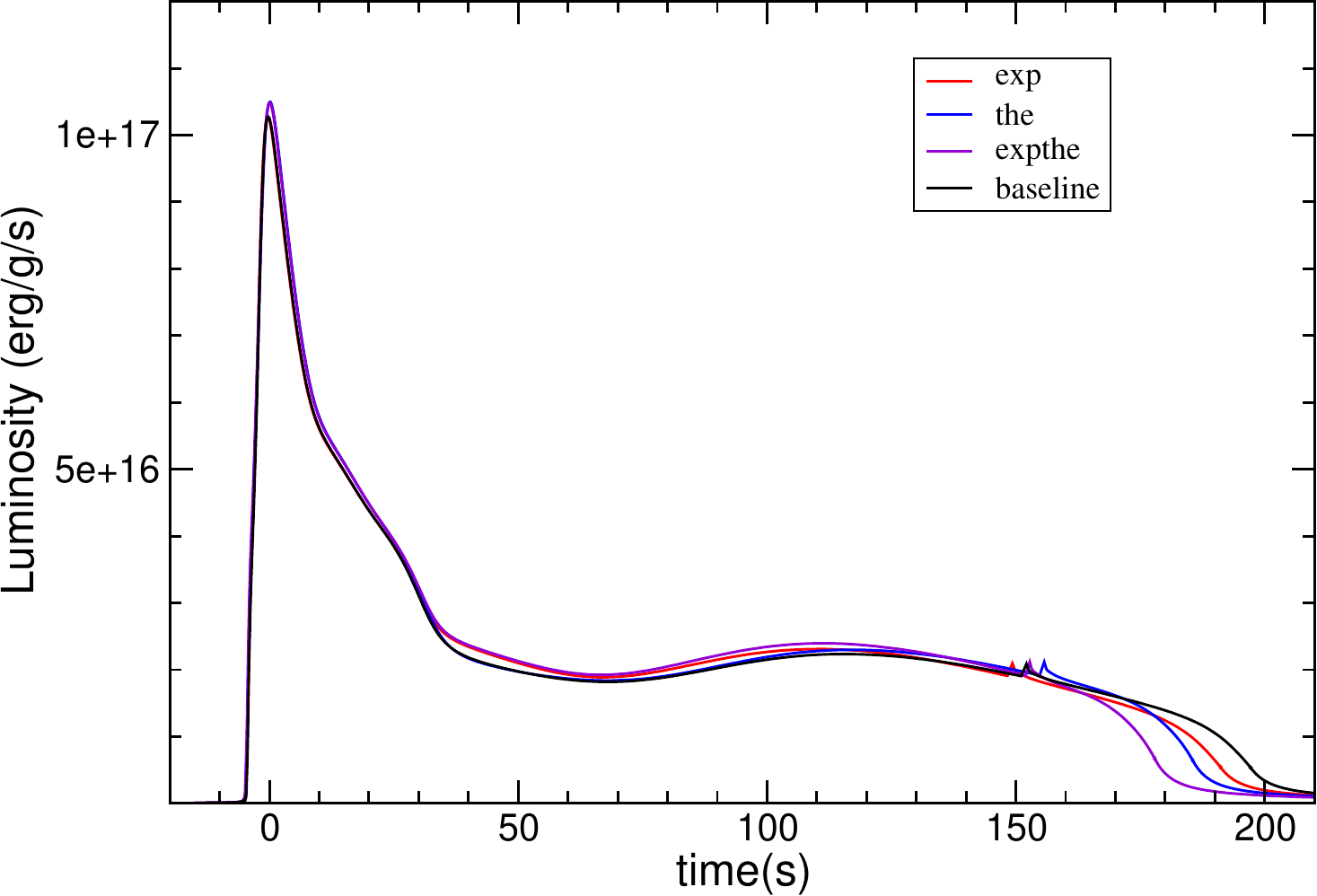}
\caption{Estimated light curves from the extreme rp-process model calculations using the \code{ONEZONE} model calculations for the baseline prior to our update (black line), updates of individual rates (red line), updates of Hauser-Feshbach rates (blue line) and all updates (purple line). \label{FigXRB_AXRB21c_lc}}
\end{figure}

In addition we explored the impact of our updates on a more extreme X-ray burst using the same one-zone model approach, but with ignition conditions determined from a 1D steady state accretion model with accreted metallicities and accretion rates chosen to maximize the extent of the rp-process \cite{schatzEndPointRp2001a} (Figs.~\ref{FigXRB_AXRB21c_ab} and \ref{FigXRB_AXRB21c_lc}). On one hand, the impact of the individual reaction rate updates on composition is much smaller, 
likely due to the higher temperatures that lead to a more extended $\alpha$p-process that bypasses many of the new reactions, and generally shifts the critical bottle-neck reactions to higher $Z$. On the other hand, similar to what we found for the conditions approximating GS1826-24, the update of the statistical model rates leads to many changes in the factor of 2-3 range in the heavier mass region with $A>56$. There is no significant impact on the light curve for this model. 

\begin{figure}
\includegraphics[width=0.5\columnwidth]{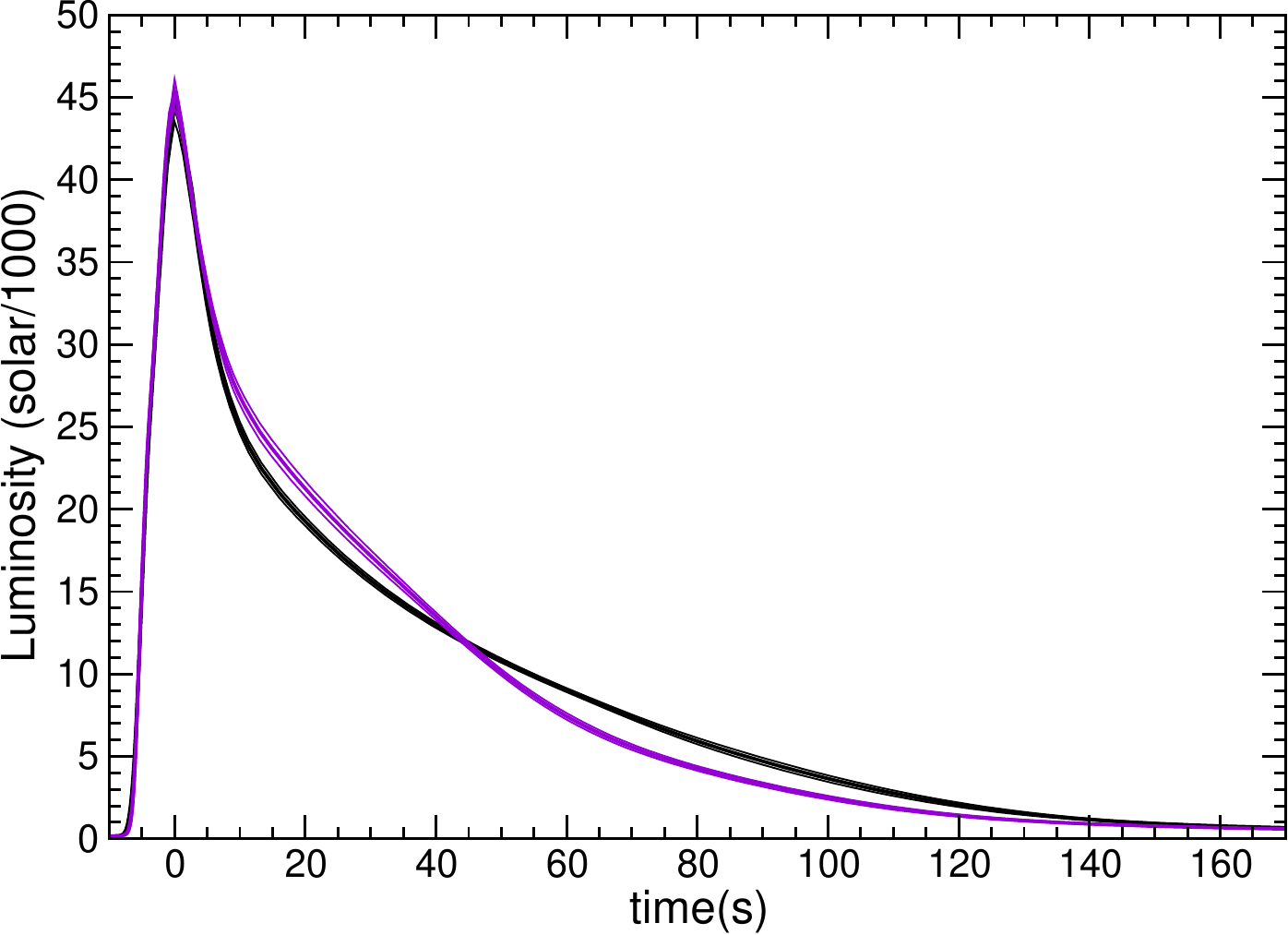}
\caption{Luminosities (light curves) in units of solar luminosity and then divided by a factor of 1000, as functions of time, calculated with the \code{MESA}. Shown are results from a multi-zone X-ray burst model using the original reaction rates prior to our update (black line), and using the updated reaction rates from this work (blue). The thick line indicates the average light curve, and the thin lines with the same color indicate the 1$\sigma$ uncertainty due to burst-to-burst variations.  \label{FigXRB_Mesa}}
\end{figure}

Finally we implemented the new updated reaction libraries in an X-ray burst model based on the \code{MESA} multi-zone 1D stellar evolution code version 9793. The model is the same as the one used to interpret observations of bursts from GS 1826-24 and is described in more detail in \cite{meiselConsistentModelingGS2018}. In the model, an $\sim$10~m thick envelope segmented into $\sim$1000~zones is heated from its base by 0.1~MeV per accreted nucleon. The local gravity is modified by a post-Newtonian correction to approximate general relativistic effects, assuming the base of the envelope is atop a 1.4~$M_{\odot}$ mass and 11.2~km radius neutron star. Convection is approximated using time-dependent mixing-length theory. The advantage of this approach is the more realistic radiation transport that allows direct comparisons with observations, and the more realistic treatment of accretion enabling the simulation of a more realistic sequence of bursts. This will also account for important compositional inertia effects \cite{taamXrayBurstsThermonuclear1980a}, where the partially burned ashes of a previous burst affects the following burst. The simulations also include the important inter-burst phase, where reactions in the CNO region alter the fuel composition leading to additional nuclear physics sensitivities. The drawback of the 1D model is the high computational cost and significant variations from burst to burst, driven in part by feedback effects from previous bursts that may not be fully statistical and thus may require unfeasibly long burst sequences to fully capture~\citep{LampeBurstTrains2016}. Here we ran a sequence of 18 bursts. The model exhibits significant burst-to-burst variations with a 1$\sigma$ relative change of 17\% (22\%) and 26\% (9\%) of the peak luminosity and the luminosity at 20~s after peak, respectively, for the updated (previous) reaction rates. The 20~s mark is important as at this time the reaction rate sensitivity of the burst tail appears to be largest. We discard the first 2 bursts from the analysis to ensure the sequence is in a steady state and average the profiles of the remaining 16 bursts following the approach described in \cite{meiselConsistentModelingGS2018}. The resulting average burst light curves and their 1$\sigma$ uncertainties due to burst-to-burst variations are shown in Fig.~\ref{FigXRB_Mesa}. For the burst tail, the impact of the reaction rate updates in the multi-zone model is very similar to what is found in the one-zone model \code{ONEZONE}. In both models there is an increase in luminosity especially in the 10-40~s range after peak by about 9\% and a more rapid consumption of fuel leading to a reduced luminosity towards the end of the burst. Unlike \code{ONEZONE}, the multi-zone calculations also show some changes in the peak luminosity, though uncertainties from burst to burst variations are particularly large there. A test calculation with \code{MESA} confirmed that changes in the same two reactions identified in the \code{ONEZONE} study, $^{23}$Al(p,$\gamma$)$^{24}$Si and $^{22}$Mg($\alpha$,p)$^{25}$Al, are responsible for most of the change. This underlines that one-zone approximations can be useful for identifying reaction rate sensitivities. 

To obtain a prediction of the composition of the steady state burst ashes of importance for the neutron star crust, we sum abundances by mass number in the deepest zones with pressure $P>\EE{1.4}{23}$~dyne/cm$^2$. At these depths hydrogen is depleted and the helium abundance is <10$^{-3}$, which is much smaller than the abundance of the $\alpha$-isotope abundances that could potentially still capture $^4$He. Therefore, no significant further modification of the ashes from X-ray burst activity is expected. We exclude from the analysis the very deepest layers with $P>\EE{3.2}{23}$~dyne/cm$^2$, which correspond to the ashes from the very first bursts of the burst sequence. Due to burst to burst variations and feedback effects in the multi-zone approach, the abundances are not perfectly constant but there are small variations from layer to layer. We therefore determine an uncertainty by calculating the standard deviation of the abundance variations over the integrated depth. 

\begin{figure*}
\includegraphics[width=0.5\columnwidth]{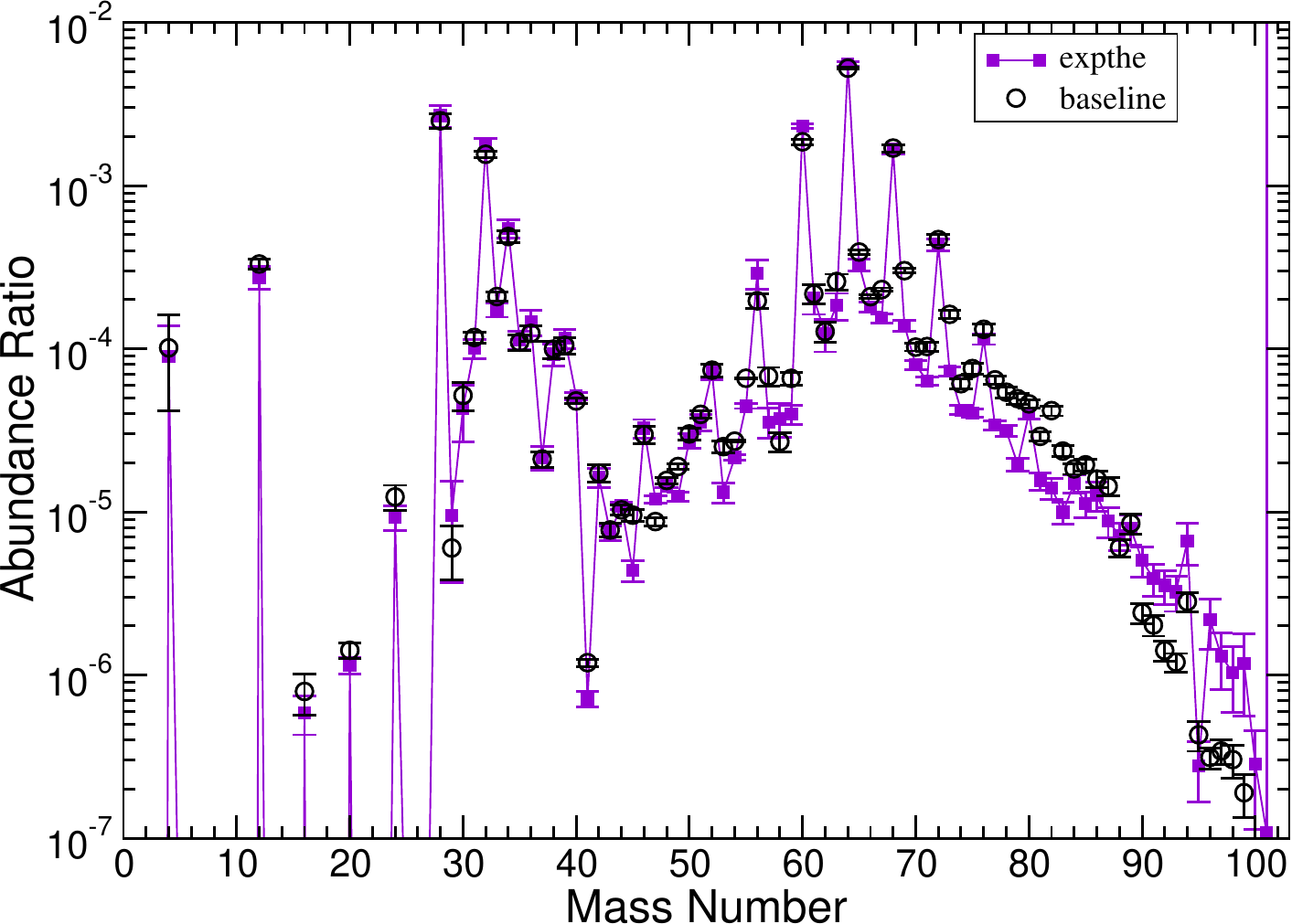}
\includegraphics[width=0.5\columnwidth]{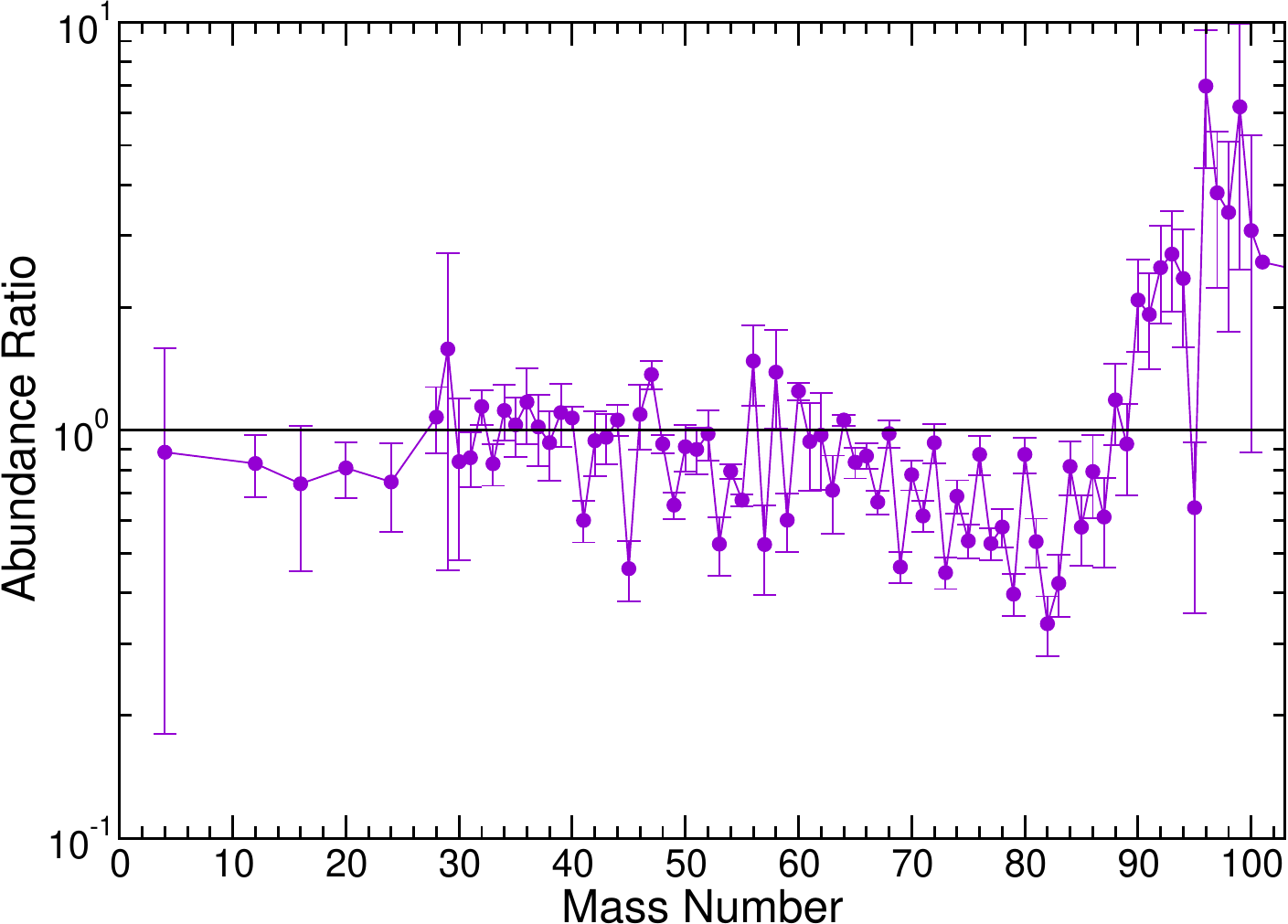}
\caption{Final abundance distribution summed by mass number at the end of the \code{MESA} calculation for the baseline prior to our update (open circles) and updated reaction library (purple squares). Shown are absolute abundances (left) and the ratio to the baseline (right). \label{FigXRB_MESA_ab}}
\end{figure*}

Overall the final abundances (Fig.~\ref{FigXRB_MESA_ab}) for $A>40$ are similar to the ones calculated with \code{ONEZONE} (Fig.~\ref{FigXRB_ONEZONE_ab}). $A=64$ is the most abundant mass chain, followed by $A=60$, and then $A=68$, with abundance decreasing for higher mass numbers.  There are large differences in the $A<40$ mass region between one zone and multi-zone calculations that have also been found in \cite{cyburtDependenceXRayBurst2016a}. This is not surprising as this mass region is shaped by He burning layers, which is not the focus of the one-zone models. The one-zone models freezeout somewhat faster after hydrogen is consumed, which can be seen from the faster drop in luminosity at the very end of the light curve. Therefore a large amount of He is left unburned. In contrast, in the multi-zone model with more realistic radiation transport the layers cool more slowly and more helium is burned after hydrogen is consumed. In addition, any remaining helium is burned in deeper layers that are re-heated by subsequent burst. The latter effect is not included in the one-zone model, that only follows a single burst. The $A<40$ mass number range composition in the one-zone model is therefore dominated by rapid freezeout from the $\alpha$p-process and thus by $A=22, 26, 30, 34$ instead of $\alpha$-chain isotopes. Comparison of the \code{MESA} burst ashes composition with the composition produced by \code{KEPLER} shown in \cite{cyburtDependenceXRayBurst2016a} is much more similar. The main difference is that \code{MESA} predicts the ashes in this region to be dominated by $^{28}$Si, while \code{KEPLER} indicates a slightly more extended He burning mostly producing $^{32}$S, likely owing to the somewhat higher temperatures (see below). The changes in burst ashes from our reaction rate upgrade tend to be smaller in \code{MESA} than in \code{ONEZONE} staying mostly within a factor of 2. In particular the occasional larger deviations are missing, likely due to averaging effects over multiple zones. An exception are the heaviest mass chains with $A>90$, where we find a strong increase with the new rates. It appears that with the updated reaction rates, the rp-process extends slightly further towards heavier nuclei. It should be noted that the layer to layer variations in the composition calculated by \code{MESA}, indicated by the error bars, make it more difficult to discern smaller changes due to nuclear physics. 

\subsection{Reaction Flows}
\label{subsec:flow}

We can now use the results from our updated multi-zone 1D model to review the current understanding of the total reaction flow pattern in X-ray bursts. Reaction flow is defined as the time integrated abundance change caused by an individual reaction. Within an individual burst, conditions and thus reaction flows vary greatly with depth (zone). It is therefore important to use a 1D multi-zone model to obtain a full picture of the nuclear reactions that occur in X-ray bursts. We employ here the \code{MESA} based model discussed above \cite{meiselConsistentModelingGS2018}.

To calculate reaction flows we use here a simple explicit postprocessing integration scheme that integrates over time and depth, weighted by zone mass. For example, we obtain the reaction flow $f_{i \rightarrow j}$ for a 2-body reaction with a light projectile p (e.g. a proton or $\alpha$-particle) that converts nucleus $i$ into nucleus $j$ using  
\begin{equation}
f_{i \rightarrow j} = \frac{1}{M} \sum_k^{N\rm time} \sum_l^{N\rm zones,k} Y_{{\rm p},l} Y_{i,l} \, \rho_l N_A <\sigma v>_l m_{lk} \Delta t_k,
\end{equation}
with $Y_{{\rm p},l}$ and $Y_{i,l}$ the abundances of projectile and target nucleus $i$, respectively, in zone $l$ at timestep $\Delta t_k$, $\rho_l$ the mass density, $N_A <\sigma v>_l$ the reaction rate, and $m_{lk}$ the total mass of zone $l$ in timestep $k$. $M=\sum_l \sum_k m_{lk}$.

\begin{figure}
\includegraphics[width=0.5\textwidth]{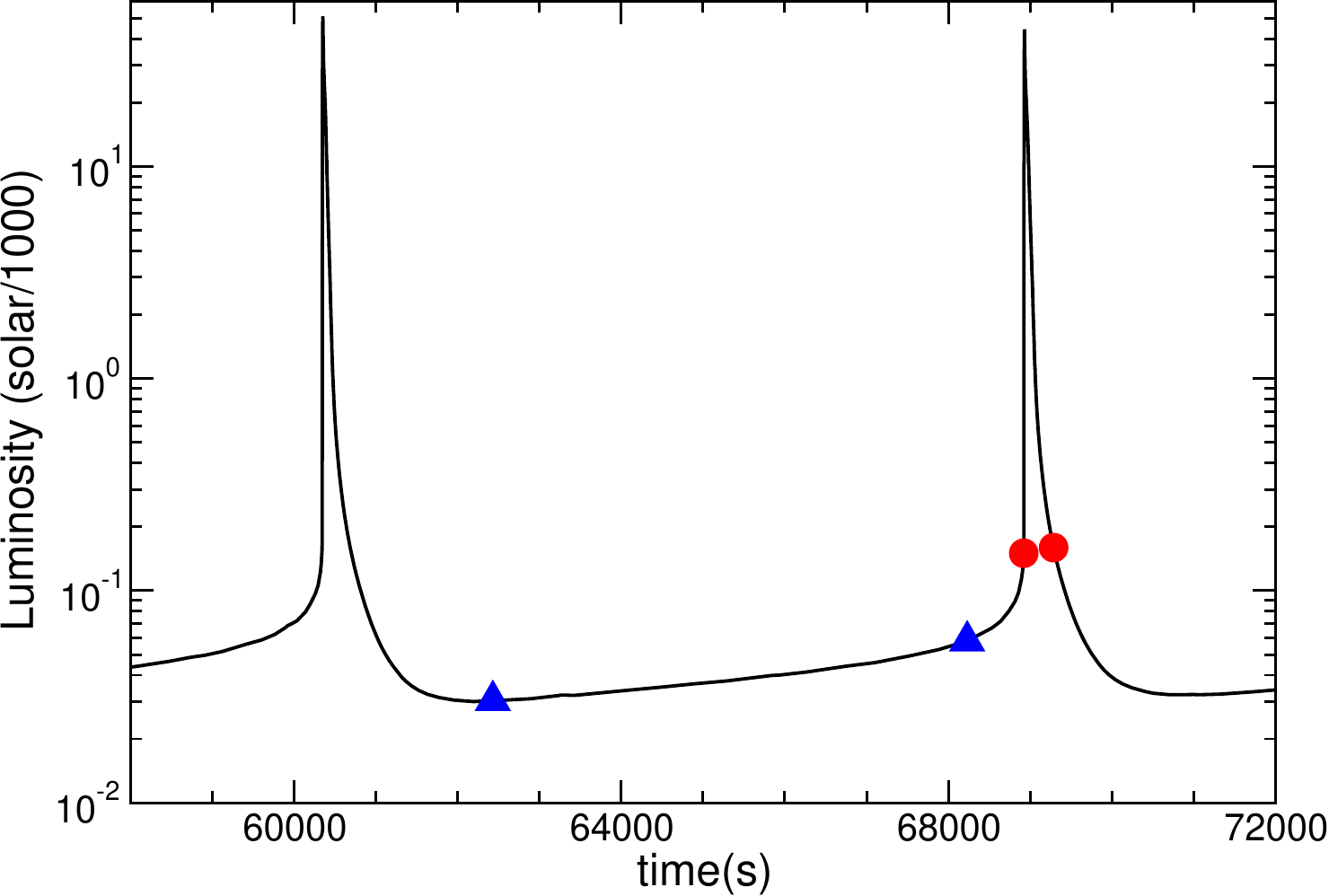}
\caption{Integration time limits for the determination of the typical reaction flow during a burst (red circles) and the inter burst phase (blue triangles) shown on the relevant part of the burst train light curve displaying the seventh and eighth burst. 
\label{fig:flow_integration}}
\end{figure}

\begin{figure*}
\includegraphics[width=0.9\textwidth]{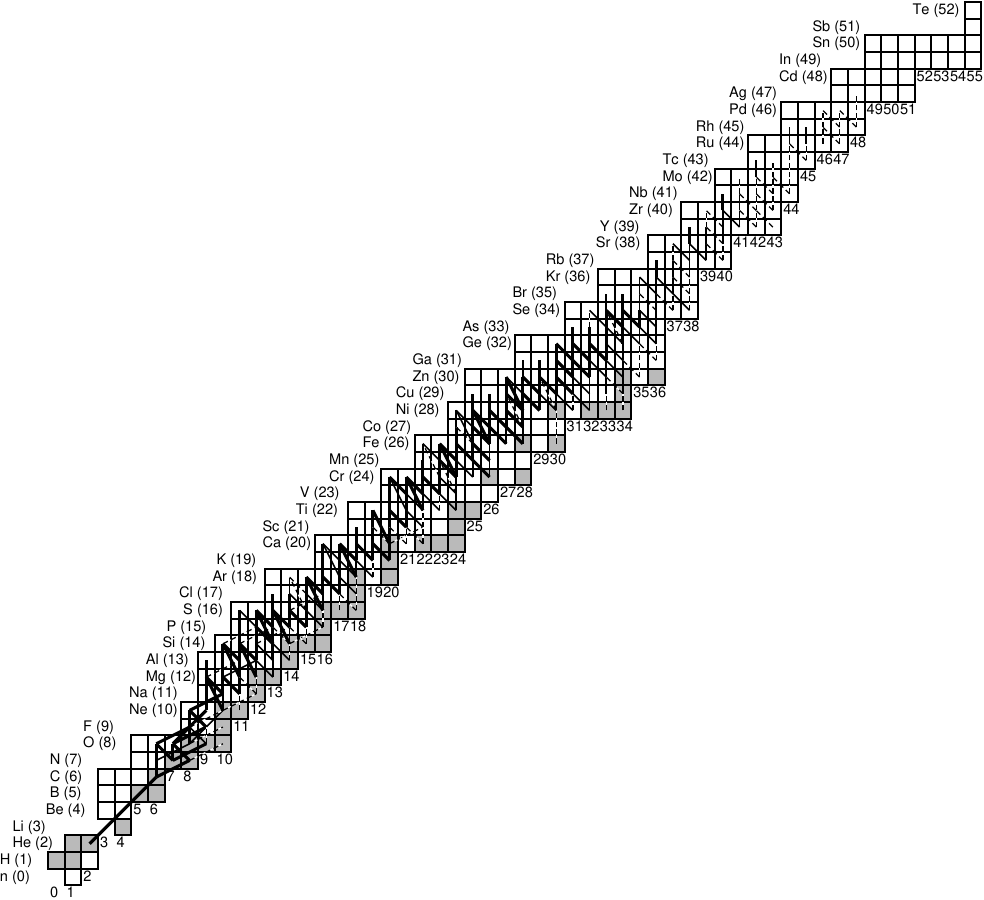}
\caption{Reaction net flows in the hydrogen burning zones of a typical burst, integrated over burst duration and depth. Thick solid, thin solid, and dashed lines indicate flows within 10\%, 1\%, and 0.1\% of the 3$\alpha$ reaction. The explicit integration scheme leads to spuriously high net flows for (p,$\gamma$) reactions with strong inverse reactions - in such cases the thick solid line indicates (p,$\gamma$)-($\gamma$,p) equilibrium, rather than a large net flow. 
\label{fig:flow_h}}
\end{figure*}

\begin{figure*}
\includegraphics[width=0.9\textwidth]{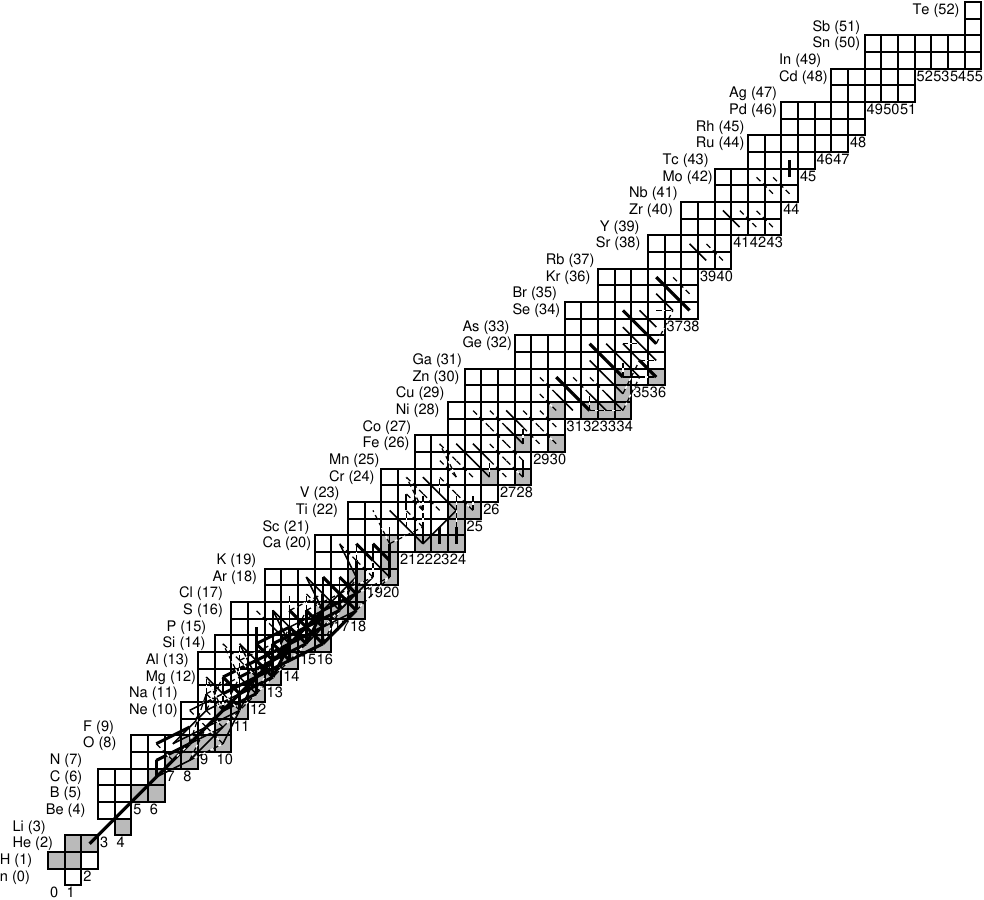}
\caption{Reaction net flows in the pure helium burning zones of a typical burst, integrated over burst duration and depth.
\label{fig:flow_he}}
\end{figure*}

\begin{figure*}
\includegraphics[width=0.9\textwidth]{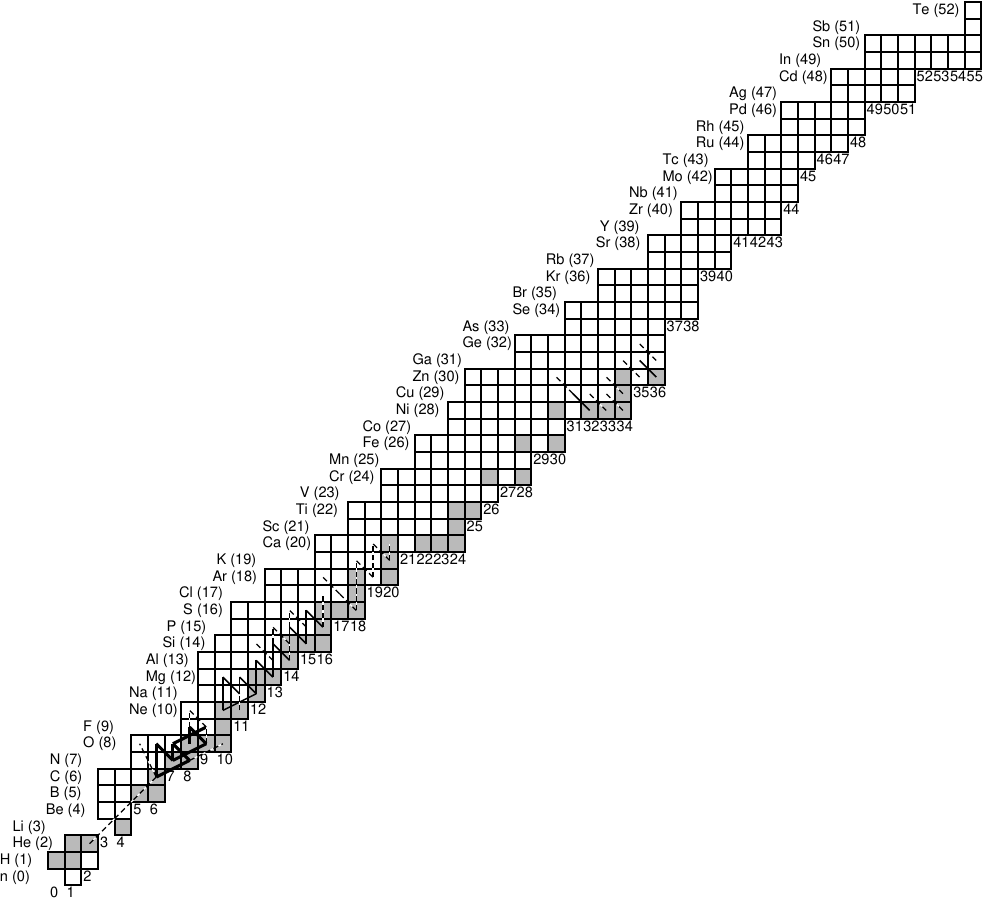}
\caption{Reaction net flows in-between bursts integrated over burst duration and depth. Thick solid, thin solid, and dashed lines indicate flows within 10\%, 1\%, and 0.1\% of the $^{14}$O $\beta$-decay. 
\label{fig:flow_interburst}}
\end{figure*}

We note that with this definition the reaction flow is simply a measure of how often a specific nuclear reaction has occurred during the burst. It does not provide a complete history or reaction sequence for how a particular isotope has been produced, as abundances in each zone can also be changed through the transport of nuclei from zone to zone, for example by convection.

In addition, each burst ignites in an environment set by the previous burst, in terms of thermal structure, and partially processed nuclear ashes. Together with the strong coupling between nuclear burning and the temperature and density evolution this can lead to burst-to-burst variations, in nature, as well as in models. To determine the typical reaction flow in our model, we select one burst in the burst train that most closely resembles the average burst profile and sum over the timesteps from beginning to end of the burst. The integration time limits are indicated in Fig.~\ref{fig:flow_integration}.

As we are primarily interested in mixed H/He burning, we first consider integration of reactions occurring in layers with hydrogen present (hydrogen mass fraction $X_H>10^{-5}$) (Fig.~\ref{fig:flow_h}). The reaction patterns follow broadly the results from previous studies that analyzed flow patterns zone by zone and as a function of time \cite{fiskerExplosiveHydrogenBurning2008, joseHydrodynamicModelsType2010a}. They are also similar to the integrated flow patterns obtained with the \code{KEPLER} model \cite{cyburtDependenceXRayBurst2016a}. Burning starts with the 3$\alpha$ reaction, followed by the hot CNO cycles, and strong breakout via both $^{18}$Ne($\alpha$,p)$^{21}$Na and $^{15}$O($\alpha$,$\gamma$)$^{19}$Ne. ($\alpha$,p) reactions as part of the $\alpha$p-process process material further into the Al region, with the last ($\alpha$,p) reaction with flow >0.1\% of the 3$\alpha$-reaction being the $^{22}$Mg($\alpha$,p)$^{25}$Al reaction. The end of the $\alpha$p process transitions into the beginning of the rp-process, rapid proton captures and slow $\beta$-decays proceeding along the proton drip line mainly up to Sr, with significant flow continuing into the Zr region, and some flow up to Ag. Well known potential waiting points in the rp-process are the $N=Z$ nuclei $^{60}$Zn, $^{64}$Ge, $^{68}$Se, and $^{72}$Kr, which have relatively long half-lives and low or negative proton capture Q-values. An rp-process bypass of these waiting points would be indicated by a $\beta^+$ decay flow of the corresponding (Z+1,N) and (Z+2,N) nuclei. We find that there is a strong bypass with comparable flows for $^{60}$Zn, and a weaker bypass of $^{64}$Ge at the 1-10\% level (see also \cite{zhouMassMeasurementsShow2023b}). On the other hand, $^{68}$Se and $^{72}$Kr are not bypassed and impede the rp-process with their full decay time. $^{56}$Ni is essentially stable on X-ray burst timescales and can also impede the rp-process, especially at high temperatures \cite{schatzRpprocessNucleosynthesisExtreme1998b}. We identify an effective bypass of $^{56}$Ni via $^{55}$Ni(2p,2$\gamma$)$^{57}$Zn($\beta^+$)$^{57}$Cu that has been discussed before \cite{ongLowlyingLevelStructure2017,valverdeHighPrecisionMassMeasurement2018}.

There are also significant differences to the integrated reaction flows obtained with the \code{KEPLER} model \cite{cyburtDependenceXRayBurst2016a}. Most importantly, \cite{cyburtDependenceXRayBurst2016a} find a much more extended $\alpha$p-process reaching into the Ca-Sc region, and significant activity in the Co-Ni and Cu-Zn cycles \cite{vanwormerReactionRatesReaction1994a} which we do not see above the 0.1\% 3$\alpha$-flow level. This is likely due to higher peak temperatures in \code{KEPLER} (and also \code{ONEZONE}) of 1.2~GK, compared to a peak temperature of 1.03~GK found here. 

Mixed H/He bursts also exhibit layers with pure He burning. This occurs at the bottom of the hydrogen burning layer, where hydrogen is consumed much faster than helium. The reason is that helium burning relies on the three-body 3$\alpha$ reaction and thus depends on He density cubed, as opposed to H burning via proton capture that is proportional to the H density. As the He density drops, the burn rate drops much faster, resulting in residual He in an essentially H-free zone. While some of the He is burned into $^{12}$C and heavier $\alpha$-chain nuclei during the burst, significant amounts of He remain, and burning continues in the next burst that reheats the layer. Fig.~\ref{fig:flow_he} shows the typical He burning reaction flow patterns in He burning layers $X_H<10^{-5}$. These include the 3$\alpha$-reaction, an $\alpha$-capture chain from $^{12}$C to $^{32}$S, as well as various ($\alpha$,p) reactions, and captures of the released protons up to the Ca region. We can also identify the  $^{12}$C(p,$\gamma$)$^{13}$N($\alpha$,p)$^{16}$O bypass of $^{12}$C($\alpha$,$\gamma$)$^{16}$O enabled by the small H abundance produced by the ($\alpha$,p) reactions that can significantly accelerate He burning \cite{weinbergExposingNuclearBurning2006b}. In addition, there are continuing $\beta$-decays of heavier nuclei previously produced by hydrogen burning via the rp-process that may contribute to heating of the He burning layers. 

Finally, we also determined the reaction flow during the interburst phase  (Fig.~\ref{fig:flow_interburst}). The corresponding integration time range is indicated in Fig.~\ref{fig:flow_integration}. There is significant nuclear burning in between bursts. This is important as it may alter the H/He ratio available at burst ignition. The reaction flow pattern is limited to hydrogen burning close to stability via the hot CNO cycles up to $^{18}$Ne as well as some modest activity in the NaNe cycle and a slow proton capture process from Ne to P following the valley of stability. The impact of the latter reactions on the bursts remains to be explored.

\section{Conclusions}  
X-ray bursts are the most frequently observed thermonuclear explosions in space and provide a unique window into the physics of neutron stars and dense matter. With the large amount of observational data accumulated and the possibility of future observations of the same systems on a daily basis, they also offer unique opportunities for probing the physics of astrophysical thermonuclear explosions and for validating associated model techniques. Nuclear physics is critical for interpreting observations and taking advantage of these opportunities. We have summarized the significant recent efforts in experimental and theoretical nuclear science that were motivated by the X-ray burst problem.  

On the experimental side these efforts highlight new opportunities at a broad range of stable and radioactive beam facilities, enabled by a very broad array of experimental techniques and detection systems. We have used this new information to update 32 reactions in the JINA REACLIB database. We find that these updates led to significant changes of more than a factor of 2 in 15 cases in the relevant temperature range for X-ray bursts. This shows that nuclear reaction rate data for X-ray bursts are still very much in flux. Experimental efforts have largely focused on the lighter mass region below Ca. Owing to these efforts, most reactions in this mass range now can be determined without using the statistical model. An exception are ($\alpha$,p) reactions beyond Si. However, there remain important open nuclear physics questions in this mass region that remain to be addressed. Furthermore, it will be important to expand experimental nuclear physics efforts to heavier nuclei beyond Ca in the future. While a variety of techniques and instruments are now available, continued efforts to increase radioactive beam intensities will be essential to accomplish the nuclear physics goals for X-ray bursts. 

On the theoretical side, we implemented updated statistical model rates using the publicly available \code{TALYS} code. We find significant changes exceeding factors of 5 in many rates. While some of these changes are due to improvements in nuclear input such as masses, other changes are due to differences in theoretical nuclear physics input such as strength functions, level densities, and optical potentials. For the latter, while there may be arguments why certain descriptions are more reliable, much of these differences may reflect uncertainties in the description of nuclear properties far from stability that ultimately need to be addressed by experiments. For comparison, we also calculated statistical model rates for cases where rates are based on more reliable experimental data or shell model calculations. We find significant differences, typically within a factor of 10, but in some cases reaching factors of 100. This provides some insight into the scale of rate changes that may be expected in the future as more statistical model reactions are replaced. 

Our new reaction library can readily be employed in X-ray burst models. Implementing the new data in one-zone and 1D multi-zone models indicated modest changes in final composition and light curves, that may nevertheless be significant in interpreting observations along the lines of  \cite{huAdvancementPhotosphericRadius2021a}. We also presented 1D multi-zone model flow patterns that are largely in line with previous studies but may indicate less prevalence of cycles in the Co-Ge region, in part due to the smaller (p,$\alpha$) rates predicted by \code{TALYS}. 

Our model comparison also indicates some differences between X-ray burst model codes, for example \code{KEPLER} and \code{MESA}. While many conclusions are robust, a particular concern for nuclear physics are differences in peak temperatures, that strongly affect the extent of the $\alpha$p-process and as such the importance of ($\alpha$,p) reactions. The impact of \code{ONEZONE}-\code{KEPLER} and \code{MESA} peak temperature differences on nuclear reaction sensitivities has been noted previously in the context of the role of photodisintegration and proton binding energies near the dripline \cite{Yandow2023}.  

There is nothing wrong with multiple independent astrophysical models with different physics descriptions and different numerical implementations being used in the community. In fact, this is important to advance understanding of X-ray bursts. Ultimately comparison of each model with observations is needed to identify the successful models. To that end, nuclear physics should provide each model the nuclear data it needs to be validated. It is not surprising that the nuclear physics needs of different models will be different. The goal of nuclear astrophysics is to provide all the needed data for all models. Nevertheless, it may be useful to compare models and explore whether some convergence can be achieved to narrow down the nuclear physics requirements. For example, \citep{JohnstonMultiEpochModeling} note that \code{KEPLER} X-ray burst calculations run prior to 2018 have an opacity artificially inflated by 50\%, leading to hotter envelope temperatures. This maybe part of the reason our study using \code{MESA} obtains lower peak temperatures.

At the moment the JINA REACLIB database does not include uncertainties. While the majority of astrophysical applications currently do not require that information, it is needed for some important use cases such as any future attempts to propagate systematically nuclear physics uncertainties to observables or to provide guidance for future nuclear physics work to address the most important uncertainties. While beyond the scope of the current work, we plan to implement uncertainties as a next step in the future. The STARLIB library \cite{sallaskaSTARLIBNextgenerationReactionrate2013} has already taken a first important step in this direction, focusing on uncertainties from charged particle reactions with narrow resonances in a certain mass region. It will be important to also include uncertainties in rates with broad resonances typically obtained with R-matrix analysis, $\beta$-decay, shell model rates, statistical model rates, stellar enhancement factors, and partition functions. A particular challenge is to account for correlations that stem, for example, from common experimental inputs (such as masses) or common theoretical assumptions or inputs across multiple rates. 

The JINA REACLIB database can be used for calculations of a broad range of nucleosynthesis processes beyond the rp- and $\alpha$p-processes in X-ray bursts. The updated reaction rate library discussed in this work will also be applicable to other processes dominated by proton and $\alpha$-induced reactions on neutron deficient nuclei in the $A < 110$ mass range such as the rp-process in Nova explosions, the $\nu$p-process in supernovae, or explosive Si burning. Updates of neutron capture and $\beta$-decay rates for the s-, r-, and i-processes, as well as updates of $\alpha,n$ reactions on neutron rich nuclei in the weak r-process, and updates of stellar fusion rates are planned for the future. 

\section*{Acknowledgements}
This work was supported by the US National Science Foundation under Award Nos. OISE-1927130 (IReNA), PHY-2209429, PHY-1430152 (JINA Center for the Evolution of the Elements) and by the U.S. Department of Energy, Office of Science, Office of Nuclear Physics under Award No. DE-SC0023128 (CeNAM). PM was supported by NKFIH (K134197).
YX acknowledges additional support from the Romanian Ministry of Research, Innovation and Digitization, CNCS-UEFIS-CDI, project number PN-IV–P1-PCE-2023-0384 within PNCDI IV, the Romanian Ministry of Research, Romanian Ministry of Research, Innovation and Digitalization under Contract No. PN 23 21 01 06, the ELI-RO project with Contract ELI-RORDI-2024-008 (AMAP), and the IAEA Coordinated Research Project on "Updating and Improving Nuclear Level Densities for Applications" (F41034) under Contract No. 28638.

\bibliography{main,HS_XRBPaper,HS_Rates}{}
\bibliographystyle{elsarticle-num-names} 

\section*{Appendix $^{14}$C($\alpha$,$\gamma$)$^{18}$O}
We follow the analysis of \cite{PhysRevC.80.045805} that includes four contributions to the $^{14}$C($\alpha$,$\gamma$)$^{18}$O reaction rate: resonant capture via the subthreshold state at E$_x$ = 6.198 MeV ($J^{\pi}$ = $1^{-}$) through the high energy tail and two resonant states at E$_x$ = 6.404 MeV ($J^{\pi}$ = $3^{-}$) and 7.12 MeV ($J^{\pi}$ = $4^{+}$) in $^{18}$O, respectively, as well as the direct capture. \cite{GORRES1992414} measured the resonance strength for the 7.12 MeV state directly and provided a low energy S-factor extrapolation for the direct capture component, based on measurements above 1.14 MeV. \cite{PhysRevC.80.045805} used the ($^6$Li,d) and ($^7$Li,t) $\alpha$-transfer reactions in inverse kinematics at Florida State University to determine the $\alpha$-width of the resonance state at 6.404~MeV and the ANC of the subthreshold resonance state at 6.198~MeV. 
The experimentally unknown sign of the interference between the predominantly p-wave direct capture and the subthreshold resonance introduces a several orders of magnitude uncertainty at very low energies relevant for temperatures below 0.03~GK, which is not relevant for X-ray burst environments. In the present evaluation, the constructive interference is considered. The numerical reaction rates are given in Table \ref{rate:set1}.

\begin{table}
\caption{Reaction rates of $^{14}$C($\alpha$,$\gamma$)$^{18}$O, $^{13}$N($\alpha$,p)$^{16}$O, $^{14}$O($\alpha$,p)$^{17}$F and $^{17}$O(p,$\alpha$)$^{14}$N. \label{rate:set1}}
\scalebox{0.89}{
\begin{tabular}{c c c c c}
\hline \hline
& $^{14}$C($\alpha$,$\gamma$)$^{18}$O & $^{13}$N($\alpha$,p)$^{16}$O & $^{14}$O($\alpha$,p)$^{17}$F & $^{17}$O(p,$\alpha$)$^{14}$N \\
$T$ & $N_{A}<\sigma\upsilon>$ & $N_{A}<\sigma\upsilon>$ & $N_{A}<\sigma\upsilon>$ & $N_{A}<\sigma\upsilon>$ \\
$GK$ & $cm^3/(mol*sec)$ & $cm^3/(mol*sec)$ & $cm^3/(mol*sec)$ & $cm^3/(mol*sec)$  \\
\hline
 0.01 & $-$        &  $-$        & $-$         & 0.7609E-24  \\
 0.02 & 0.5152E-40 &  0.2163E-38 & $-$         & 0.3770E-16  \\
 0.03 & 0.1391E-33 &  0.2051E-31 & 0.1229E-38  & 0.3902E-11  \\
 0.04 & 0.9791E-28 &  0.5292E-27 & 0.2515E-33  & 0.1277E-08  \\
 0.05 & 0.1992E-23 &  0.1065E-23 & 0.1052E-29  & 0.4018E-07  \\
 0.06 & 0.1425E-20 &  0.5385E-21 & 0.5560E-27  & 0.3822E-06  \\
 0.07 & 0.1504E-18 &  0.7621E-19 & 0.8163E-25  & 0.1792E-05  \\
 0.08 & 0.4822E-17 &  0.3904E-17 & 0.5010E-23  & 0.5404E-05  \\
 0.09 & 0.7005E-16 &  0.9475E-16 & 0.1635E-21  & 0.1289E-04  \\
 0.10 & 0.5861E-15 &  0.1326E-14 & 0.3307E-20  & 0.2998E-04  \\
 0.12 & 0.1368E-13 &  0.8238E-13 & 0.4758E-18  & 0.2194E-03  \\
 0.14 & 0.1252E-12 &  0.1860E-11 & 0.2549E-16  & 0.1532E-02  \\
 0.16 & 0.6419E-12 &  0.2216E-10 & 0.6882E-15  & 0.7614E-02  \\
 0.18 & 0.2244E-11 &  0.1748E-09 & 0.1130E-13  & 0.2775E-01  \\
 0.20 & 0.6027E-11 &  0.1056E-08 & 0.1275E-12  & 0.7878E-01  \\
 0.25 & 0.3559E-10 &  0.4848E-07 & 0.1738E-10  & 0.4876E+00  \\
 0.30 & 0.2160E-09 &  0.1165E-05 & 0.8058E-09  & 0.1685E+01  \\
 0.35 & 0.9220E-08 &  0.1635E-04 & 0.1977E-07  & 0.6314E+01  \\
 0.40 & 0.2889E-06 &  0.1514E-03 & 0.3473E-06  & 0.2649E+02  \\
 0.45 & 0.4277E-05 &  0.1032E-02 & 0.4645E-05  & 0.9456E+02  \\
 0.50 & 0.3646E-04 &  0.5488E-02 & 0.4432E-04  & 0.2729E+03  \\
 0.55 & 0.2079E-03 &  0.2348E-01 & 0.2999E-03  & 0.6595E+03  \\
 0.60 & 0.8768E-03 &  0.8301E-01 & 0.1505E-02  & 0.1388E+04  \\
 0.65 & 0.2936E-02 &  0.2494E+00 & 0.5911E-02  & 0.2625E+04  \\
 0.70 & 0.8202E-02 &  0.6539E+00 & 0.1910E-01  & 0.4559E+04  \\
 0.75 & 0.1984E-01 &  0.1532E+01 & 0.5281E-01  & 0.7391E+04  \\
 0.80 & 0.4272E-01 &  0.3266E+01 & 0.1288E+00  & 0.1133E+05  \\
 0.85 & 0.8356E-01 &  0.6445E+01 & 0.2834E+00  & 0.1659E+05  \\
 0.90 & 0.1510E+00 &  0.1192E+02 & 0.5709E+00  & 0.2337E+05  \\
 0.95 & 0.2552E+00 &  0.2085E+02 & 0.1066E+01  & 0.3185E+05  \\
 1.00 & 0.4077E+00 &  0.3485E+02 & 0.1866E+01  & 0.4223E+05  \\
 1.20 & 0.1745E+01 &  0.1930E+03 & 0.1077E+02  & 0.1058E+06  \\
 1.40 & 0.4756E+01 &  0.7443E+03 & 0.3789E+02  & 0.2106E+06  \\
 1.60 & 0.9826E+01 &  0.2296E+04 & 0.1044E+03  & 0.3630E+06  \\
 1.80 & 0.1693E+02 &  0.6050E+04 & 0.2575E+03  & 0.5676E+06  \\
 2.00 & 0.2575E+02 &  0.1410E+05 & 0.6093E+03  & 0.8283E+06  \\
 2.50 & 0.5244E+02 &  0.7719E+05 & 0.4332E+04  & 0.1748E+07  \\
 3.00 & 0.8172E+02 &  0.2728E+06 & 0.2011E+05  & 0.3091E+07  \\
 4.00 & 0.1516E+03 &  0.1505E+07 & 0.1515E+06  & 0.7204E+07  \\
 5.00 & 0.3078E+03 &  0.4452E+07 & 0.5083E+06  & 0.1334E+08  \\
 6.00 & 0.7540E+03 &  0.9471E+07 & 0.1112E+07  & 0.2137E+08  \\
 7.00 & 0.1859E+04 &  0.1676E+08 & 0.1901E+07  & 0.3078E+08  \\
 8.00 & 0.4137E+04 &  0.2669E+08 & 0.2791E+07  & 0.4077E+08  \\
 9.00 & 0.8116E+04 &  0.4002E+08 & 0.3704E+07  & 0.5035E+08  \\
10.00 & 0.1411E+05 &  0.5801E+08 & 0.4553E+07  & 0.5856E+08  \\
\hline
\end{tabular}
}
\end{table}

\section*{Appendix $^{13}$N($\alpha$,p)$^{16}$O}
We adopt the reaction rate evaluation of \cite{meyerEvaluation13N16O2020}, who carried out $^{13}$C($^7$Li,t)$^{17}$O $\alpha$-transfer measurements at Orsay to determine $\alpha$-widths of states in $^{17}$O that are isospin mirrors of resonance states in $^{17}$F. \cite{meyerEvaluation13N16O2020} combined their results with previous work identifying states in $^{17}$F, mostly via proton inelastic scattering and measurements of the inverse $^{16}$O(p,$\alpha$)$^{13}$N reaction. In cases where mirror assignments were not possible, $\alpha$-widths were estimated. An R-matrix analysis revealed 6 resonances that contribute the most to the reaction rate at temperatures below 2~GK (at 221, 741, 959, 1213, 1542, and 1664~keV). 

For temperatures below 2~GK the $^{13}$N($\alpha$,p)$^{16}$O reaction can in principle also be determined from direct measurements of the inverse $^{16}$O(p,$\alpha$)$^{13}$N reaction, as in that temperature regime contributions from excited states in $^{16}$O should be negligible \cite{jayatissaFirstDirectMeasurement2022}. The corresponding traditional Gamow window spans up to 2.2~MeV center of mass energy in  $^{13}$N($\alpha$,p)$^{16}$O, corresponding to resonance energies up to 7.4 MeV in $^{16}$O(p,$\alpha$)$^{13}$N. Direct cross section data for $^{16}$O(p,$\alpha$)$^{13}$N are limited at these low energies, but some data are available down to 6.5~MeV \cite{mccamisStellarRates16O1973}, corresponding to temperatures around 0.6~GK. These data were available for the $^{16}$O(p,$\alpha$)$^{13}$N reaction evaluation in the CF88 rate compilation \cite{caughlanThermonuclearReactionRates1988}, and the tabulated reverse rate agrees reasonably well with the here adopted $^{13}$N($\alpha$,p)$^{16}$O rate in the temperature range of 0.6-2~GK \cite{meyerEvaluation13N16O2020}. This indicates consistency, though the $^{16}$O(p,$\alpha$)$^{13}$N data do not show any resonances below 7.4 MeV, likely due to the sparse energy sampling. 

While not relevant for X-ray bursts, the here adopted rate from \cite{meyerEvaluation13N16O2020} for temperatures above 2~GK is based on \code{TALYS} statistical model calculations from \starlib. Recently, a first direct measurement of the $^{13}$N($\alpha$,p)$^{16}$O reaction \cite{jayatissaFirstDirectMeasurement2022} provided first cross section data in the higher energy range above 3.3~MeV center of mass energy, relevant for temperatures above about 4~GK. While broadly consistent with \code{TALYS} cross sections within theoretical and experimental uncertainties, the experimental data tend to be somewhat smaller than \code{TALYS} predictions. The authors attribute this to the population of states above the $\alpha$-threshold in $^{16}$O, which subsequently decay by $\alpha$-emission into $^{12}$C. Such events are part of the \code{TALYS} rate, but not included in the experimental data. To properly account for this effect in network calculations, a $^{13}$N($\alpha$,p+$\alpha$)$^{12}$C branching would have to be explicitly included in the reaction network. This may be relevant for higher temperature scenarios such as explosive helium burning in core collapse supernovae. The reaction rate is listed in Table \ref{rate:set1}. No SEF is used for this reaction.

\section*{Appendix $^{14}$O($\alpha$,p)$^{17}$F}
For temperatures below 1-2~GK, the rate is dominated by a resonance corresponding to the 6.15~MeV state in $^{18}$Ne and direct capture. Direct capture and the 6.15~MeV resonance interfere with an unknown sign, which introduces uncertainties of an order of magnitude or more for temperatures below 0.3~GK, and uncertainties of less than 20\% for temperatures above 0.5~GK \cite{hahnStructure18Mathrm1996}. For higher temperatures, a number of higher lying resonances contribute as well. We largely follow the evaluation of \cite{huExaminationRoleO142014a} that includes resonances corresponding to states in the 5.1 - 8.11~MeV range. The resonance properties are mostly based on the resonance parameter compilation of \cite{almaraz-calderonLevelStructure182012}, which in turn provides some improvements to the resonance parameters listed in the earlier compilation of \cite{hahnStructure18Mathrm1996}. The main difference in \cite{huExaminationRoleO142014a} compared to earlier work \cite{almaraz-calderonLevelStructure182012} is an improved calculation of the low energy interference using R-matrix theory, which increases the rate below 0.5~GK by an order of magnitude, and the use of the about a factor of 2 larger $\alpha$-spectroscopic factor from \cite{fortuneEnsuremathAlphaWidth2012} for the important 6.15~MeV state. \cite{huExaminationRoleO142014a} also include new experimental data for an improved proton width of the 6.3~MeV resonance. Here we use the reaction rate from \cite{huExaminationRoleO142014a} as a basis, but update with new information on resonances corresponding to higher lying states at $E_{\rm x}=$7.35~MeV, 7.58~MeV, 7.72~MeV, and 8.1~MeV from a thick target direct measurement in inverse kinematics at the CRIB facility at the University of Tokyo \cite{kimMeasurement14Mathrm2015}. These resonances affect the reaction rate at temperatures beyond 2~GK. The 7.72~MeV resonance has not been included previously, and we simply calculate its contribution and add it to the rate of \cite{huExaminationRoleO142014a}. For the $E_{\rm x}=$7.35~MeV, 7.58~MeV, and 8.1~MeV resonances, we subtract their contributions from the total rate given by \cite{huExaminationRoleO142014a}, and then add the contributions calculated with the new parameters from \cite{kimMeasurement14Mathrm2015} using the narrow resonance formalism, so as to not affect the low energy calculation of \cite{huExaminationRoleO142014a}. No SEF is used for this reaction. The evaluated reaction rates are given in Table \ref{rate:set1}.

\section*{Appendix $^{17}$O(p,$\alpha$)$^{14}$N}
For T $\leq$ 1.0 GK, the $^{17}$O(p,$\alpha$)$^{14}$N reaction rate reported by \cite{lunacollaborationImprovedDirectMeasurement2016} is adopted in the present evaluation. The rate is dominated by four resonances and mostly based on the detailed evaluation of \cite{bucknerHighintensitybeamStudy172015} but includes an updated resonance strength for the 65~keV resonance from a direct measurement at the LUNA accelerator at the Laboratori Nazionali del Gran Sasso (LNGS) of INFN, Italy underground laboratory that increases the rate at temperatures below 0.1~GK by a factor of 2. Above T = 1.0 GK we use the slope predicted by the \code{NON-SMOKER} Hauser-Feshbach model \cite{rauscherTABLESNUCLEARCROSS2001} and normalize it to the rate at 1~GK. The reaction rate is listed in Table \ref{rate:set1}.

\section*{Appendix $^{17}$F(p,$\gamma$)$^{18}$Ne}
The reaction rate at X-ray burst temperatures is dominated by a 3$^+$ resonance at 599.8~keV, for which the resonance strength has been measured directly using a $^{17}$F radioactive beam at the ORNL DRS recoil separator \cite{chippsFirstDirectMeasurement2009b}, albeit with a $\approx$100\% error bar. At low temperatures direct capture is expected to dominate. The relevant spectroscopic factors have been determined from $^{17}$F(d,n)$^{18}$Ne proton transfer reaction measurements in inverse kinematics at Florida State University \cite{kuvinMeasurement17F18Ne2017}, who extracted asymptotic normalization coefficients (ANC) of the bound states in $^{18}$Ne. We adopt the new direct capture rate and the compiled resonance parameters from \cite{kuvinMeasurement17F18Ne2017} for T $\leq$ 0.9 GK. For T $>$ 0.9 GK, the \code{NON-SMOKER} Hauser-Feshbach reaction rate \cite{rauscherTABLESNUCLEARCROSS2001}, normalized to the adopted rate at T = 0.9 GK, is used. The numerical results are listed in Table \ref{rate:set2}.

\begin{table}
\caption{Reaction rates of $^{17}$F(p,$\gamma$)$^{18}$Ne, $^{18}$F(p,$\alpha$)$^{15}$O, $^{18}$F(p,$\gamma$)$^{19}$Ne, and $^{19}$F(p,$\alpha$)$^{16}$O. \label{rate:set2}}
\scalebox{0.89}{
\begin{tabular}{ c c c c c}
\hline \hline
& $^{17}$F(p,$\gamma$)$^{18}$Ne & $^{18}$F(p,$\alpha$)$^{15}$O & $^{18}$F(p,$\gamma$)$^{19}$Ne & $^{19}$F(p,$\alpha$)$^{16}$O \\
$T$ & $N_{A}<\sigma\upsilon>$ & $N_{A}<\sigma\upsilon>$ & $N_{A}<\sigma\upsilon>$ & $N_{A}<\sigma\upsilon>$  \\
$GK$ & $cm^3/(mol*sec)$ & $cm^3/(mol*sec)$ & $cm^3/(mol*sec)$ & $cm^3/(mol*sec)$ \\
\hline
 0.008 & 0.3429E-30  &  0.1970E-24  & 0.3854E-35 &  0.9553E-25   \\
 0.010 & 0.2167E-27  &  0.2757E-21  & 0.6644E-30 &  0.2830E-22   \\
 0.015 & 0.7987E-23  &  0.9258E-16  & 0.8560E-23 &  0.3518E-18   \\
 0.020 & 0.5934E-20  &  0.8286E-13  & 0.3875E-19 &  0.1364E-15   \\
 0.030 & 0.2287E-16  &  0.8988E-10  & 0.2421E-15 &  0.2654E-12   \\
 0.040 & 0.4399E-14  &  0.3347E-08  & 0.2447E-13 &  0.3525E-10   \\
 0.050 & 0.1606E-12  &  0.3598E-07  & 0.4853E-12 &  0.1250E-08   \\
 0.060 & 0.2619E-11  &  0.2507E-06  & 0.4688E-11 &  0.1981E-07   \\
 0.070 & 0.2418E-10  &  0.1433E-05  & 0.3160E-10 &  0.1810E-06   \\
 0.080 & 0.1456E-09  &  0.6736E-05  & 0.1652E-09 &  0.1112E-05   \\
 0.090 & 0.7034E-09  &  0.2770E-04  & 0.6976E-09 &  0.5187E-05   \\
 0.100 & 0.2531E-08  &  0.1144E-03  & 0.2468E-08 &  0.2041E-04   \\
 0.120 & 0.2331E-07  &  0.1071E-02  & 0.2061E-07 &  0.2529E-03   \\
 0.140 & 0.1315E-06  &  0.5194E-02  & 0.1169E-06 &  0.2456E-02   \\
 0.160 & 0.5454E-06  &  0.1975E-01  & 0.5357E-06 &  0.1817E-01   \\
 0.180 & 0.1807E-05  &  0.6474E-01  & 0.2343E-05 &  0.1074E+00   \\
 0.200 & 0.4822E-05  &  0.1878E+00  & 0.1006E-04 &  0.5212E+00   \\
 0.250 & 0.3928E-04  &  0.1751E+01  & 0.2336E-03 &  0.1139E+02   \\
 0.300 & 0.1851E-03  &  0.1055E+02  & 0.2285E-02 &  0.9755E+02   \\
 0.350 & 0.7149E-03  &  0.4593E+02  & 0.1190E-01 &  0.4615E+03   \\
 0.400 & 0.2448E-02  &  0.1531E+03  & 0.4153E-01 &  0.1427E+04   \\
 0.450 & 0.8316E-02  &  0.3999E+03  & 0.1048E+00 &  0.3337E+04   \\
 0.500 & 0.2507E-01  &  0.8344E+03  & 0.2250E+00 &  0.6862E+04   \\
 0.550 & 0.6619E-01  &  0.1574E+04  & 0.4907E+00 &  0.1201E+05   \\
 0.600 & 0.1530E+00  &  0.2810E+04  & 0.1123E+01 &  0.1856E+05   \\
 0.650 & 0.3146E+00  &  0.4668E+04  & 0.1843E+01 &  0.2663E+05   \\
 0.700 & 0.5855E+00  &  0.7293E+04  & 0.2701E+01 &  0.3743E+05   \\
 0.750 & 0.1003E+01  &  0.1084E+05  & 0.3873E+01 &  0.4862E+05   \\
 0.800 & 0.1605E+01  &  0.1547E+05  & 0.5306E+01 &  0.6307E+05   \\
 0.850 & 0.2424E+01  &  0.2134E+05  & 0.7348E+01 &  0.7972E+05   \\
 0.900 & 0.3490E+01  &  0.2858E+05  & 0.9163E+01 &  0.9531E+05   \\
 0.950 & 0.4823E+01  &  0.3736E+05  & 0.1207E+02 &  0.1175E+06   \\
 1.000 & 0.6439E+01  &  0.4769E+05  & 0.1442E+02 &  0.1323E+06   \\
 1.200 & 0.1579E+02  &  0.1085E+06  & 0.3006E+02 &  0.2223E+06   \\
 1.400 & 0.2929E+02  &  0.2041E+06  & 0.5129E+02 &  0.3690E+06   \\
 1.600 & 0.4576E+02  &  0.3389E+06  & 0.7773E+02 &  0.6420E+06   \\
 1.800 & 0.6394E+02  &  0.5151E+06  & 0.1083E+03 &  0.1072E+07   \\
 2.000 & 0.8277E+02  &  0.7333E+06  & 0.1420E+03 &  0.1819E+07   \\
 2.500 & 0.1287E+03  &  0.1454E+07  & 0.2355E+03 &  0.4160E+07   \\
 3.000 & 0.1703E+03  &  0.2392E+07  & 0.3364E+03 &  0.7429E+07   \\
 4.000 & 0.2454E+03  &  0.4695E+07  & 0.5456E+03 &  0.1777E+08   \\
 5.000 & 0.3253E+03  &  0.7198E+07  & 0.7319E+03 &  0.3141E+08   \\
 6.000 & 0.4061E+03  &  0.9565E+07  & 0.9394E+03 &  0.4877E+08   \\
 7.000 & 0.5066E+03  &  0.1160E+08  & 0.1166E+04 &  0.6580E+08   \\
 8.000 & 0.6290E+03  &  0.1324E+08  & 0.1372E+04 &  0.8376E+08   \\
 9.000 & 0.7809E+03  &  0.1446E+08  & 0.1582E+04 &  0.9960E+08   \\
10.000 & 1.0075E+03  &  0.1530E+08  & 0.1802E+04 &  0.1132E+09   \\
\hline
\end{tabular}
}
\end{table}

\section*{Appendix $^{18}$F(p,$\alpha$)$^{15}$O}
R-matrix calculations of the cross section are based on a mix of relatively sparse and uncertain direct cross section measurements, and indirect studies of various excited states around the proton threshold in $^{19}$Ne that provide resonant contributions. Possible unknown interference between resonances as well as incomplete level information results in significant uncertainties (see the most recent analysis of \cite{hall19NeLevelStructure2020}). Since the evaluation of \cite{iliadisChargedparticleThermonuclearReaction2010b} previously recommended in JINA REACLIB, a number of indirect studies provided additional information on excited states: $^{18}$F(d,n) measurements at ORNL identified a subthreshold resonance \cite{adekolaFirstProtontransferStudy2011a}, for which spin and parity 1/2$^+$ are later determined using the $^{20}$Ne(p,d)$^{19}$Ne reaction with the JENSA gas jet target at ORNL \cite{bardayanFirstScienceResult2015a, bardayanSpectroscopicStudy20Ne2017}. In addition several $^{19}$F($^3$He,t)$^{19}$Ne measurements at Garching and Argonne National Laboratory provided additional level information \cite{lairdGRayEmissionNovae2013, parikhSpectroscopy19NeThermonuclear2015}, including improved constraints on low lying 3/2$^+$ resonances \cite{hallKey19NeStates2019,hall19NeLevelStructure2020}. Cross section data were also obtained using the Trojan horse method via the 
three-body reaction $^{2}$H($^{18}$F,$\alpha$$^{15}$O)n \cite{cherubiniFirstApplicationTrojan2015, lacognataTrojanHorseApproach2017}. R-matrix calculations including the Trojan horse data \cite{lacognataTrojanHorseApproach2017} found results consistent with the R-matrix analysis of \cite{bardayanSpectroscopicStudy20Ne2017}. The latest analysis \cite{hall19NeLevelStructure2020} differs from \cite{bardayanSpectroscopicStudy20Ne2017} mainly in a greatly reduced uncertainty of the reaction rate below 0.3~GK, and provides an upper and a lower reaction rate based on different resonance interference assumptions. We use here an average rate. In the 0.1-0.4~GK temperature range, this rate is up to 60\% larger than the previously recommended rate of \cite{iliadisChargedparticleThermonuclearReaction2010b}, while for higher temperature the rates agree and we continue to to use \cite{iliadisChargedparticleThermonuclearReaction2010b} for temperatures up to 10~GK. The numerical results of $^{18}$F(p,$\alpha$)$^{15}$O reaction rate are listed in Table \ref{rate:set2}.

\section*{Appendix $^{18}$F(p,$\gamma$)$^{19}$Ne}
Above 0.2~GK the rate is thought to be dominated by a 330~keV resonance \cite{iliadisChargedparticleThermonuclearReaction2010b}. \cite{akersMeasurementRadiativeProton2016} performed a direct measurement of the resonance strength of 
the 665~keV resonance, potentially contributing to the rate below 0.2~GK, using the DRAGON recoil separator at TRIUMF and demonstrated, albeit with only two counts, that the resonance strength is significantly lower than previous upper limits making its contribution to the rate negligible. We adopt the analysis of \cite{akersMeasurementRadiativeProton2016}, who for the other contributions to the low temperature rate employed the resonance parameters of \cite{bardayanFirstScienceResult2015a, bardayanSpectroscopicStudy20Ne2017}, creating consistency with our recommended $^{18}$F(p,$\alpha$) reaction rate. The new rate is significantly lower below 0.2~GK compared to the previously recommended Monte Carlo rate from \cite{iliadisChargedparticleThermonuclearReaction2010b} due to the lower contribution of the 665~keV resonance that was previously treated using an upper limit. For example, at 0.15~GK the new rate is a factor of 4 lower than the lower limit provided by \cite{iliadisChargedparticleThermonuclearReaction2010b}.
For T $>$ 0.5 GK, the reaction rate is dominated by the resonance at $E_{\rm cm}$ = 330 keV \cite{longlandChargedparticleThermonuclearReaction2010}.The final $^{18}$F(p,$\gamma$)$^{19}$Ne reaction rates are obtained from combining the rates in these two temperature ranges, and are listed in Table \ref{rate:set2}.

\section*{Appendix $^{19}$F(p,$\alpha$)$^{16}$O}

For the $^{19}$F(p,$\alpha$)$^{16}$O reaction, transitions to $^{16}$O in the ground state (p,$\alpha_{0}$), the excited state at E$_x$ = 6.05 MeV (pair emitting, p,$\alpha_{\pi}$), and the sum of transitions to all other $\gamma$-emitting excited states (p,$\alpha_{\gamma}$) are considered separately.

The reaction rates of $^{19}$F(p,$\alpha_{0}$)$^{16}$O plus $^{19}$F(p,$\alpha_{\pi}$)$^{16}$O are taken from~\cite{lombardoNewAnalysis192019}, in which a comprehensive R-matrix analysis of the two channels is performed based on available experimental data. For $^{19}$F(p,$\alpha_{\gamma}$)$^{16}$O, the reaction rate obtained from a recent direct low energy cross section measurement at the underground laboratory JUNA \cite{zhangDirectMeasurementAstrophysical2021,zhangDirectMeasurementAstrophysical2022a} that extended cross section data to significantly lower energies is used for T $\leq$ 1.0 GK. For T $>$ 1.0 GK we adopt the NACRE reaction rate. The reaction rates of $^{19}$F(p,$\alpha_{total}$)$^{16}$O are listed in Table \ref{rate:set2}.

\section*{Appendix $^{19}$F($\alpha$,p)$^{22}$Ne}
\cite{PhysRevC.77.035801} performed a comprehensive analysis of the $^{19}$F($\alpha$,p)$^{22}$Ne reaction rate using their own direct cross section measurements performed at the Nuclear Science Laboratory of the University of Notre Dame, R-matrix analysis, and, for higher energies, Hauser-Feshbach calculations. \cite{Pizzone_2017,DAgata_2018} carried out additional indirect measurements at the the Ruer Bošković Institute (Zagreb, Croatia) using the Trojan Horse method via the $^{6}$Li($^{19}$F,p$^{22}$Ne)d reaction to obtain data in the energy range below 660~keV that has not been reached by direct measurements. As a result they find an increase of the astrophysical reaction rate below 0.6~GK by up to a factor of 3.91 compared to \cite{PhysRevC.77.035801}. We adopted their enhancement and otherwise the reaction rate of \cite{PhysRevC.77.035801}. The reaction rates of $^{19}$F($\alpha$,p)$^{22}$Ne are listed in Table \ref{rate:set3}.

\begin{table}
\caption{Reaction rates of $^{19}$F($\alpha$,p)$^{22}$Ne, $^{19}$Ne(p,$\gamma$)$^{20}$Na, $^{23}$Na(p,$\gamma$)$^{24}$Mg, $^{23}$Na($\alpha$,p)$^{26}$Mg. \label{rate:set3}}
\scalebox{0.89}{
\begin{tabular}{c c c c c}
\hline \hline
& $^{19}$F($\alpha$,p)$^{22}$Ne & $^{19}$Ne(p,$\gamma$)$^{20}$Na & $^{23}$Na(p,$\gamma$)$^{24}$Mg & $^{23}$Na($\alpha$,p)$^{26}$Mg  \\
$T$ & $N_{A}<\sigma\upsilon>$ & $N_{A}<\sigma\upsilon>$ & $N_{A}<\sigma\upsilon>$ & $N_{A}<\sigma\upsilon>$ \\
$GK$ & $cm^3/(mol*sec)$ & $cm^3/(mol*sec)$ & $cm^3/(mol*sec)$ & $cm^3/(mol*sec)$ \\
\hline
0.010 &   $-$           &  0.3125E-30   & 0.8663E-32 &  $-$          \\
0.015 &   $-$           &  0.2084E-25   & 0.1325E-26 &  $-$          \\
0.020 &   $-$           &  0.2281E-22   & 0.2468E-23 &  $-$          \\
0.030 &   $-$           &  0.1458E-18   & 0.3124E-19 &  $-$          \\
0.040 &   $-$           &  0.3626E-16   & 0.1277E-16 &  $-$          \\
0.050 &   $-$           &  0.1824E-14   & 0.1836E-14 &  $-$          \\
0.060 &   0.4954E-34    &  0.3605E-13   & 0.1505E-12 &  $-$          \\
0.070 &   0.1801E-28    &  0.3897E-12   & 0.4325E-11 &  0.3302E-38   \\
0.080 &   0.4142E-25    &  0.2770E-11   & 0.5609E-10 &  0.3496E-34   \\
0.090 &   0.7545E-23    &  0.1451E-10   & 0.4954E-09 &  0.4458E-31   \\
0.100 &   0.3492E-21    &  0.6035E-10   & 0.4538E-08 &  0.1342E-28   \\
0.120 &   0.9664E-19    &  0.6302E-09   & 0.3485E-06 &  0.7318E-25   \\
0.140 &   0.6696E-17    &  0.4094E-08   & 0.1152E-04 &  0.3812E-22   \\
0.160 &   0.2194E-15    &  0.1948E-07   & 0.1719E-03 &  0.4725E-20   \\
0.180 &   0.4309E-14    &  0.8322E-07   & 0.1419E-02 &  0.2275E-18   \\
0.200 &   0.5556E-13    &  0.4278E-06   & 0.7628E-02 &  0.5628E-17   \\
0.250 &   0.9690E-11    &  0.3248E-04   & 0.1526E+00 &  0.2579E-14   \\
0.300 &   0.4710E-09    &  0.7673E-03   & 0.1089E+01 &  0.2465E-12   \\
0.350 &   0.9302E-08    &  0.7811E-02   & 0.4337E+01 &  0.9389E-11   \\
0.400 &   0.1049E-06    &  0.4146E-01   & 0.1206E+02 &  0.1774E-09   \\
0.450 &   0.8098E-06    &  0.1548E+00   & 0.2646E+02 &  0.2121E-08   \\
0.500 &   0.4758E-05    &  0.4328E+00   & 0.4927E+02 &  0.1868E-07   \\
0.550 &   0.2303E-04    &  0.1003E+01   & 0.8156E+02 &  0.1175E-06   \\
0.600 &   0.1021E-03    &  0.2015E+01   & 0.1237E+03 &  0.6457E-06   \\
0.650 &   0.4058E-03    &  0.3631E+01   & 0.1757E+03 &  0.2754E-05   \\
0.700 &   0.1414E-02    &  0.6017E+01   & 0.2369E+03 &  0.1091E-04   \\
0.750 &   0.4331E-02    &  0.9321E+01   & 0.3068E+03 &  0.3604E-04   \\
0.800 &   0.1188E-01    &  0.1402E+02   & 0.3845E+03 &  0.1105E-03   \\
0.850 &   0.2995E-01    &  0.1914E+02   & 0.4693E+03 &  0.3098E-03   \\
0.900 &   0.7237E-01    &  0.2644E+02   & 0.5605E+03 &  0.8024E-03   \\
0.950 &   0.1749E+00    &  0.3363E+02   & 0.6574E+03 &  0.1939E-02   \\
1.000 &   0.4168E+00    &  0.4354E+02   & 0.7594E+03 &  0.4407E-02   \\
1.200 &   0.4075E+01    &  0.8889E+02   & 0.1211E+04 &  0.7162E-01   \\
1.400 &   0.1773E+02    &  0.1460E+03   & 0.1725E+04 &  0.6435E+00   \\
1.600 &   0.7236E+02    &  0.2063E+03   & 0.2304E+04 &  0.3813E+01   \\
1.800 &   0.2299E+03    &  0.2638E+03   & 0.2964E+04 &  0.1665E+02   \\
2.000 &   0.6047E+03    &  0.3212E+03   & 0.3719E+04 &  0.5568E+02   \\
2.500 &   0.3865E+04    &  0.4110E+03   & 0.6063E+04 &  0.6368E+03   \\
3.000 &   0.1474E+05    &  0.4675E+03   & 0.8807E+04 &  0.3620E+04   \\
4.000 &   0.9179E+05    &  0.5239E+03   & 0.1318E+05 &  0.3997E+05   \\
5.000 &   0.3081E+06    &  0.5725E+03   & 0.1574E+05 &  0.1894E+06   \\
6.000 &   0.7370E+06    &  0.6518E+03   & 0.1854E+05 &  0.5706E+06   \\
7.000 &   0.1430E+07    &  0.7776E+03   & 0.2099E+05 &  0.1300E+07   \\
8.000 &   0.2413E+07    &  0.9579E+03   & 0.2195E+05 &  0.2464E+07   \\
9.000 &   0.3682E+07    &  0.1198E+04   & 0.2106E+05 &  0.4122E+07   \\
10.000&   0.5214E+07    &  0.1501E+04   & 0.1857E+05 &  0.6285E+07   \\
\hline
\end{tabular}
}
\end{table}




\section*{Appendix $^{19}$Ne(p,$\gamma$)$^{20}$Na}
Two reaction channels, capture on the ground state $^{19}$Ne$_{g.s.}$(p,$\gamma$)$^{20}$Na and capture on the first excited 0.238~MeV target state $^{19}$Ne$_{e.x.}$(p,$\gamma$)$^{20}$Na are considered. Capture on the third excited $^{19}$Ne state at 0.275~MeV is negligible. 

\begin{figure}[tbp]
\includegraphics[width=0.64\textwidth]{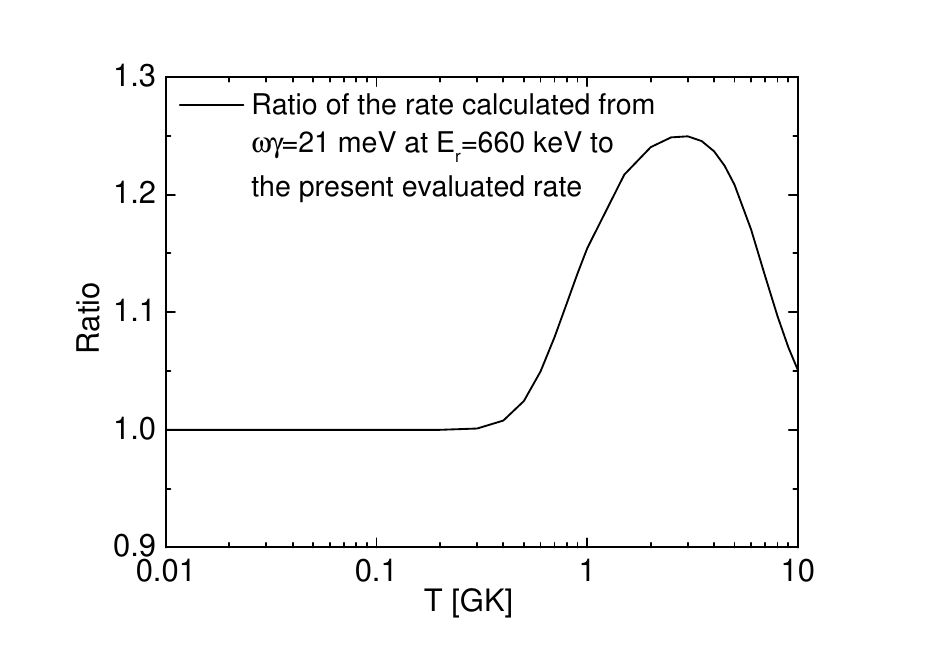}
\caption{Comparison of the $^{19}$Ne(p,$\gamma$)$^{20}$Na reaction rate calculated from 1$^+$ assignment ($\omega$$\gamma$ = 21~meV) for $E_r$ = 660~keV \cite{belargeExperimentalInvestigation192016} to the recommended one.\label{fig1:Ne19pg}}
\end{figure}

For $^{19}$Ne$_{g.s.}$(p,$\gamma$)$^{20}$Na, four resonances at $E_r$ = 0.456,
0.660, 0.797, and 0.887 MeV, as well as the direct capture contribution, determine the reaction rate.
For $E_r$ = 0.456 MeV we adopt the resonance strength from the direct measurement with the DRAGON recoil separator at TRIUMF \cite{wilkinsonDirectMeasurementKey2017}. The spin of the $E_r$ = 0.661 MeV resonance is not certain. ENSDF currently assigns 3$^+$ to the corresponding $E_x=$~2.849~MeV state based on (p,n) and two ($^3$He,t) measurements. This is in line with earlier reaction rate evaluations \cite{lammLevelStructure20Na1990,vancraeynest19MathrmNe1998}. However, a more recent (d,n) experiment seems to indicate a preference for a 1$^+$ assignment \cite{belargeExperimentalInvestigation192016}. More experimental data would be desirable to clarify the nature of this state. We adopt here the 3$^+$ assignment and the resonance strength of \cite{vancraeynest19MathrmNe1998}. Using the 1$^+$ assignment and the larger resonance strength of \cite{belargeExperimentalInvestigation192016} (21~meV) has only a modest impact, resulting in a reaction rate that is identical up to around 1~GK and is then lower by up to 30\% for higher temperatures (see Figure \ref{fig1:Ne19pg}). For the higher lying resonances as well as for direct capture we use the 
parameters from \cite{vancraeynest19MathrmNe1998}. 

For $^{19}$Ne$_{e.x.}$(p,$\gamma$)$^{20}$Na only the resonances at $E_r$ = 0.218 and 0.422~MeV, corresponding to the same states as the ground state resonances at $E_r$ = 0.456 and 0.660~MeV, respectively, contribute significantly. For $E_r$ = 0.218 MeV we adopt the resonance strength of \cite{belargeExperimentalInvestigation192016}, who derived it from their measurement of the proton branchings via a (d,n) transfer reaction at the RESOLUT facility at Florida State University in combination with the direct resonance strength measurement for the ground state rate from  \cite{wilkinsonDirectMeasurementKey2017}. For the $E_r$ = 0.218 MeV resonance we follow \cite{vancraeynest19MathrmNe1998}.
The total reaction rate of $^{19}$Ne(p,$\gamma$)$^{20}$Na is listed in Table \ref{rate:set3}.

\section*{Appendix $^{23}$Na($\alpha$,p)$^{26}$Mg}
A recent active target direct measurement in inverse kinematics using the MUSIC detector at the ATLAS facility at Argonne National Laboratory \cite{avilaExperimentalStudyAstrophysically2016} resolved discrepancies among previous measurements and confirmed the excellent agreement of experimental cross section data with statistical model calculations using \code{TALYS} with the McFadden/Satchler $\alpha$-nucleus potential \cite{mohrCrossSectionsAinduced2015a}. The experiment measured the angle and final excitation energy integrated cross section.  We calculated the $^{23}$Na($\alpha$,p)$^{26}$Mg reaction rate from the statistical model cross section displayed in \cite{avilaExperimentalStudyAstrophysically2016} and apply a SEF. The results are listed in Table \ref{rate:set3}.

\section*{Appendix $^{23}$Na(p,$\gamma$)$^{24}$Mg}
The rate below 1~GK is dominated by direct capture and three resonances. For the three resonances we adopt the updated strengths from \cite{boeltzigDirectMeasurementsLowenergy2019}, who performed direct measurements at the LUNA underground accelerator. They reported the first direct measurement of the resonance at $E_r^{CM}$=135.4~keV (their $E_r^{Lab}$=140~keV)  (\cite{cesarattoMeasurement138KeV2013} reported their tentative detection at the LENA facility as an upper limit), a more accurate resonance strength for the  $E_r^{CM}$=242.2~keV resonance (their $E_r^{Lab}$=251~keV) and a confirmation of the previously measured resonance strengths of the   $E_r^{CM}$=298.1~keV resonance (their $E_r^{Lab}$=309~keV). For the first resonance, we adopt the updated resonance energy $E_r^{CM}$=133(3)~keV from a measurement using the ($^3$He,d) transfer reaction at the Enge split-pole spectrograph at TUNL \cite{marshallNewEnergy133keV2021}, which agrees within uncertainties with the energy reported in  \cite{boeltzigDirectMeasurementsLowenergy2019} but is 5~keV lower than what was adopted in some earlier compilations (e.g. \cite{cesarattoMeasurement138KeV2013}) based on older measurements (see the overview in  \cite{boeltzigDirectMeasurementsLowenergy2019}). For higher lying resonances that start playing a significant role above 1~GK we adopt the properties in \cite{iliadisChargedparticleThermonuclearReaction2010b}. For direct capture we use the updated S-factors from \cite{boeltzigInvestigationDirectCapture2022} that are based on direct measurements at the University of Notre Dame and an R-matrix analysis. While we included significant updates for direct capture and the lowest lying relevant resonances affecting the low temperature behavior of the reaction rate, for temperatures relevant for X-ray bursts ($T>0.2$~GK) our rate is identical to the one tabulated in \cite{cesarattoMeasurement138KeV2013}. The reaction rate of $^{23}$Na(p,$\gamma$)$^{24}$Mg is listed in Table \ref{rate:set3}.

\section*{Appendix $^{24}$Mg($\alpha$,$\gamma$)$^{28}$Si}
For $T \leq$ 1.25 GK, we follow the updated resonance analysis of \cite{adsleyStatus24Mg28Si2020}, who re-evaluate experimental data from various direct measurements with data from inelastic scattering of alphas, protons, and electrons on $^{28}$Si performed at iThemba LABS in South Africa. In the relevant temperature range for X-ray bursts ($T > 0.5$~GK) the new rate is largely identical to the \starlib  \,reaction rate, though there are some deviations at lower temperatures that may impact other applications. Above $T = 1.25$~GK, the \code{STARLIB} reaction rate is used. The numerical results are given in Table \ref{rate:set4}.

\begin{table}
\caption{Reaction rates of $^{24}$Mg($\alpha$,$\gamma$)$^{28}$Si, $^{25}$Mg(p,$\gamma$)$^{26}$Al$_{g.s.}$, and $^{25}$Mg(p,$\gamma$)$^{26}$Al$_{isomer}$. \label{rate:set4}}
\scalebox{0.89}{
\begin{tabular}{c c c c}
\hline \hline
& $^{24}$Mg($\alpha$,$\gamma$)$^{28}$Si & $^{25}$Mg(p,$\gamma$)$^{26}$Al$_{g.s.}$ & $^{25}$Mg(p,$\gamma$)$^{26}$Al$_{isomer}$ \\
$T$ & $N_{A}<\sigma\upsilon>$ & $N_{A}<\sigma\upsilon>$ & $N_{A}<\sigma\upsilon>$ \\
$GK$ & $cm^3/(mol*sec)$ & $cm^3/(mol*sec)$ & $cm^3/(mol*sec)$ \\
\hline
0.010  &  $-$          &   0.1339E-31   &   0.3551E-32   \\
0.015  &  $-$          &   0.8694E-24   &   0.2043E-24   \\
0.020  &  $-$          &   0.3914E-19   &   0.9181E-20   \\
0.030  &  $-$          &   0.1497E-14   &   0.3521E-15   \\
0.040  &  $-$          &   0.2635E-12   &   0.6590E-13   \\
0.050  &  $-$          &   0.6138E-11   &   0.2057E-11   \\
0.060  &  $-$          &   0.6129E-10   &   0.3100E-10   \\
0.070  &  $-$          &   0.3900E-09   &   0.2600E-09   \\
0.080  &  $-$          &   0.1751E-08   &   0.1343E-08   \\
0.090  &  $-$          &   0.5941E-08   &   0.4854E-08   \\
0.100  &  0.8710E-37   &   0.1652E-07   &   0.1363E-07   \\
0.120  &  0.1426E-29   &   0.1116E-06   &   0.7096E-07   \\
0.140  &  0.1182E-25   &   0.1408E-05   &   0.3964E-06   \\
0.160  &  0.3599E-23   &   0.1992E-04   &   0.3469E-05   \\
0.180  &  0.2559E-21   &   0.1818E-03   &   0.2820E-04   \\
0.200  &  0.8067E-20   &   0.1090E-02   &   0.1666E-03   \\
0.250  &  0.6900E-17   &   0.2737E-01   &   0.4399E-02   \\
0.300  &  0.1430E-14   &   0.2330E+00   &   0.4022E-01   \\
0.350  &  0.1465E-12   &   0.1078E+01   &   0.1995E+00   \\
0.400  &  0.6832E-11   &   0.3427E+01   &   0.6761E+00   \\
0.450  &  0.1566E-09   &   0.8498E+01   &   0.1778E+01   \\
0.500  &  0.2137E-08   &   0.1771E+02   &   0.3916E+01   \\
0.550  &  0.1967E-07   &   0.3250E+02   &   0.7569E+01   \\
0.600  &  0.1293E-06   &   0.5418E+02   &   0.1325E+02   \\
0.650  &  0.6512E-06   &   0.8386E+02   &   0.2147E+02   \\
0.700  &  0.2653E-05   &   0.1224E+03   &   0.3271E+02   \\
0.750  &  0.9128E-05   &   0.1703E+03   &   0.4740E+02   \\
0.800  &  0.2787E-04   &   0.2279E+03   &   0.6589E+02   \\
0.850  &  0.7317E-04   &   0.2953E+03   &   0.8848E+02   \\
0.900  &  0.1773E-03   &   0.3726E+03   &   0.1154E+03   \\
0.950  &  0.3934E-03   &   0.4594E+03   &   0.1468E+03   \\
1.000  &  0.8072E-03   &   0.5556E+03   &   0.1828E+03   \\
1.200  &  0.7847E-02   &   0.1026E+04   &   0.3725E+03   \\
1.400  &  0.3899E-01   &   0.1614E+04   &   0.6314E+03   \\
1.600  &  0.1467E+00   &   0.2292E+04   &   0.9493E+03   \\
1.800  &  0.4009E+00   &   0.3039E+04   &   0.1314E+04   \\
2.000  &  0.8566E+00   &   0.3834E+04   &   0.1713E+04   \\
2.500  &  0.3754E+01   &   0.5942E+04   &   0.2788E+04   \\
3.000  &  0.1071E+02   &   0.8090E+04   &   0.3871E+04   \\
4.000  &  0.4324E+02   &   0.1210E+05   &   0.5772E+04   \\
5.000  &  0.1031E+03   &   0.1538E+05   &   0.7163E+04   \\
6.000  &  0.1778E+03   &   0.1780E+05   &   0.8065E+04   \\
7.000  &  0.2552E+03   &   0.1938E+05   &   0.8577E+04   \\
8.000  &  0.3289E+03   &   0.2023E+05   &   0.8805E+04   \\
9.000  &  0.4059E+03   &   0.2049E+05   &   0.8837E+04   \\
10.000 &  0.5007E+03   &   0.2029E+05   &   0.8738E+04   \\
\hline
\end{tabular}
}
\end{table}

\section*{Appendix $^{25}$Mg(p,$\gamma$)$^{26}$Al}
$^{26}$Al has a $T_{1/2}$=717,000 yr ground state, and a $T_{1/2}$=6.346~s excited isomeric state. The ground and isomeric state are thought to be in thermal equilibrium for $T>0.4$~GK \cite{mischAstromersNuclearIsomers2020a} and our X-ray burst model therefore includes only a fully thermalized $^{26}$Al. However, for other applications it is critical to track the ground and isomeric state separately. 
\code{JINA REACLIB} therefore includes separate $^{25}$Mg(p,$\gamma$)$^{26}$Al$_{\rm g.s}$ and $^{25}$Mg(p,$\gamma$)$^{26}$Al$_{\rm isomer}$ rates that are added to obtain the total $^{25}$Mg(p,$\gamma$)$^{26}$Al rate feeding a thermalized $^{26}$Al. The rates are calculated using the strengths of individual narrow resonances, and the factor $f_0$ denoting the feeding fraction to the $^{26}$Al ground state. At X-ray burst temperatures the resonances at 304~keV and above dominate, while for other astrophysical sites lower lying resonances are important. We largely use the resonance parameters from \starlib \,with the following updates: for the 92.2~keV, 189.5~keV and 304~keV resonances we use new resonance strengths from direct measurements at the LUNA underground accelerator \cite{strieder25Mg26AlReaction2012, limataNewExperimentalStudy2010} following the evaluation of \cite{stranieroImpactRevised25Mg2013}. 
We also use the corresponding updated $f_{0}$ values, except for the 92.2~keV resonance, where we adopt the more precise value obtained from a recent $\gamma$-spectroscopy experiment at Argonne National Laboratory using a fusion evaporation reaction with the Gammasphere detector array \cite{kankainenDecayKey92keV2021a}. For the 57.7 keV resonance, a recent indirect measurement using the $^{25}$Mg($^{7}$Li,$^6$He)$^{26}$Al transfer reaction at the HI-13 tandem accelerator in Beijing \cite{liIndirectMeasurement572020} claimed a more precise resonance strength of \EEerr{2.93}{0.50}{-13}. 
The central value is almost identical with the \code{STARLIB} adopted value of \EEerr{2.8}{1.1}{-13}, nevertheless we adopt the value from \cite{liIndirectMeasurement572020}. 
The SEF is considered for both channels. The evaluated reaction rates of $^{25}$Mg(p,$\gamma$)$^{26}Al_{g.s}$ and $^{25}$Mg(p,$\gamma$)$^{26}Al_{\rm isomer}$ are presented in Table \ref{rate:set4}.

\section*{Appendix $^{23}$Al(p,$\gamma$)$^{24}$Si}
The reaction rate at X-ray burst temperatures is dominated by a single resonance corresponding to the second 2$^+$ state in $^{24}$Si. \cite{PhysRevLett.79.3845} determined an excitation energy of $E_{\rm x}=$3441(10)~keV using the $^{28}$Si($^{4}$He,$^{8}$He) transfer reaction. This places the resonance at a relatively low energy that is just high enough to dominate the rate, resulting in an unusually strong resonance energy dependence of the reaction rate due to the steep energy dependence of the proton penetrability. As a result the relatively modest remaining uncertainties in excitation energy (10~keV) and $^{24}$Si mass (24~keV at the time) resulted in almost 3 orders of magnitude of reaction rate uncertainty. A subsequent $^{23}$Al(d,n) neutron transfer measurement carried out in inverse kinematics at NSCL at Michigan State University provided an improved  $E_{\rm x}=$3449(5)~keV \cite{PhysRevLett.122.232701} in agreement with \cite{PhysRevLett.79.3845}, but in disagreement with a previous $E_{\rm x}$ measurement from 2n knockout \cite{PhysRevC.74.021303}. This measurement also provided evidence for an additional resonance corresponding to the third excited state in $^{24}$Si at $E_{\rm x}=$3471(6)~keV, either the second 0$^+$ or first 4$^+$ excited state of $^{24}$Si, as well as spectroscopic factors for both resonances. A more recent 2n knockout measurement \cite{PhysRevC.101.031303} performed at NSCL again confirmed the $E_{\rm x}$ from \cite{PhysRevLett.79.3845} and \cite{PhysRevLett.122.232701}. This experiment also resolved the spin assignment ambiguity of the $E_{\rm x}=$3449(5)~keV state noting that its non-observation contradicts the large expected 2n knockout cross section of the 4$^+$ state, thus assigning 0$^+$. This implies an unusually strong Thomas Ehrmann shift of more than 1~MeV from the known $E_{\rm x}$=4766.6~keV in the $^{24}$Ne mirror. A recent Penning Trap mass measurement performed at LEBIT at Michigan State University of $^{24}$Si \cite{PhysRevC.106.L012801} drastically reduced the Q-value uncertainty with $Q=$3283.2(37) keV. For the resonant rate contribution we follow the analysis of \cite{PhysRevC.106.L012801} Table I, who adopt for the two lowest lying resonances at $E_{x}$ = 3471 keV ($J^{\pi}$ = $0^{+}$) and 3449 keV ($J^{\pi}$ = $2^{+}$) the most precise excitation energy measurements and the spectroscopic factors of \cite{PhysRevLett.122.232701}, together with the spin assignment from  \cite{PhysRevC.101.031303}. For higher lying states shell model calculations are used \cite{PhysRevLett.122.232701}.  The non-resonant contribution is taken from Ref. \cite{PhysRevC.86.015806} obtained from proton breakup measurements at GANIL. The numerical reaction rates are given in Table \ref{rate:set5}.

\begin{table}
\caption{Reaction rates of $^{23}$Al(p,$\gamma$)$^{24}$Si, $^{24}$Al(p,$\gamma$)$^{25}$Si, $^{25}$Al(p,$\gamma$)$^{26}$Si. \label{rate:set5}}
\scalebox{0.89}{
\begin{tabular}{c c c c}
\hline \hline
& $^{23}$Al(p,$\gamma$)$^{24}$Si & $^{24}$Al(p,$\gamma$)$^{25}$Si & $^{25}$Al(p,$\gamma$)$^{26}$Si \\
$T$ & $N_{A}<\sigma\upsilon>$ & $N_{A}<\sigma\upsilon>$ & $N_{A}<\sigma\upsilon>$ \\
$GK$ & $cm^3/(mol*sec)$ & $cm^3/(mol*sec)$ & $cm^3/(mol*sec)$ \\
\hline
 0.020  &  0.3013E-28  &   $-$         &  0.1350E-39   \\
 0.030  &  0.1385E-23  &   0.5846E-37  &  0.4139E-27   \\
 0.040  &  0.9213E-18  &   0.9200E-28  &  0.7098E-21   \\
 0.050  &  0.1004E-13  &   0.6322E-22  &  0.3858E-17   \\
 0.060  &  0.4701E-11  &   0.7035E-18  &  0.1183E-14   \\
 0.070  &  0.3668E-09  &   0.5259E-15  &  0.7022E-13   \\
 0.080  &  0.9380E-08  &   0.8647E-13  &  0.1495E-11   \\
 0.090  &  0.1144E-06  &   0.5044E-11  &  0.1608E-10   \\
 0.100  &  0.8327E-06  &   0.1301E-09  &  0.1073E-09   \\
 0.120  &  0.1579E-04  &   0.1946E-07  &  0.1838E-08   \\
 0.140  &  0.1249E-03  &   0.6926E-06  &  0.1405E-07   \\
 0.160  &  0.5738E-03  &   0.1003E-04  &  0.7268E-07   \\
 0.180  &  0.1842E-02  &   0.7957E-04  &  0.4282E-06   \\
 0.200  &  0.4607E-02  &   0.4076E-03  &  0.3263E-05   \\
 0.250  &  0.2287E-01  &   0.7423E-02  &  0.2393E-03   \\
 0.300  &  0.6338E-01  &   0.4958E-01  &  0.4402E-02   \\
 0.350  &  0.1267E+00  &   0.1854E+00  &  0.3419E-01   \\
 0.400  &  0.2076E+00  &   0.4901E+00  &  0.1550E+00   \\
 0.450  &  0.2986E+00  &   0.1006E+01  &  0.4920E+00   \\
 0.500  &  0.3928E+00  &   0.1839E+01  &  0.1219E+01   \\
 0.550  &  0.4850E+00  &   0.2794E+01  &  0.2528E+01   \\
 0.600  &  0.5717E+00  &   0.4404E+01  &  0.4589E+01   \\
 0.650  &  0.6509E+00  &   0.5749E+01  &  0.7529E+01   \\
 0.700  &  0.7218E+00  &   0.8241E+01  &  0.1141E+02   \\
 0.750  &  0.7841E+00  &   0.1012E+02  &  0.1626E+02   \\
 0.800  &  0.8386E+00  &   0.1334E+02  &  0.2201E+02   \\
 0.850  &  0.8860E+00  &   0.1605E+02  &  0.2861E+02   \\
 0.900  &  0.9277E+00  &   0.1959E+02  &  0.3594E+02   \\
 0.950  &  0.9652E+00  &   0.2350E+02  &  0.4388E+02   \\
 1.000  &  0.1000E+01  &   0.2704E+02  &  0.5232E+02   \\
 1.200  &  0.1151E+01  &   0.4765E+02  &  0.8865E+02   \\
 1.400  &  0.1403E+01  &   0.7107E+02  &  0.1247E+03   \\
 1.600  &  0.1845E+01  &   0.9675E+02  &  0.1570E+03   \\
 1.800  &  0.2524E+01  &   0.1238E+03  &  0.1841E+03   \\
 2.000  &  0.3452E+01  &   0.1441E+03  &  0.2060E+03   \\
 2.500  &  0.6749E+01  &   0.2161E+03  &  0.2432E+03   \\
 3.000  &  0.1118E+02  &   0.2908E+03  &  0.2667E+03   \\
 4.000  &  0.2310E+02  &   0.4265E+03  &  0.3185E+03   \\
 5.000  &  0.4043E+02  &   0.5516E+03  &  0.4191E+03   \\
 6.000  &  0.6549E+02  &   0.6751E+03  &  0.5915E+03   \\
 7.000  &  0.1002E+03  &   0.7752E+03  &  0.8427E+03   \\
 8.000  &  0.1458E+03  &   0.8586E+03  &  0.1173E+04   \\
 9.000  &  0.2029E+03  &   0.9160E+03  &  0.1580E+04   \\
10.000  &  0.2715E+03  &   0.9741E+03  &  0.2060E+04   \\
\hline
\end{tabular}
}
\end{table}

\section*{Appendix $^{24}$Al(p,$\gamma$)$^{25}$Si}
\cite{PhysRevC.97.054307} performed in-beam $\gamma$-spectroscopy of $^{25}$Si via a 1n knockout reaction at NSCL at Michigan State University and identified 2 states above the proton threshold, a 9/2$^+$ state at 3695(14)~keV and a 1/2$^+$ state at 3802(11)~keV. This result confirmed older previous conference proceedings reports of a possible resonance at 3.7~MeV identified in  ($^{3}$He,$^{6}$He) transfer reaction studies \cite{KUBONO1997195}. We adopt the resonance parameters in \cite{PhysRevC.97.054307} obtained with a Q-value of 3.414~MeV, who complement their excitation energy measurements with shell model calculations for resonance widths and additional excited states. The reaction rate is found to be dominated by the 9/2$^+$ resonance. The numerical reaction rates are given in Table \ref{rate:set5}.

\section*{Appendix $^{25}$Al(p,$\gamma$)$^{26}$Si}
We follow largely \cite{liangSimultaneousMeasurementDelayed2020} in considering 4 resonances to contribute to the $^{25}$Al(p,$\gamma$)$^{26}$Si rate. The rate for X-ray burst conditions is largely dominated by a single 3$^+$ resonance, for which we determine a resonance energy of 415.4(8)~keV from the ENSDF recommended $^{26}$Si excitation energy \cite{BASUNIA20161} and the AME20 mass table \cite{wangAME2020Atomic2021}. This resonance energy is in agreement with but more precise than the directly determined resonance energy from the detected proton energy in $^{26}$P $\beta$-delayed proton emission measurements at Lanzhou (418(8)~keV) \cite{liangSimultaneousMeasurementDelayed2020} and Michigan State University (412(2)~keV) \cite{bennettClassicalNovaContributionMilky2013}. The resonance strength can be determined from the measured proton width by combining (d,n) proton transfer reaction measurements in inverse kinematics at Florida State University \cite{peplowskiLowestProtonResonance2009} and the ratio of proton to $\gamma$-width obtained from the proton and $\gamma$-intensity ratio of the 5929~keV state in $^{26}$Si measured by $\beta$-delayed proton emission. For the proton intensity we adopt the value from \cite{liangSimultaneousMeasurementDelayed2020}, who confirm the lower proton intensity measured earlier by \cite{janiakEnsuremathBetaDelayed2017}, which is significantly lower than the value reported by \cite{thomasBetadecayProperties25Si2004} from a measurement in GANIL. For the $\gamma$-intensity we use the result from \cite{bennettClassicalNovaContributionMilky2013} measured at NSCL at Michigan State University. The resulting resonance strength is 32~meV, which is significantly larger than the values reported in the review of \cite{PhysRevC.93.035801} owing to the lower proton  branching. For the other three resonances we adopt the shell model calculated resonance strengths from \cite{liangSimultaneousMeasurementDelayed2020}. Their contribution is mostly negligible, with the 432~keV 4$^+$ resonance contributing up to 10\% at the highest relevant temperature of 2~GK.  The numerical reaction rates are given in Table \ref{rate:set5}.

\section*{Appendix $^{26}$Al(p,$\gamma$)$^{27}$Si}
The ground and isomeric state reaction rates of $^{26}$Al$_{g.s.}$(p,$\gamma$)$^{27}$Si and $^{26}$Al$_{\rm m}$(p,$\gamma$)$^{27}$Si are respectively evaluated. These two sets of rates are used to compute the total thermalized $^{26}$Al(p,$\gamma$)$^{27}$Si reaction rate, considering the thermal populations of the ground state and isomeric state of $^{26}$Al (Eq. (4) in Ref. \cite{PhysRevC.72.065804}).

\begin{figure}[tbp]
\includegraphics[width=0.64\textwidth]{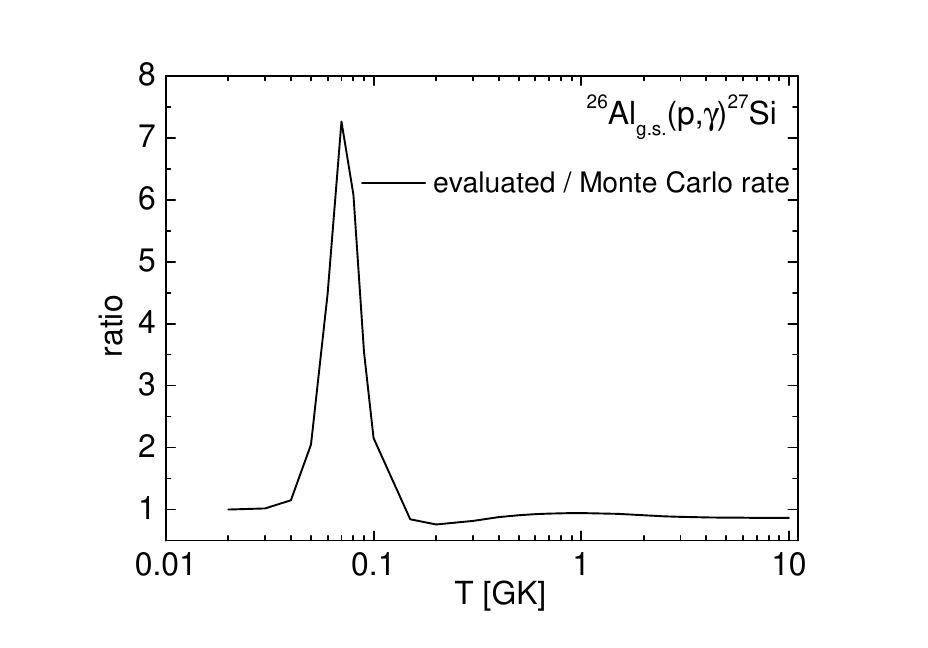}
\caption{The ratio of the new calculated reaction rate to the Monte Carlo reaction rate for $^{26}$Al$_{g.s.}$(p,$\gamma$)$^{27}$Si.
\label{fig1:Al26pg}}
\end{figure}

For the $^{26}$Al$_{g.s.}$(p,$\gamma$)$^{27}$Si rate, \cite{Lotay2020, margerinInverseKinematicStudy2015} recently performed (d,p) neutron transfer measurements on $^{26}$Al$_{g.s.}$ using a 6~MeV/A $^{26}$Al beam at TRIUMF detecting reaction protons with the TUDA array, in order to determine spectroscopic factors for resonances in the $^{27}$Al isospin mirror. They extracted spectroscopic factors using a global Koning-Delaroche optical model parameterization \cite{KONING2003231} for three resonances at $E_r=$127, 189, and 276~keV. For the 127 and 276~keV resonances we adopt their resulting resonance strengths that also show excellent agreement with shell model calculations. For the 189~keV resonance, there are, in addition two direct measurements of the resonance strength available. \cite{caltechthesis7945} used a proton beam impinging on a $^{26}$Al radioactive target and obtained $\omega \gamma=$55(9)~$\mu$eV, however the result was not published in a refereed journal. More recently, \cite{ruizMeasurement184KeV2006} obtained a significantly lower $\omega \gamma=$33(7)~$\mu$eV using a $^{26}$Al radioactive beam and the DRAGON recoil separator at TRIUMF. The indirect (d,p) measurement from \cite{Lotay2020} reports an even slightly lower $\omega \gamma=$26(7)~$\mu$eV. Given that \cite{ruizMeasurement184KeV2006} and \cite{Lotay2020} agree within error bars and point to a significantly lower resonance strength than \cite{caltechthesis7945}, we chose a weighted average of \cite{ruizMeasurement184KeV2006} and \cite{Lotay2020} only. However, to account for possible underestimated systematic errors in the indirect approach, for example from the use of a global optical potential, and the assumption of mirror symmetry, we doubled their reported error of 25\%, resulting in a weighted average of $\omega \gamma=$33~$\mu$eV. For all other resonances we use the resonance strengths recommended in \cite{iliadisChargedparticleThermonuclearReaction2010a}. For the resonances at 68~keV and 369~keV these strengths are compatible with the upper limits from \cite{Lotay2020}. Figure \ref{fig1:Al26pg} shows the ratio of the present calculation to the compiled Monte Carlo reaction rate \cite{iliadisChargedparticleThermonuclearReaction2010a}. The differences between the two rates are caused by introducing the new resonance parameters of \cite{Lotay2020} in the present evaluation.

For $^{26}$Al$_{\rm m}$(p,$\gamma$)$^{27}$Si we take into account the resonances and upper limits for resonance strengths in Table II in \cite{PhysRevLett.126.042701}. These limits were obtained using a $^{26}$Si(d,p)$^{27}$Si neutron transfer reaction at NSCL at Michigan State University, taking advantage of the fact that the $^{26}$Si ground state is part of the same isospin triplet as $^{26}$Al$_{\rm isomer}$. These limits are compatible with, but much more stringent than the limits obtained from $^{26}$Al$_{\rm m}$(d,p)$^{27}$Al measurements with a $^{26}$Al$_{\rm isomer}$ beam at Argonne National Laboratory \cite{almaraz-calderonStudy26272017}. The rate obtained with these upper limits is then compared with the NACRE evaluation for each temperature, and the smaller value is chosen as the recommended rate for $^{26}$Al$_{\rm m}$(p,$\gamma$)$^{27}$Si.

The total thermalized $^{26}$Al(p,$\gamma$)$^{27}$Si rate is finally calculated by adding the rates for $^{26}$Al$_{g.s.}$(p,$\gamma$)$^{27}$Si and $^{26}$Al$_{\rm m}$(p,$\gamma$)$^{27}$Si weighted with the Boltzmann population of the respective levels as a function of temperature. The numerical results are given in Table \ref{rate:set6}.

\begin{table}
\caption{Reaction rates of $^{26}$Al$_{g.s.}$(p,$\gamma$)$^{27}$Si, $^{26}$Al$_{isomer}$(p,$\gamma$)$^{27}$Si, and $^{26}$Al$_{total}$(p,$\gamma$)$^{27}$Si. \label{rate:set6}}
\scalebox{0.89}{
\begin{tabular}{c c c c}
\hline \hline
& $^{26}$Al$_{g.s.}$(p,$\gamma$)$^{27}$Si & $^{26}$Al$_{isomer}$(p,$\gamma$)$^{27}$Si & $^{26}$Al$_{total}$(p,$\gamma$)$^{27}$Si \\
$T$ & $N_{A}<\sigma\upsilon>$ & $N_{A}<\sigma\upsilon>$ & $N_{A}<\sigma\upsilon>$ \\
$GK$ & $cm^3/(mol*sec)$ & $cm^3/(mol*sec)$ & $cm^3/(mol*sec)$ \\
\hline
 0.015   &   0.5228E-29   &   $-$           &   0.5228E-29  \\
 0.020   &   0.4755E-25   &   0.1177E-36    &   0.4755E-25  \\
 0.030   &   0.1514E-18   &   0.2281E-24    &   0.1514E-18  \\
 0.040   &   0.8435E-15   &   0.2226E-18    &   0.8435E-15  \\
 0.050   &   0.2546E-12   &   0.7313E-15    &   0.2546E-12  \\
 0.060   &   0.1776E-10   &   0.1563E-12    &   0.1776E-10  \\
 0.070   &   0.4176E-09   &   0.7366E-11    &   0.4176E-09  \\
 0.080   &   0.4719E-08   &   0.1401E-09    &   0.4719E-08  \\
 0.090   &   0.3329E-07   &   0.1510E-08    &   0.3329E-07  \\
 0.100   &   0.1737E-06   &   0.1123E-07    &   0.1737E-06  \\
 0.120   &   0.2634E-05   &   0.2944E-06    &   0.2634E-05  \\
 0.140   &   0.2240E-04   &   0.3723E-05    &   0.2240E-04  \\
 0.160   &   0.1236E-03   &   0.2758E-04    &   0.1236E-03  \\
 0.180   &   0.5019E-03   &   0.1384E-03    &   0.5019E-03  \\
 0.200   &   0.1620E-02   &   0.5484E-03    &   0.1650E-02  \\
 0.250   &   0.1819E-01   &   0.6209E-02    &   0.1819E-01  \\
 0.300   &   0.1163E+00   &   0.3729E-01    &   0.1163E+00  \\
 0.350   &   0.4961E+00   &   0.1887E+00    &   0.4960E+00  \\
 0.400   &   0.1547E+01   &   0.9204E+00    &   0.1547E+01  \\
 0.450   &   0.3808E+01   &   0.3095E+01    &   0.3806E+01  \\
 0.500   &   0.7846E+01   &   0.8467E+01    &   0.7838E+01  \\
 0.550   &   0.1414E+02   &   0.1858E+02    &   0.1411E+02  \\
 0.600   &   0.2300E+02   &   0.3445E+02    &   0.2290E+02  \\
 0.650   &   0.3457E+02   &   0.5625E+02    &   0.3432E+02  \\
 0.700   &   0.4884E+02   &   0.9139E+02    &   0.4831E+02  \\
 0.750   &   0.6567E+02   &   0.1147E+03    &   0.6476E+02  \\
 0.800   &   0.8491E+02   &   0.1490E+03    &   0.8351E+02  \\
 0.850   &   0.1063E+03   &   0.1849E+03    &   0.1044E+03  \\
 0.900   &   0.1298E+03   &   0.2067E+03    &   0.1272E+03  \\
 0.950   &   0.1550E+03   &   0.2570E+03    &   0.1517E+03  \\
 1.000   &   0.1818E+03   &   0.2681E+03    &   0.1777E+03  \\
 1.200   &   0.3030E+03   &   0.4096E+03    &   0.2937E+03  \\
 1.400   &   0.4410E+03   &   0.4911E+03    &   0.4233E+03  \\
 1.600   &   0.5901E+03   &   0.5424E+03    &   0.5653E+03  \\
 1.800   &   0.7444E+03   &   0.5724E+03    &   0.7230E+03  \\
 2.000   &   0.8983E+03   &   0.6237E+03    &   0.8997E+03  \\
 2.500   &   0.1253E+04   &   0.6507E+03    &   0.1464E+04  \\
 3.000   &   0.1537E+04   &   0.6360E+03    &   0.2031E+04  \\
 4.000   &   0.1877E+04   &   0.5661E+03    &   0.3072E+04  \\
 5.000   &   0.1995E+04   &   0.4893E+03    &   0.3957E+04  \\
 6.000   &   0.1988E+04   &   0.4219E+03    &   0.4757E+04  \\
 7.000   &   0.1919E+04   &   0.3663E+03    &   0.5324E+04  \\
 8.000   &   0.1821E+04   &   0.3207E+03    &   0.6086E+04  \\
 9.000   &   0.1715E+04   &   0.2806E+03    &   0.6592E+04  \\
10.000   &   0.1607E+04   &   0.2522E+03    &   0.7065E+04  \\
\hline
\end{tabular}
}
\end{table}

\section*{Appendix $^{26}$Si(p,$\gamma$)$^{27}$P}

The reaction rate is dominated by a single resonance at $E_r=$318~keV corresponding to the first excited 3/2$^+$ state at 1125~keV in $^{27}$P. The resonance energy had recently been revised upwards by 64~keV based on measurements of the proton energies in a $^{27}$S $\beta$-delayed proton $\gamma$ spectroscopy experiment at Lanzhou \cite{ribllcollaborationEnsuremathBetaDecay2019}. 
Combining this result with the well known $^{26}$Si mass and the known excitation energy of the first excited state in $^{27}$P from $\gamma$-ray spectroscopy leads to a revised $^{27}$P mass, that has been adopted in AME20 and deviates significantly from previous estimates, including estimates using isobaric multiplets. 
The new resonance energy agrees well with the Coulomb dissociation experiment performed at RIKEN \cite{toganoResonanceStates272011} (315(17)~keV). However, it disagrees with the Coulomb dissociation experiment of \cite{r3bcollaborationCoulombDissociation272016} at R3B, GSI (267(20)~keV) which would be more in line with the previous estimate.
We adopt in this work the resonance strength of $\omega \gamma=$2.9~meV from \cite{ribllcollaborationEnsuremathBetaDecay2019}, who derive it using their measured $\gamma$-to-proton branching ratio, and the $\gamma$-width estimated by \cite{iliadisChargedparticleThermonuclearReaction2010b}. This resonance strength is about a factor of 2 smaller than the strength derived based on a recent (d,p) transfer reaction measurement at the Czech Academy of Science \cite{dagata26MathrmSi2021}. In that experiment, ANC and proton width of the isospin mirror are extracted from data, and combined with the same $\gamma$-to-proton branching ratio from \cite{ribllcollaborationEnsuremathBetaDecay2019}, resulting in a correspondingly larger $\gamma$-width.
However, the here adopted resonance strength is in reasonable agreement with previous Coulomb Dissociation experiments that obtain $\omega \gamma$ = 0.7 meV \cite{r3bcollaborationCoulombDissociation272016} and $\omega \gamma$ = 2 meV \cite{toganoResonanceStates272011}. Direct capture starts to dominate at temperatures below 0.1~GK, where the rate is too slow to matter in X-ray bursts. Nevertheless, we include the direct capture component obtained from \cite{dagata26MathrmSi2021}. For completeness we include two higher lying resonances corresponding to the second and third excited 5/2$^+$ states in $^{27}$P. The resonance energies of $E_r=762$~keV and 1054~keV have been measured using $\beta$-delayed protons \cite{ribllcollaborationEnsuremathBetaDecay2019}. For the resonance strengths, we adopt the average of all previous measurements provided by \cite{ribllcollaborationEnsuremathBetaDecay2019} $\omega \gamma=$0.65~meV, and an average of the Coulomb Dissociation experiments \cite{r3bcollaborationCoulombDissociation272016,toganoResonanceStates272011} $\omega \gamma=$1~meV, respectively. The $^{26}$Si(p,$\gamma$)$^{27}$P reaction rates corrected by the SEF are given in Table \ref{rate:set7}.

\begin{table}
\caption{Reaction rates of $^{26}$Si(p,$\gamma$)$^{27}$P, $^{29}$Si(p,$\gamma$)$^{30}$P, $^{29}$P(p,$\gamma$)$^{30}$S, and $^{30}$P(p,$\gamma$)$^{31}$S. \label{rate:set7}}
\scalebox{0.89}{
\begin{tabular}{c c c c c}
\hline \hline
& $^{26}$Si(p,$\gamma$)$^{27}$P & $^{29}$Si(p,$\gamma$)$^{30}$P & $^{29}$P(p,$\gamma$)$^{30}$S & $^{30}$P(p,$\gamma$)$^{31}$S \\
$T$ & $N_{A}<\sigma\upsilon>$ & $N_{A}<\sigma\upsilon>$ & $N_{A}<\sigma\upsilon>$ & $N_{A}<\sigma\upsilon>$ \\
$GK$ & $cm^3/(mol*sec)$ & $cm^3/(mol*sec)$ & $cm^3/(mol*sec)$ & $cm^3/(mol*sec)$  \\
\hline
0.020   &   0.1799E-28  & 0.6866E-29   &  $-$         &  0.3364E-31   \\
0.030   &   0.1170E-23  & 0.3665E-20   &  $-$         &  0.1017E-21   \\
0.040   &   0.1260E-20  & 0.7674E-16   &  0.5720E-33  &  0.4939E-17   \\
0.050   &   0.1796E-18  & 0.2904E-13   &  0.6943E-26  &  0.2975E-14   \\
0.060   &   0.7863E-17  & 0.1504E-11   &  0.3487E-21  &  0.2022E-12   \\
0.070   &   0.1607E-15  & 0.2513E-10   &  0.7669E-18  &  0.3996E-11   \\
0.080   &   0.2120E-14  & 0.2070E-09   &  0.2397E-15  &  0.3727E-10   \\
0.090   &   0.4317E-13  & 0.1063E-08   &  0.2047E-13  &  0.2213E-09   \\
0.100   &   0.1511E-11  & 0.3947E-08   &  0.7115E-12  &  0.1041E-08   \\
0.120   &   0.5067E-09  & 0.3422E-07   &  0.1381E-09  &  0.1868E-07   \\
0.140   &   0.3243E-07  & 0.4845E-06   &  0.5776E-08  &  0.2554E-06   \\
0.160   &   0.7156E-06  & 0.8176E-05   &  0.9269E-07  &  0.2242E-05   \\
0.180   &   0.7778E-05  & 0.8486E-04   &  0.7909E-06  &  0.1299E-04   \\
0.200   &   0.5159E-04  & 0.5618E-03   &  0.4406E-05  &  0.5404E-04   \\
0.250   &   0.1479E-02  & 0.1718E-01   &  0.1016E-03  &  0.7097E-03   \\
0.300   &   0.1317E-01  & 0.1721E+00   &  0.8397E-03  &  0.3908E-02   \\
0.350   &   0.6057E-01  & 0.9092E+00   &  0.4238E-02  &  0.1340E-01   \\
0.400   &   0.1852E+00  & 0.3200E+01   &  0.1520E-01  &  0.3627E-01   \\
0.450   &   0.4326E+00  & 0.8536E+01   &  0.4055E-01  &  0.8980E-01   \\
0.500   &   0.8387E+00  & 0.1869E+02   &  0.9620E-01  &  0.2153E+00   \\
0.550   &   0.1422E+01  & 0.3535E+02   &  0.1827E+00  &  0.4939E+00   \\
0.600   &   0.2183E+01  & 0.5983E+02   &  0.3335E+00  &  0.1057E+01   \\
0.650   &   0.3108E+01  & 0.9293E+02   &  0.5216E+00  &  0.2085E+01   \\
0.700   &   0.4173E+01  & 0.1349E+03   &  0.8081E+00  &  0.3796E+01   \\
0.750   &   0.5349E+01  & 0.1853E+03   &  0.1126E+01  &  0.6431E+01   \\
0.800   &   0.6606E+01  & 0.2436E+03   &  0.1542E+01  &  0.1025E+02   \\
0.850   &   0.7916E+01  & 0.3088E+03   &  0.2037E+01  &  0.1555E+02   \\
0.900   &   0.9253E+01  & 0.3798E+03   &  0.2616E+01  &  0.2266E+02   \\
0.950   &   0.1060E+02  & 0.4554E+03   &  0.3279E+01  &  0.3197E+02   \\
1.000   &   0.1193E+02  & 0.5344E+03   &  0.4031E+01  &  0.4391E+02   \\
1.200   &   0.1690E+02  & 0.8639E+03   &  0.7945E+01  &  0.1259E+03   \\
1.400   &   0.2108E+02  & 0.1178E+04   &  0.1335E+02  &  0.2736E+03   \\
1.600   &   0.2453E+02  & 0.1453E+04   &  0.2017E+02  &  0.4874E+03   \\
1.800   &   0.2755E+02  & 0.1693E+04   &  0.2817E+02  &  0.7582E+03   \\
2.000   &   0.3049E+02  & 0.1841E+04   &  0.3708E+02  &  0.1086E+04   \\
2.500   &   0.4010E+02  & 0.2537E+04   &  0.6154E+02  &  0.1980E+04   \\
3.000   &   0.5766E+02  & 0.3210E+04   &  0.8661E+02  &  0.2777E+04   \\
4.000   &   0.1141E+03  & 0.4633E+04   &  0.1298E+03  &  0.3692E+04   \\
5.000   &   0.3078E+03  & 0.5914E+04   &  0.1579E+03  &  0.3916E+04   \\
6.000   &   0.5899E+03  & 0.6938E+04   &  0.1758E+03  &  0.3921E+04   \\
7.000   &   0.9931E+03  & 0.7648E+04   &  0.1904E+03  &  0.3941E+04   \\
8.000   &   0.1514E+04  & 0.8175E+04   &  0.2054E+03  &  0.4266E+04   \\
9.000   &   0.2139E+04  & 0.8483E+04   &  0.2182E+03  &  0.4878E+04   \\
10.000  &   0.2869E+04  & 0.8667E+04   &  0.2195E+03  &  0.6033E+04   \\
\hline
\end{tabular}
}
\end{table}

\section*{Appendix $^{29}$Si(p,$\gamma$)$^{30}$P}

The reaction rate is derived from 13 resonances below $E_r$ = 500 keV compiled in \cite{lotaySpectroscopy30Mathrm2020} who updated with new data from a $\gamma$-spectroscopy experiment, as well as 73 resonances above $E_r$ = 500 keV using the resonance properties from \cite{iliadisChargedparticleThermonuclearReaction2010b}. The calculated reaction rates, corrected with an SEF, are given in Table \ref{rate:set7}.

\section*{Appendix $^{29}$P(p,$\gamma$)$^{30}$S}
The rate provided in \cite{almaraz-calderonErratumLevelStructure2013} is adopted here, which is largely based on their earlier compilation 
\cite{almaraz-calderonLevelStructure302012} that summarizes previous work and new information on $^{30}$S levels from $^{28}$Si($^3$He,n) and $^{32}$S(p,t) transfer reaction measurements performed at the University of Notre Dame and RCNP at Osaka University, respectively. \cite{almaraz-calderonErratumLevelStructure2013} does include an additional resonance at $E_r$ = 820 keV ($3^+$ state) identified in a ($^3$He,n) $\gamma$-ray spectroscopy experiment at Argonne National Laboratory using the Gammasphere detector array \cite{lotayLevelStructure302012} and uses an updated Q-value. \cite{lotayLevelStructure302012} confirmed spins and assignments of the important $^{30}$S states to states in the $^{30}$Si isospin mirror, the latter being used to obtain the proton spectroscopic factors. Based on the current information on 25 states above the proton threshold in $^{30}$S, the rate at X-ray burst temperatures is dominated by a $E_r$ = 410 keV 2$^+$ resonance, with a few additional resonances coming into play above around 1.5~GK. For temperatures below 0.3~GK the $E_r$ = 289 keV 3$^+$ resonance becomes dominant. An SEF is applied and the results are given in Table \ref{rate:set7}.

\section*{Appendix $^{30}$P(p,$\gamma$)$^{31}$S}

Ten resonances are considered for this reaction rate. We follow largely the compilation of \cite{kankainenMeasurementKeyResonance2017}, with updates from \cite{budnerConstraining30Mathrm2022}, and information for additional states from shell model calculations \cite{brownShellmodelStudiesAstrophysical2014}. For the resonances at $E_r=$196~keV and 261.9~keV we use the resonance strengths derived from (d,n$\gamma$) transfer reaction measurements at NSCL at Michigan State University \cite{kankainenMeasurementKeyResonance2017}. The same experiment also provided upper limits for the resonance strengths of the 
$E_r=$226.7~keV and 411.3~keV resonances, which we adopt as values. We note that for the important 411.3~keV resonance, shell model calculations indicate a resonance strength about a factor of 30 above the experimental upper limit. For the important $E_r=$259.6~keV resonance we use the resonance strengths obtained from $\beta$-delayed proton measurements \cite{budnerConstraining30Mathrm2022}. For the other resonances at $E_r=$116~keV, 124.7~keV, 471~keV, 589~keV, and 618~keV we adopt the resonance strengths from shell model calculations \cite{brownShellmodelStudiesAstrophysical2014}. While these shell model calculations had indicated that the 246~keV 9/2$^-$ resonance made a non-negligible contribution, this resonance is now negligible given the dominant contribution of the $E_r=$259.6~keV resonance that had not been predicted by the shell model. 
The highest energy of these considered resonances is 618 keV, which corresponds to a $T = 0.8$ GK Gamow window. Therefore, for $T > 0.8$ GK we use the slope of the \code{NON-SMOKER} Hauser-Feshbach reaction rate \cite{rauscherTABLESNUCLEARCROSS2001} normalized to the rate at $T = 0.8$ GK. The SEF corrected rate is listed in Table \ref{rate:set7}.

\section*{Appendix $^{33}$Cl(p,$\gamma$)$^{34}$Ar}

The $^{33}$Cl(p,$\gamma$)$^{34}$Ar reaction rate is calculated from the updated resonance parameters given in \cite{PhysRevLett.124.252702}, who performed $\gamma$-spectroscopy of $^{34}$Ar populated by a fusion evaporation reaction at Argonne National Laboratory using the GRETINA detector array. The measurement results in more precise excitation energies for 4 previously known states above the proton threshold, as well as the discovery of a fifth resonance at a resonance energy of 397 keV corresponding to a new state in $^{34}$Ar at 5060.8 keV. \cite{PhysRevLett.124.252702} determined proton spectroscopic factors and $\gamma$-widths for these 5 resonances from mirror nucleus information, complemented with shell model calculations. The new 397 keV resonance is found to dominate the reaction rate above 0.2 GK. The previously recommended \code{NON-SMOKER} Hauser-Feshbach rate \cite{rauscherTABLESNUCLEARCROSS2001} agrees well with the new resonant rate at 0.8~GK and is used for temperatures above 0.8~GK to account for additional higher lying resonances. The calculated reaction rates are given in Table \ref{rate:set8}.

\begin{table}
\caption{Reaction rates of $^{33}$Cl(p,$\gamma$)$^{34}$Ar, $^{34}$Cl$_{g.s.}$(p,$\gamma$)$^{35}$Ar, $^{34}$Cl$_{\rm isomer}$(p,$\gamma$)$^{35}$Ar, and $^{34}$Cl(p,$\gamma$)$^{35}$Ar. \label{rate:set8}}
\scalebox{0.89}{
\begin{tabular}{c c c c c}
\hline \hline
& $^{33}$Cl(p,$\gamma$)$^{34}$Ar & $^{34}$Cl$_{g.s.}$(p,$\gamma$)$^{35}$Ar & $^{34}$Cl$_{\rm isomer}$(p,$\gamma$)$^{35}$Ar & $^{34}$Cl(p,$\gamma$)$^{35}$Ar \\
$T$ & $N_{A}<\sigma\upsilon>$ & $N_{A}<\sigma\upsilon>$ & $N_{A}<\sigma\upsilon>$ & $N_{A}<\sigma\upsilon>$ \\
$GK$ & $cm^3/(mol*sec)$ & $cm^3/(mol*sec)$ & $cm^3/(mol*sec)$ & $cm^3/(mol*sec)$ \\
\hline
 0.020  &  $-$          &  0.8546E-28  &   0.3579E-28  &  0.8546E-28  \\
 0.030  &  0.4559E-33   &  0.1341E-22  &   0.1839E-22  &  0.1341E-22  \\
 0.040  &  0.2501E-25   &  0.4675E-20  &   0.1607E-19  &  0.4675E-20  \\
 0.050  &  0.1292E-20   &  0.1453E-18  &   0.1658E-17  &  0.1453E-18  \\
 0.060  &  0.2290E-17   &  0.1375E-17  &   0.5269E-16  &  0.1375E-17  \\
 0.070  &  0.5558E-15   &  0.1295E-16  &   0.1237E-14  &  0.1295E-16  \\
 0.080  &  0.3611E-13   &  0.8798E-15  &   0.3384E-13  &  0.8798E-15  \\
 0.090  &  0.9409E-12   &  0.3801E-13  &   0.6140E-12  &  0.3801E-13  \\
 0.100  &  0.1276E-10   &  0.7731E-12  &   0.6630E-11  &  0.7731E-12  \\
 0.120  &  0.6281E-09   &  0.6851E-10  &   0.2488E-09  &  0.6851E-10  \\
 0.140  &  0.1005E-07   &  0.1628E-08  &   0.3950E-08  &  0.1628E-08  \\
 0.160  &  0.8258E-07   &  0.1718E-07  &   0.4375E-07  &  0.1718E-07  \\
 0.180  &  0.4701E-06   &  0.1080E-06  &   0.3760E-06  &  0.1081E-06  \\
 0.200  &  0.2268E-05   &  0.4973E-06  &   0.2415E-05  &  0.4999E-06  \\
 0.250  &  0.7387E-04   &  0.1364E-04  &   0.8417E-04  &  0.1423E-04  \\
 0.300  &  0.1031E-02   &  0.2385E-03  &   0.1069E-02  &  0.2673E-03  \\
 0.350  &  0.7017E-02   &  0.2293E-02  &   0.7572E-02  &  0.2582E-02  \\
 0.400  &  0.2926E-01   &  0.1349E-01  &   0.3602E-01  &  0.1552E-01  \\
 0.450  &  0.8745E-01   &  0.5553E-01  &   0.1274E+00  &  0.6484E-01  \\
 0.500  &  0.2069E+00   &  0.1760E+00  &   0.3594E+00  &  0.2111E+00  \\
 0.550  &  0.4133E+00   &  0.4580E+00  &   0.8522E+00  &  0.5437E+00  \\
 0.600  &  0.7276E+00   &  0.1024E+01  &   0.1765E+01  &  0.1241E+01  \\
 0.650  &  0.1164E+01   &  0.2030E+01  &   0.3284E+01  &  0.2422E+01  \\
 0.700  &  0.1729E+01   &  0.3661E+01  &   0.5609E+01  &  0.4364E+01  \\
 0.750  &  0.2444E+01   &  0.6111E+01  &   0.8935E+01  &  0.7266E+01  \\
 0.800  &  0.3212E+01   &  0.9576E+01  &   0.1344E+02  &  0.1134E+02  \\
 0.850  &  0.4790E+01   &  0.1424E+02  &   0.1927E+02  &  0.1676E+02  \\
 0.900  &  0.7313E+01   &  0.2027E+02  &   0.2653E+02  &  0.2370E+02  \\
 0.950  &  0.1032E+02   &  0.2781E+02  &   0.3530E+02  &  0.3224E+02  \\
 1.000  &  0.1447E+02   &  0.3697E+02  &   0.4561E+02  &  0.4184E+02  \\
 1.200  &  0.4971E+02   &  0.9110E+02  &   0.1016E+03  &  0.1000E+03  \\
 1.400  &  0.1033E+03   &  0.1732E+03  &   0.1769E+03  &  0.1807E+03  \\
 1.600  &  0.1712E+03   &  0.2796E+03  &   0.2637E+03  &  0.2766E+03  \\
 1.800  &  0.2545E+03   &  0.4040E+03  &   0.3543E+03  &  0.3795E+03  \\
 2.000  &  0.3810E+03   &  0.5397E+03  &   0.4430E+03  &  0.4672E+03  \\
 2.500  &  0.7065E+03   &  0.1022E+04  &   0.6347E+03  &  0.7132E+03  \\
 3.000  &  0.1027E+04   &  0.1476E+04  &   0.7713E+03  &  0.9129E+03  \\
 4.000  &  0.1425E+04   &  0.2064E+04  &   0.9050E+03  &  0.1113E+04  \\
 5.000  &  0.1605E+04   &  0.2327E+04  &   0.9262E+03  &  0.1147E+04  \\
 6.000  &  0.1736E+04   &  0.2518E+04  &   0.8955E+03  &  0.1154E+04  \\
 7.000  &  0.1889E+04   &  0.2833E+04  &   0.8439E+03  &  0.1151E+04  \\
 8.000  &  0.2211E+04   &  0.3461E+04  &   0.7860E+03  &  0.1187E+04  \\
 9.000  &  0.2648E+04   &  0.4697E+04  &   0.7285E+03  &  0.1312E+04  \\
10.000  &  0.3517E+04   &  0.7179E+04  &   0.6743E+03  &  0.1616E+04  \\
\hline
\end{tabular}
}
\end{table}

\section*{Appendix $^{34}$Cl(p,$\gamma$)$^{35}$Ar}

$^{34}$Cl has a low lying excited state at 146 keV, which is a long-lived isomer under terrestrial conditions with a half-life of 32 min. The reaction rates on ground ($^{34}$Cl$_{g.s.}$(p,$\gamma$)$^{35}$Ar) and isomeric ($^{34}$Cl$_{\rm m}$(p,$\gamma$)$^{35}$Ar) states are therefore evaluated separately, and in addition, a thermalized $^{34}$Cl(p,$\gamma$)$^{35}$Ar reaction rate is calculated by adding the $^{34}$Cl$_{g.s.}$(p,$\gamma$)$^{35}$Ar and $^{34}$Cl$_{\rm m}$(p,$\gamma$)$^{35}$Ar rates weighted by their respective thermal population (Eq. (4) in \cite{PhysRevC.72.065804}). According to calculation \code{TALYS}, the contributions from the higher lying thermally populated excited states in $^{34}$Cl are negligible.

As experimental information on the relevant states in $^{35}$Ar remains too sparse for a reaction rate calculation, we adopt the shell model calculations from \cite{PhysRevC.102.025801}. They compare calculations with spectroscopic factors of states in the $^{35}$Cl mirror nucleus obtained from ($^{3}$He,d) transfer reaction measurements at Garching \cite{gillespieFirstMeasurement342017} and ($\alpha$,p) reaction studies at the TUNL Enge split-pole spectrometer \cite{setoodehniaExperimentalStudyCl2019}, as well as information on excited states in $^{35}$Ar from $^{36}$Ar(d,t)$^{35}$Ar reaction measurements at Garching \cite{fryDiscovery34MCl2015} and find the impact of differences on the final rate to be within the quoted rate uncertainties obtained from a Monte Carlo approach. We calculate the $^{34}$Cl$_{g.s.}$(p,$\gamma$)$^{35}$Ar reaction rate with the parameters of the resonant states listed in Table II in \cite{PhysRevC.102.025801}, and the  $^{34}$Cl$_{\rm m}$(p,$\gamma$)$^{35}$Ar reaction rate with the parameters of the resonant states listed in Table III in \cite{PhysRevC.102.025801}.

Note that the highest energy of the listed resonances is E$_r$ = 1.16 MeV, corresponding to the upper end of the Gamow window at T $\simeq$ 1.8 GK. Therefore, above T = 1.8 GK the \code{NON-SMOKER} Hauser-Feshbach reaction rates \cite{rauscherTABLESNUCLEARCROSS2001}, normalized to the calculation result at T = 1.8 GK for the ground state rate, are used. All the numerical reaction rates are given in Table \ref{rate:set8}.

\cite{mischAstromersNuclearIsomers2020a} calculated transition rates and concluded that the $^{34}$Cl isomeric state is in thermal equilibrium with the ground state for temperatures above 0.23~GK, which is in the range of typical burst ignition temperatures. In addition, for a rate $N_{A}<\sigma\upsilon>$ to matter, it must be faster than the typical rp-process time scale $\tau$=10~s for typical densities $\rho$=10$^5$~g/cm$^3$ and hydrogen mass fractions $X=0.7$. This translates into a requirement of  $N_{A}<\sigma\upsilon>$ $>$ (Y$\rho$$\tau$)$^{-1}$ = 1.4$\times$10$^{-6}$, which requires temperatures higher than about 0.23~GK for the rate to matter in X-ray bursts (see Table~\ref{rate:set9}). We therefore use in our burst calculation the thermalized reaction rate, though a non thermal population may be possible near ignition and should be explored in future burst calculations.

\section*{Appendix $^{36}$Cl(p,$\alpha$)$^{33}$S}
The rate of the positive Q-value reaction  $^{36}$Cl(p,$\alpha$)$^{33}$S  is evaluated considering experimental data for its reverse reaction $^{33}$S($\alpha$,p)$^{36}$Cl \cite{andersonRemeasurement33Cl2017}. The data cover the 0.8-1.5~MeV/nucleon energy range and were obtained from direct inverse kinematics measurements performed at the University of Notre Dame using an activation technique. The $^{36}$Cl(p,$\alpha$)$^{33}$S cross sections are obtained from the experimental $^{33}$S($\alpha$,p)$^{36}$Cl cross sections using the detailed balance theorem, taking only the ground state contribution into account. \code{TALYS} calculations are then used to reproduce the deduced $^{36}$Cl(p,$\alpha$)$^{33}$S ground state cross section by adjusting the parameters of proton and $\alpha$ optical model potentials. The resulting optical model potentials are then used to compute the astrophysical reaction rate of $^{36}$Cl(p,$\alpha$)$^{33}$S, and the results are given in Table \ref{rate:set9}.

\begin{table}
\caption{Reaction rates of $^{36}$Cl(p,$\alpha$)$^{33}$S, $^{35}$K(p,$\gamma$)$^{36}$Ca and $^{42}$Ti(p,$\gamma$)$^{43}$V. \label{rate:set9}}
\scalebox{0.89}{
\begin{tabular}{c c c c}
\hline \hline
& $^{36}$Cl(p,$\alpha$)$^{33}$S & $^{35}$K(p,$\gamma$)$^{36}$Ca & $^{42}$Ti(p,$\gamma$)$^{43}$V \\
$T$ & $N_{A}<\sigma\upsilon>$ & $N_{A}<\sigma\upsilon>$ & $N_{A}<\sigma\upsilon>$ \\
$GK$ & $cm^3/(mol*sec)$ & $cm^3/(mol*sec)$ & $cm^3/(mol*sec)$ \\
\hline
 0.030  &   0.6075E-28 &  0.4801E-32   &   $-$         \\
 0.040  &   0.6971E-23 &  0.2969E-28   &   0.8337E-33  \\
 0.050  &   0.1187E-19 &  0.1585E-25   &   0.9961E-30  \\
 0.060  &   0.2419E-17 &  0.1719E-23   &   0.4712E-27  \\
 0.070  &   0.1364E-15 &  0.7170E-22   &   0.3346E-24  \\
 0.080  &   0.3390E-14 &  0.1604E-20   &   0.5285E-22  \\
 0.090  &   0.4784E-13 &  0.2152E-19   &   0.2699E-20  \\
 0.100  &   0.4962E-12 &  0.2274E-18   &   0.7116E-19  \\
 0.120  &   0.1669E-10 &  0.1070E-15   &   0.1391E-16  \\
 0.140  &   0.2860E-09 &  0.3663E-13   &   0.9808E-15  \\
 0.160  &   0.2916E-08 &  0.3002E-11   &   0.3544E-13  \\
 0.180  &   0.2055E-07 &  0.9075E-10   &   0.7885E-12  \\
 0.200  &   0.1024E-06 &  0.1365E-08   &   0.1092E-10  \\
 0.250  &   0.3078E-05 &  0.1708E-06   &   0.1658E-08  \\
 0.300  &   0.4130E-04 &  0.4063E-05   &   0.4977E-07  \\
 0.350  &   0.3237E-03 &  0.3771E-04   &   0.5214E-06  \\
 0.400  &   0.1860E-02 &  0.1952E-03   &   0.3067E-05  \\
 0.450  &   0.7553E-02 &  0.6868E-03   &   0.1165E-04  \\
 0.500  &   0.2771E-01 &  0.1848E-02   &   0.3418E-04  \\
 0.550  &   0.7847E-01 &  0.4096E-02   &   0.8014E-04  \\
 0.600  &   0.2161E+00 &  0.7861E-02   &   0.1639E-03  \\
 0.650  &   0.4915E+00 &  0.1352E-01   &   0.2993E-03  \\
 0.700  &   0.1105E+01 &  0.2133E-01   &   0.4901E-03  \\
 0.750  &   0.2183E+01 &  0.3145E-01   &   0.7736E-03  \\
 0.800  &   0.4218E+01 &  0.4388E-01   &   0.1120E-02  \\
 0.850  &   0.7578E+01 &  0.5856E-01   &   0.1573E-02  \\
 0.900  &   0.1300E+02 &  0.7530E-01   &   0.2150E-02  \\
 0.950  &   0.2186E+02 &  0.9387E-01   &   0.2748E-02  \\
 1.000  &   0.3412E+02 &  0.1140E+00   &   0.3629E-02  \\
 1.200  &   0.1761E+03 &  0.2050E+00   &   0.8032E-02  \\
 1.400  &   0.6309E+03 &  0.3025E+00   &   0.1683E-01  \\
 1.600  &   0.1791E+04 &  0.4015E+00   &   0.3528E-01  \\
 1.800  &   0.4292E+04 &  0.5079E+00   &   0.7381E-01  \\
 2.000  &   0.8605E+04 &  0.6355E+00   &   0.1537E+00  \\
 2.500  &   0.3926E+05 &  0.1138E+01   &   0.6161E+00  \\
 3.000  &   0.1232E+06 &  0.2035E+01   &   0.1809E+01  \\
 4.000  &   0.6268E+06 &  0.5364E+01   &   0.6665E+01  \\
 5.000  &   0.1905E+07 &  0.1044E+02   &   0.1325E+02  \\
 6.000  &   0.4282E+07 &  0.1742E+02   &   0.1871E+02  \\
 7.000  &   0.7921E+07 &  0.2625E+02   &   0.2259E+02  \\
 8.000  &   0.1281E+08 &  0.3686E+02   &   0.2485E+02  \\
 9.000  &   0.1883E+08 &  0.4958E+02   &   0.2491E+02  \\
10.000  &   0.2573E+08 &  0.6335E+02   &   0.2278E+02  \\
\hline
\end{tabular}
}
\end{table}

\section*{Appendix $^{35}$K(p,$\gamma$)$^{36}$Ca}
The $^{35}$K(p,$\gamma$)$^{36}$Ca reaction rate is dominated by resonant capture through the first excited 2$^+$ state in $^{36}$Ca up to a temperature of around 2~GK. \cite{lalanneEvaluation35K36Ca2021} performed a $^{37}$Ca(p,d)$^{36}$Ca measurement at GANIL using the CRYPTA liquid hydrogen target and determined the proton branching of this state, corresponding to a resonance at 445~keV. Combining the measurement with a shell model calculation for the $\gamma$-partial width results in an updated resonance strength. The numerical results, including the stellar enhancement factor, are given in Table \ref{rate:set9}.

\section*{Appendix $^{42}$Ti(p,$\gamma$)$^{43}$V}
We follow the evaluation of \citet{Hou2023}, who determined the reaction rate from the 8 lowest energy resonances that can be populated by $l=0,1,2,3$ orbital angular momentum proton capture. Resonance energies are determined from the known excitation energies in the $^{43}$Ca mirror nucleus and an updated Q-value based on a recent mass measurement at the HIRFL cooler storage ring in Lanzhou \cite{Yan13}. No Coulomb shifts are applied, instead, a 100~keV uncertainty is assumed. Proton spectroscopic factors are taken from experimental (p,d) stripping reaction data for $^{43}$Ca when available, otherwise the shell model is used. $\gamma$-widths were determined from the known level lifetimes in $^{43}$Ca. A direct capture component is included as well. A SEF is applied but only impacts the rate above 4~GK. The numerical astrophysical reaction rates of $^{42}$Ti(p,$\gamma$)$^{43}$V are given in Table \ref{rate:set9}.

\section*{Appendix $^{55}$Ni(p,$\gamma$)$^{56}$Cu}
Excitation energies of the three lowest energy resonances were determined from in-beam $\gamma$-ray spectroscopy using a radioactive $^{56}$Ni beam impinging on a CD$_2$ target at the NSCL at Michigan State University \cite{ongLowlyingLevelStructure2017}. Resonance states in $^{56}$Cu were populated via the (d,2n) reaction channel. Experimental data on excitation energies were complemented with GXPF1A shell model calculations for $\gamma$- and proton-widths, as well as for resonance properties of higher lying resonances. More recently, the mass of $^{56}$Cu was measured precisely using the LEBIT Penning trap at Michigan State University leading to a revised reaction Q-value of 582(6)~keV, and consequently revised resonance energies, and, using updated shell model calculations, revised proton and $\gamma$-widths \cite{valverdeHighPrecisionMassMeasurement2018}. We adopt the recommended rate from  \cite{valverdeHighPrecisionMassMeasurement2018}, corrected with a SEF, and tabulated in Table \ref{rate:set10}.

\begin{table}
\caption{Reaction rates of $^{55}$Ni(p,$\gamma$)$^{56}$Cu, $^{56}$Ni(p,$\gamma$)$^{57}$Cu, $^{57}$Cu(p,$\gamma$)$^{58}$Zn, and $^{65}$As(p,$\gamma$)$^{66}$Se. \label{rate:set10}}
\scalebox{0.89}{
\begin{tabular}{c c c c c}
\hline \hline
& $^{55}$Ni(p,$\gamma$)$^{56}$Cu & $^{56}$Ni(p,$\gamma$)$^{57}$Cu & $^{57}$Cu(p,$\gamma$)$^{58}$Zn & $^{65}$As(p,$\gamma$)$^{66}$Se \\
$T$ & $N_{A}<\sigma\upsilon>$ & $N_{A}<\sigma\upsilon>$ & $N_{A}<\sigma\upsilon>$ & $N_{A}<\sigma\upsilon>$ \\
$GK$ & $cm^3/(mol*sec)$ & $cm^3/(mol*sec)$ & $cm^3/(mol*sec)$ & $cm^3/(mol*sec)$ \\
\hline
 0.050   &   0.5939E-30   &   $-$          &   0.1687E-35   &   0.6803E-39  \\
 0.060   &   0.5864E-26   &   0.1736E-34   &   0.4033E-30   &   0.1199E-33  \\
 0.070   &   0.4024E-23   &   0.1461E-29   &   0.2720E-26   &   0.8221E-30  \\
 0.080   &   0.5245E-21   &   0.8521E-26   &   0.1979E-23   &   0.6703E-27  \\
 0.090   &   0.2266E-19   &   0.7927E-23   &   0.3273E-21   &   0.1261E-24  \\
 0.100   &   0.4531E-18   &   0.1601E-20   &   0.1921E-19   &   0.8364E-23  \\
 0.120   &   0.3923E-16   &   0.2859E-17   &   0.8387E-17   &   0.4759E-20  \\
 0.140   &   0.1005E-14   &   0.5614E-15   &   0.6257E-15   &   0.5376E-18  \\
 0.160   &   0.1854E-13   &   0.3279E-13   &   0.1554E-13   &   0.2341E-16  \\
 0.180   &   0.3636E-12   &   0.7891E-12   &   0.1867E-12   &   0.5005E-15  \\
 0.200   &   0.5263E-11   &   0.9936E-11   &   0.1370E-11   &   0.5889E-14  \\
 0.250   &   0.7589E-09   &   0.9045E-09   &   0.6781E-10   &   0.5517E-12  \\
 0.300   &   0.2173E-07   &   0.1742E-07   &   0.2099E-08   &   0.1244E-10  \\
 0.350   &   0.2521E-06   &   0.1390E-06   &   0.3627E-07   &   0.2140E-09  \\
 0.400   &   0.1744E-05   &   0.6430E-06   &   0.3311E-06   &   0.3994E-08  \\
 0.450   &   0.7920E-05   &   0.2073E-05   &   0.1866E-05   &   0.4753E-07  \\
 0.500   &   0.2931E-04   &   0.5201E-05   &   0.7452E-05   &   0.3344E-06  \\
 0.550   &   0.8305E-04   &   0.1089E-04   &   0.2325E-04   &   0.1780E-05  \\
 0.600   &   0.2141E-03   &   0.1994E-04   &   0.6078E-04   &   0.6976E-05  \\
 0.650   &   0.4533E-03   &   0.3294E-04   &   0.1400E-03   &   0.2247E-04  \\
 0.700   &   0.9195E-03   &   0.5025E-04   &   0.2945E-03   &   0.6176E-04  \\
 0.750   &   0.1626E-02   &   0.7196E-04   &   0.5789E-03   &   0.1491E-03  \\
 0.800   &   0.2756E-02   &   0.9792E-04   &   0.1079E-02   &   0.3243E-03  \\
 0.850   &   0.4413E-02   &   0.1279E-03   &   0.1920E-02   &   0.6492E-03  \\
 0.900   &   0.6741E-02   &   0.1616E-03   &   0.3279E-02   &   0.1216E-02  \\
 0.950   &   0.9901E-02   &   0.1991E-03   &   0.5382E-02   &   0.2159E-02  \\
 1.000   &   0.1407E-01   &   0.2409E-03   &   0.8509E-02   &   0.3666E-02  \\
 1.200   &   0.4537E-01   &   0.5362E-03   &   0.3826E-01   &   0.2148E-01  \\
 1.400   &   0.1147E+00   &   0.1864E-02   &   0.1139E+00   &   0.8162E-01  \\
 1.600   &   0.2488E+00   &   0.7519E-02   &   0.2559E+00   &   0.2295E+00  \\
 1.800   &   0.4833E+00   &   0.2475E-01   &   0.4742E+00   &   0.5341E+00  \\
 2.000   &   0.8889E+00   &   0.6490E-01   &   0.7679E+00   &   0.1100E+01  \\
 2.500   &   0.2753E+01   &   0.3592E+00   &   0.1765E+01   &   0.4657E+01  \\
 3.000   &   0.6630E+01   &   0.1089E+01   &   0.2954E+01   &   0.1171E+02  \\
 4.000   &   0.2356E+02   &   0.4010E+01   &   0.5124E+01   &   0.3070E+02  \\
 5.000   &   0.6243E+02   &   0.8001E+01   &   0.6455E+01   &   0.4338E+02  \\
 6.000   &   0.1486E+03   &   0.1205E+02   &   0.6970E+01   &   0.4574E+02  \\
 7.000   &   0.3173E+03   &   0.1535E+02   &   0.6940E+01   &   0.4152E+02  \\
 8.000   &   0.5944E+03   &   0.1718E+02   &   0.6607E+01   &   0.3535E+02  \\
 9.000   &   0.1010E+04   &   0.1741E+02   &   0.6134E+01   &   0.2970E+02  \\
10.000   &   0.1558E+04   &   0.1645E+02   &   0.5614E+01   &   0.2580E+02  \\
\hline
\end{tabular}
}
\end{table}

\section*{Appendix $^{56}$Ni(p,$\gamma$)$^{57}$Cu}
For T $\leq$ 2 GK, the reaction rate is dominated by 4 resonances at E$_{\rm r}$ = 338, 418, 1708 and 1835 keV corresponding to the well known 4 first excited states in $^{57}$Cu. Recently, proton spectroscopic factors have been determined experimentally using angle integrated cross sections for (d,n) and (d,p) (populating the isospin mirror) transfer reactions in inverse kinematics performed at NSCL at Michigan State University \cite{kahlSingleparticleShellStrengths2019}. These are important for the two lowest lying resonances, where resonance strengths are dominated by proton widths. We adopt the resonance parameters from \cite{kahlSingleparticleShellStrengths2019}, which combine experimental data and shell model calculations to determine $\gamma$-widths. For T $>$ 2 GK, the slope of the \code{NON-SMOKER} \cite{rauscherTABLESNUCLEARCROSS2001} Hauser-Feshbach reaction rate is used, normalized to the rate at T = 2 GK. A SEF is applied to correct the highest temperature data points above 5~GK. Results are given in Table \ref{rate:set10}.

\section*{Appendix $^{57}$Cu(p,$\gamma$)$^{58}$Zn}
We adopt direct capture S-factors and resonance properties from \cite{PhysRevLett.113.032502}, who determined excitation energies of the 6 lowest lying resonant states in $^{58}$Zn using $\gamma$-ray spectroscopy using the d($^{57}$Cu,$^{58}$Zn)n reaction in inverse kinematics at NSCL at Michigan State University. Resonance energies were determined using the Q-value of 2279(50)~keV given in \cite{PhysRevLett.113.032502}, which agrees within uncertainties with the AME20 Q-value (2270(50)) \cite{wangAME2020Atomic2021}. Spectroscopic factors, proton- and $\gamma$-widths are calculated using fp-shell model calculations with the GXPF1A interaction. The calculated reaction rate, corrected with a SEF, are tabulated in Table \ref{rate:set10}.

\section*{Appendix $^{65}$As(p,$\gamma$)$^{66}$Se}
\cite{Lam_2022} reports a shell model calculation of the $^{65}$As(p,$\gamma$)$^{66}$Se reaction rate below T = 2 GK, which is adopted in the present evaluation. Capture on the ground state and the thermally excited first excited state in $^{65}$As is included. This calculation follows \cite{lamReactionRates64Ge2016} but uses an updated Q-value of $Q=2.469$~MeV that impacts resonance energies and associated proton penetrability. The new Q-value was determined from new $^{65}$As and $^{66}$Se masses that were determined from the experimentally well known masses of the mirror nuclei $^{65}$Ge and $^{66}$Ge, respectively, and Coulomb shift calculations using a relativistic Hartree-Fock Bogoliubov approach with axially symmetric deformation. This approach does not take advantage of the experimentally known $^{65}$As mass \cite{tuDirectMassMeasurements2011a} from a storage ring experiment adopted in AME2020, which is 184~keV larger and has a 85~keV uncertainty. 
Recently, a new measurement of the mass of $^{65}$As and a first measurement of the mass of $^{66}$Se using the same storage ring approach \cite{zhouMassMeasurementsShow2023} indicate the predicted masses are both about 50~keV too low, while the resulting central value for the new experimental Q-value agrees with the prediction within a few keV (though a future mass evaluation would have to consider a weighted average of the $^{65}$As mass measurements). The shell model rate adopted here is therefore, within uncertainties, still in agreement with the new mass measurements. The reaction rate uncertainties are large and estimated to be at least an order of magnitude.  Above T = 2 GK, the Hauser-Feshbach (HF) reaction rate from \code{NON-SMOKER} \cite{rauscherTABLESNUCLEARCROSS2001} is used. Note that the HF reaction rate is normalized to the shell model calculation at T = 2 GK. The adopted reaction rate is given in Table \ref{rate:set10}.

\end{document}